\pdfoutput=1   
\documentclass[12pt, letterpaper, breaklinks=true]{article}

\usepackage{enumerate}
\usepackage{ragged2e}
\usepackage{amssymb,array, amsmath,amsfonts, 
xcolor,graphicx, geometry, float}
\usepackage[T1]{fontenc}
\usepackage[utf8]{inputenc}
\usepackage{xparse}
\usepackage{libertine}
\usepackage{setspace}

\definecolor{darkred}{rgb}{0.65, 0, 0} 
\usepackage[
    colorlinks=true,        
    linkcolor=darkred,      
    citecolor=darkred,      
    urlcolor=darkred,       
    pdfpagelabels=false     
]{hyperref}
\hypersetup{pdfview=FitH}   
\hypersetup{
  pdftitle={Machine Learning Classification and Portfolio Construction: Does the Loss Function Matter?},
  pdfauthor={Yang Bai and Kuntara Pukthuanthong},
  pdfsubject={JEL Classification: C14, C38, C55, G11},
  pdfkeywords={Machine learning, classification, regression, loss function, portfolio construction, controlled experiment, asset pricing}
}

\allowdisplaybreaks[4]
\NewDocumentCommand{\abs}{o m}{
  \left\lvert
  \IfNoValueTF{#1}
    {#2}
    {\vphantom{#1}\smash{#2}}
  \right\rvert
}
\usepackage{threeparttable, tabularx}

\newcolumntype{Y}{>{\centering\arraybackslash}X}
\usepackage{dcolumn} \newcolumntype{d}[1]{D..{#1}}\newcolumntype{L}{D{.}{.}{2,5}}
\usepackage{caption}
\usepackage{natbib}
\setcitestyle{aysep={}}
\usepackage{booktabs}
\usepackage{appendix} 
\usepackage{bookmark} 
\usepackage[labelfont=bf]{caption} 
\usepackage{indentfirst} 
\usepackage{pdflscape} 
\usepackage{chngcntr} 
\usepackage{threeparttable}
\floatstyle{plaintop}
\restylefloat{table}
\usepackage{array, makecell}
\usepackage{multicol}
\usepackage{multirow}
\usepackage{lscape}
\usepackage{placeins}
\usepackage{float}
\usepackage{csquotes}
\graphicspath{{JEL_HF Paper_graphics/}{JEL_HF Paper_tcache/}{JEL_HF Paper_gcache/}}
\DeclareGraphicsExtensions{.pdf,.eps,.ps,.png,.jpg,.jpeg} 
\providecommand{\U}[1]{\protect \rule{.1in}{.1in}}

\makeatletter
\renewcommand{\@seccntformat}[1]{\csname the#1\endcsname.\quad}
\makeatother

\makeatletter
\renewcommand{\and}{\end{tabular}\\[6pt]\begin{tabular}[t]{c}}
\makeatother
\begin{document}
 \emergencystretch 3em 

\title{\Large\bfseries Machine Learning Classification and Portfolio Construction: \\Does the Loss Function Matter?\thanks{\footnotesize We thank the editors Luis Garcia-Feijoo and Daniel Giamouridis, two anonymous referees, Ahmed Guecioueur (discussant), Massimiliano Caporin (discussant), Keith Black (discussant), and seminar and conference participants at the University of Missouri, the 2021 AFA Annual Meeting, the 19th Crowell Memorial Prize Seminar, the 2020 University of Miami Winter Conference on Machine Learning and Business, the 2021 World Finance Conference, the 2021 SFA Annual Meeting, California State University, Fullerton, and Montclair State University for comments and suggestions. An earlier version of the paper was awarded third place in the 19th Crowell Memorial Prize. All errors are our own.}
}

\author{\large{Yang Bai}\thanks{Assistant Professor of Finance, College of Business and Economics, California State University, Fullerton, California, \href{mailto:yabai@fullerton.edu}{yabai@fullerton.edu}.}\and
\large{Kuntara Pukthuanthong}\thanks{Professor of Finance and Robert J. Trulaske, Jr. Professor of Finance, Robert J. Trulaske, Sr. College of Business, University of Missouri, Columbia, Missouri, \href{mailto:pukthuanthongk@missouri.edu}{pukthuanthongk@missouri.edu}.}
}

\date{\vspace{4pt}\normalsize\textit{Financial Analysts Journal, Forthcoming} \\[16pt]\normalsize August 21, 2026 \\[3pt]
\normalsize First Draft: May 31, 2020}
\clearpage\maketitle
\thispagestyle{empty} 
\vspace{-36pt}
\begin{abstract}
\vspace{-6pt}
\singlespacing{
\noindent Classification outperforms regression across matched machine learning models in portfolio construction. A stacking ensemble of gradient boosted trees, random forest, and neural network yields a value-weighted annualized Sharpe ratio of 2.08 for classification and 1.39 for regression. This outperformance strengthens with class granularity and persists across subsamples and after transaction costs. Spanning tests show that classification retains economically large alphas after we control for regression, whereas regression alphas shrink substantially once we control for classification. These results indicate that classification extracts more return information than matched regression. Our diagnostics trace classification's advantage to more precise separation of return deciles.

\vspace{24pt}

}
\end{abstract}
\vfill
\pagebreak
\setcounter{page}{1} 

\onehalfspacing
\normalsize
\section{Introduction}
Most studies in the financial machine learning literature on portfolio construction rely on regression-based models \citep{gu2020empirical, chen2023deep, freyberger2020dissecting, drobetz2021empirical, rossi2018predicting, allena2025confident, howard2024choices}. These models minimize the mean squared error between predicted and realized returns. However, the payoff of a long-short decile strategy depends on which decile portfolio each stock is assigned to. This raises the fundamental question: Is regression a better choice for portfolio construction than classification? Cross-study comparisons cannot resolve this question because studies in the literature differ substantially in data coverage, sample period, and model specifications. The modeling results are also sensitive to these design choices \citep{howard2024choices}. To answer the question, we conduct a controlled experiment following the data construction of \citet{green2017characteristics} and the modeling framework of \citet{gu2020empirical}. Using matched models that differ only in the loss function, we evaluate the performance of classification and regression in portfolio construction.

Contrary to the literature's preference for regression, we find that classification outperforms regression. Specifically, we implement four machine learning models, comprising gradient boosted trees (GBT), random forest (RF), neural network (NN), and a stacking ensemble of the three base models. Each model tunes the most important hyperparameters through grid search.\footnote{For example, instead of estimating five separate NN models, one for each number of hidden layers, our NN selects the number of hidden layers internally in every training period. The hyperparameter grid and the selected values are reported in the appendix.} We supplement the machine learning models with a set of linear benchmarks.\footnote{The linear benchmarks we use include ordinary least squares regression for machine learning regression models, logistic regression for binary classification, and multinomial logistic regression for multiclass classification models. In theory, both logistic regression and multinomial logistic regression are considered generalized linear models because of the distributional assumptions of the exponential family. Therefore, logistic regression and multinomial logistic regression are adopted as the linear benchmarks for the corresponding classification tasks. Hereafter, we refer to the classification benchmarks collectively as ``logistic regression'' for simplicity.} The classification stacking ensemble of GBT, RF, and NN delivers an equal-weighted annualized Sharpe ratio of 3.77 and a value-weighted annualized Sharpe ratio of 2.08. The matched regression stacking ensemble reaches 3.12 in equal-weighted terms and 1.39 in value-weighted terms.

In value-weighted portfolios, the classification stacking alpha in the Fama-French six-factor model is 4.1\% per month, while the matched regression stacking alpha is 1.8\%, and all five classification models earn larger value-weighted six-factor alphas than their matched regression counterparts. The classification advantage strengthens with class granularity in the multiclass sensitivity analysis. It holds among stocks with large market capitalization (large cap) and across subperiods, leading in three of the four subperiods and essentially tied in the remaining one. It also survives turnover-based transaction costs of 30 basis points per trade together with a 50-basis-point annual borrowing cost on short positions \citep{davolio2002market,novymarx2016taxonomy}.

The linear benchmarks provide the cleanest isolation of the loss-function effect. In the value-weighted long-short portfolio, the logistic regression achieves a Sharpe ratio of 1.57 and a six-factor alpha of 2.6\% per month, whereas the ordinary least squares (OLS) regression achieves a Sharpe ratio of 1.03 and an alpha of 1.1\%. This gap highlights the efficiency of cross-entropy as a loss function for portfolio construction, since the comparison between the logistic regression and the OLS regression does not involve any additional model complexity beyond the loss function itself.

The economic intuition points to an advantage rooted in the training objective. Cross-entropy trains the model to estimate the probability distribution over return states, directly optimizing the decile assignment that the portfolio relies on, namely distinguishing one decile from another. Mean squared error trains the model to estimate return magnitudes, a more difficult problem that requires a much higher level of prediction precision. Our diagnostics are consistent with this mechanism. We find that regression assigns stocks to deciles with lower precision and recall than classification, a key limitation for portfolio formation. This pattern suggests that the regression objective devotes less attention to the boundaries between return states that drive portfolio payoffs, because it is rewarded for fitting the level of returns within a decile rather than for separating the deciles from one another.

The regression stacking ensemble shows a row-normalized confusion matrix with diagonal entries for the 8 middle deciles that cluster between 10.58\% and 11.37\%, not meaningfully distinguishable from the 10\% random baseline. The regression ensemble does classify the tail deciles somewhat better than chance, with bottom-class recall and precision around 19.4\% and top-class recall and precision around 16.9\%. However, the increased hit rates at the tails appear to come with substantial misclassification across other deciles. We report the diagnostic details in the Online Supplemental Material (OSM).

Classification has a long record in portfolio construction. Early applications using recursive partitioning \citep{sorensen2000decision,zhu2011hybrid,zhu2012benefits} and more recent machine learning models \citep{aw2022factor,livnat2021algorithms,rasekhschaffe2019machine} demonstrate that classification can produce economically meaningful portfolio returns, yet no study has directly compared classification with regression under a common design \citep{allena2025confident,chen2023deep,drobetz2021empirical,freyberger2020dissecting,gu2020empirical,howard2024choices,rossi2018predicting}. Our paper fills this gap with the first within-study controlled comparison that isolates the loss function as a key driver of the performance difference between classification and regression in portfolio construction.

\section{Data, Models, and Experimental Design}
\label{sec:data_methods}
In this section, we describe the data, the matched models, and the experimental design used to isolate the loss function as the only difference between the classification and regression strategies. We first introduce the sample and the predictor set, then define the classification and regression objectives, and finally detail the training protocol and the portfolio construction.

\subsection{Data}
\label{subsec:data}
We obtain stock return data from the Center for Research in Security Prices and financial data from Compustat. Our sample comprises monthly observations of U.S. common stocks listed on the New York Stock Exchange, American Stock Exchange, and National Association of Securities Dealers Automated Quotations from July 1962 to December 2024.\footnote{Recent Compustat data include observations from 1951. However, Compustat was established in 1962. We err on the side of caution and follow \citet{fama1992cross}, using data only since 1962.} We retain only common stocks listed on the three major exchanges for both training and testing, and we exclude observations with missing current returns.\footnote{We keep observations with Center for Research in Security Prices share codes 10, 11, or 12 and exchange codes 1, 2, or 3.} For factor-model tests and the risk-free rate, we obtain the data from Kenneth French's data library \citep{carhart1997persistence,fama1992cross,fama2015five}.

The predictor set consists of 102 firm characteristics reconstructed following \citet{green2017characteristics}, plus the two-digit Standard Industrial Classification industry indicators. We transform each numeric characteristic each month into a cross-sectional rank normalized to $[-1,1]$, consistent with prior literature \citep{freyberger2020dissecting,gu2020empirical,howard2024choices}.\footnote{The normalization choice does not change the conclusion. Under z-score normalization, the performance gap between classification and regression is even larger than under the rank normalization.} The OSM reports the summary statistics of the firm characteristics.

\subsection{Classification Objective under Cross-Entropy Loss}
\label{subsec:cls_objective}
Our main classification results frame the cross-sectional return prediction problem as a 10-class return-decile classification. At each date, stocks are sorted into 10 return deciles based on realized returns, and classification models are trained to predict each stock's return decile in the following month. The top decile forms the portfolio that we buy, while the bottom decile forms the portfolio that we short sell. We also report 2-class, 3-class, and 5-class specifications to show how performance varies with class granularity.

The classification loss function is the categorical cross-entropy. In the 10-class setting, for stock $i$ in month $t$ with realized one-hot decile label $y_{i,t}(d)$ and predicted probability $q_{i,t}(d)$ over the return deciles $D = \{1, 2, \ldots, 10\}$, the classification loss is
\begin{align}
\mathcal{L}_{\text{cls}} = -\sum_{i,t}\sum_{d \in D} y_{i,t}(d)\ln q_{i,t}(d),
\end{align}
\noindent where $y_{i,t}(d)$ is the realized-decile indicator that equals 1 when stock $i$ realizes decile $d$ in month $t$ and 0 otherwise, and $q_{i,t}(d)$ is the model's predicted probability that stock $i$ belongs to decile $d$ in month $t$. Minimizing cross-entropy directly trains the model to assign high probability to the realized return decile.

Each month, the classification model assigns $|D|=10$ predicted probabilities to each stock, one for each return decile, and the probabilities sum to one. We consider two portfolio construction approaches. In our main approach, which we term the soft construction, each class portfolio is formed independently. All stocks are ranked by their predicted probability of belonging to class $d$, $q_{i,t}(d)$, and the top-decile stocks by that score enter class $d$'s portfolio. Because the ranking is performed separately for each class $d$, a stock can appear in multiple class portfolios or in none. We construct the soft long leg from stocks in the top decile of $q_{i,t}(|D|)$ and the soft short leg from stocks in the top decile of $q_{i,t}(1)$.\footnote{We exclude stocks that simultaneously qualify for both the long and short legs of the soft construction, as they represent cases where the model assigns high probabilities to both the top and bottom deciles. In unreported results, including these stocks does not influence our conclusion.} In our multiclass sensitivity analysis, we also report a hard construction, which assigns each stock to a single class based on the highest predicted probability, following a common practice in machine learning prediction.\footnote{We borrow the terms ``soft'' and ``hard'' from the machine learning literature, which distinguishes classifiers by their reliance on class-level probabilities versus single-label predictions \citep{wahba2002soft, liu2011hard}. In that literature, a hard classifier, in the strict sense, produces no intermediate class probabilities, while our usage refers only to the portfolio construction. From that perspective, every model we estimate performs soft classification, since each relies on the intermediate class-probability vector.}

Our choice to focus on the soft construction is motivated by the machine learning literature. Cross-entropy is a strictly proper scoring rule, under which honest and accurate probabilities are the unique best response from the training process \citep{gneiting2007strictly}. The hard construction with an additional prediction step based on argmax discards what the optimization calibrates. In other words, the predicted probability vector carries information beyond the single most likely class, as the distillation experiments of \citet{hinton2015distilling} demonstrate. Unless otherwise noted, all classification results, except for the soft-versus-hard construction comparison in the multiclass sensitivity analysis, are from the soft construction.

\subsection{Regression Objective under Mean Squared Error Loss}
\label{subsec:reg_objective}
Our regression approach follows the standard framework of \citet{gu2020empirical}. Regression models use the same model specifications, data, and features as the classification models, but they minimize the mean squared error between predicted and realized returns:
\begin{align}
\mathcal{L}_{\text{reg}} = \frac{1}{S}\sum_{i,t} (r_{i,t} - \widehat{r}_{i,t})^2,
\end{align}
\noindent where $S$ is the number of stock-month observations in the training sample, $r_{i,t}$ is stock $i$'s realized excess return in month $t$, and $\widehat{r}_{i,t}$ is the predicted return. For the decile-portfolio construction, the strategy places stocks into predicted-return deciles each month, buys the highest decile, and shorts the lowest decile. This is the standard long-short construction in the regression literature \citep{fama1992cross}.

\subsection{Models}
\label{subsec:model_classes}
We implement five matched model pairs, comprising the four machine learning models and a linear benchmark, with each model trained under both the classification and the regression objectives. We provide a brief overview of the models here and refer readers to the OSM for formal specifications.

\subsubsection{Gradient Boosted Trees}
\label{subsubsec:gbt}
GBT builds an ensemble by iteratively adding trees that minimize the residual loss of the current ensemble \citep{friedman2001greedy}. For classification, the output is passed through the softmax function to produce class-probability predictions across the $|D|$ return-decile classes, so that the predicted probability for class $d$ is $q_{i,t}(d) = \frac{\exp(z_{i,t,d})}{ \sum_{d' \in D} \exp(z_{i,t,d'})}$, where $z_{i,t,d}$ is the output for stock $i$ in month $t$ and decile $d$ with the sum running over all deciles $d' \in D$ \citep{hastie2009elements}. For regression, the objective switches to mean squared error loss.

\subsubsection{Random Forest}
\label{subsubsec:rf}
An RF builds on decision trees with bootstrap aggregating (bagging) \citep{breiman2001random}. It aggregates predictions across all trees. In our modeling process, the classification prediction is the average of per-tree class-probability votes, while the regression prediction is the average predicted return. At each node, a random subset of features is sampled before the best split is selected, which decorrelates the individual trees and reduces variance. Unlike the other algorithms, the RF does not need an output transformation to produce class probabilities in classification.

\subsubsection{Neural Network}
\label{subsubsec:nn}
Our NN is the standard fully connected multilayer perceptron, following \citet{gu2020empirical}. The input layer contains the firm characteristics and the industry indicators. Each hidden layer applies the $\tanh$ activation function elementwise to a linear combination of the previous layer's outputs. For classification, the output layer applies the softmax function to produce decile-class probabilities. For regression, the output layer contains a single linear neuron producing a magnitude forecast. The NN requires an explicit choice of optimizer, and we estimate every specification under two widely used options, Adam and Adadelta. Throughout, we report the Adam results, estimated with a learning rate of 0.01, because Adam improves the regression side of the comparison more than the classification side, which makes our reported gap the conservative one. Under Adadelta, the classification advantage is wider in every comparison. The OSM reports the side-by-side comparison for the stacking ensembles and the NNs.

\subsubsection{Stacking Ensemble}
\label{subsubsec:stacking}
The stacking ensemble combines the three base models, namely GBT, RF, and NN, using simple equal-weight averaging. In the main classification setting, each base model produces a 10-dimensional probability vector for each stock, and the ensemble averages those predicted probabilities across the base models before assigning stocks to portfolios. In the regression setting, the base models produce scalar return predictions, and the ensemble averages those predicted returns before sorting the stocks.

\subsubsection{Linear Benchmarks}
\label{subsubsec:linear}
For the linear benchmarks, we use logistic regression for the classification models and OLS regression for the regression models. The linear benchmarks are particularly informative. If classification outperforms regression in the linear benchmark comparison, the advantage must originate from the efficiency of the loss function rather than from model complexity.

\subsection{Training and Hyperparameter Tuning}
\label{subsec:design}
The retraining schedule expands the training set by 12 months each year. Every June, all data from July 1962 through that June are used for fitting. The following 12 months serve as a holdout period for tuning hyperparameters. Out-of-sample (OOS) predictions are generated for the subsequent 12 months after the holdout period. For tree models, the key hyperparameter is the maximum tree depth. Hyperparameters for the NN model include the number of hidden layers and the number of neurons. The hyperparameter search grid is reported in Appendix Table~\ref{tab:arch_search}. Each tree-based candidate is capped at 1000 trees, and each NN candidate is trained for up to 100 epochs. Appendix Table~\ref{tab:selected_models} reports the hyperparameters most frequently selected across the training windows.

\subsection{Portfolio Construction and Performance Evaluation}
\label{subsec:portfolio}
Portfolio construction is identical across models and across the two loss functions. For classification, we use the soft construction described in Section~\ref{subsec:cls_objective}. In our main decile-portfolio results, the long leg consists of stocks ranking in the top decile by their predicted probability of top-decile membership, $q_{i,t}(10)$, while the short leg consists of stocks ranking in the top decile by their predicted probability of bottom-decile membership, $q_{i,t}(1)$. Because the long and short legs are formed independently, a stock can appear in both simultaneously, in which case it is excluded from both. Our multiclass robustness tests vary the granularity of the classification target across the 2-class, 3-class, and 5-class settings, sorting on the predicted probabilities of the top and bottom classes. These tests additionally compare this soft construction against the hard construction. For regression, the signal is the predicted return itself. The highest predicted-return decile forms the long leg and the lowest predicted-return decile forms the short leg.

Both value-weighted and equal-weighted portfolios are formed, and we assess portfolio performance with summary statistics and factor-model alphas. We report summary statistics including the annualized return, the annualized standard deviation, and the annualized Sharpe ratio. For individual long or short legs, the Sharpe ratio is defined as
\begin{align}
SR_p = \frac{\mathbb{E}(\tilde{r}_p-r_f)}{\sigma(\tilde{r}_p)} \times \sqrt{12},
\end{align}

\noindent where $\tilde{r}_p$ is the portfolio return, $r_f$ is the risk-free rate so that $\tilde{r}_p-r_f$ is the excess return, $\sigma(\cdot)$ denotes the standard deviation, and the factor $\sqrt{12}$ annualizes the monthly ratio. For long-short portfolios, the Sharpe ratio is defined as
\begin{align}
SR_p = \frac{\mathbb{E}(\tilde{r}_\text{long-short})}{\sigma(\tilde{r}_\text{long-short})} \times \sqrt{12},
\end{align}
\noindent where the long-short portfolios are treated as zero-investment portfolios \citep{gu2020empirical}.

Factor-model alphas are estimated via time-series regression with Newey-West heteroskedasticity- and autocorrelation-consistent standard errors at 6 lags \citep{newey1987simple}.\footnote{The results from 12 lags do not change our conclusion.} When we estimate the factor models, we use the excess return as the dependent variable for the long and short legs. The excess return is defined as the difference between the portfolio return and the risk-free rate. For the long-short portfolios, we use their raw return directly for factor-model alpha estimation.\footnote{Using the excess return definition instead of the zero-investment portfolio definition does not change the conclusion.} Because we fit the factor models on monthly data, we report the alpha at the monthly frequency. Factor-model alphas are reported under common factor models, including the Capital Asset Pricing Model (CAPM) \citep{lintner1965security,sharpe1964capital} and the Fama-French three-, five-, and six-factor specifications, namely the FF3F, FF5F, and FF6F models, where the FF6F model augments the FF5F model with the momentum factor \citep{carhart1997persistence,fama1993common,fama2015five}.

\section{Main Results on Classification versus Regression}
\label{sec:results}
In this section, we report the main results of the controlled experiment. The first training window is from July 1962 to June 1983. After the initial window, the training sample expands each time we retrain the model. The OOS evaluation period spans July 1983 to December 2024, covering 498 monthly observations. All classification models are matched against their regression counterparts, with only the loss function varying.

\subsection{Portfolio Performance}
\label{subsec:portfolio_performance}
Table~\ref{tab:portfolio_perform} reports summary statistics, Sharpe ratios, and factor-model alphas for all 10 long-short portfolios. In value-weighted portfolios, every classification model carries a higher annualized standard deviation than its matched regression counterpart, consistent with classification's heavier reliance on volatility characteristics. Classification models also deliver higher annualized value-weighted returns, ranging from 33.5\% to 50.4\%, whereas regression models deliver 10.9\% to 27.1\%.

Out of 50 paired comparisons across the Sharpe ratios and the factor alphas for the two weighting schemes, classification outperforms regression in 49 cases. Under classification, the stacking portfolio reaches an equal-weighted Sharpe ratio of 3.77 and a value-weighted Sharpe ratio of 2.08. The matched regression stacking ensemble produces an equal-weighted Sharpe ratio of 3.12 and a value-weighted Sharpe ratio of 1.39. For perspective, the value-weighted Sharpe ratio of the market buy-and-hold portfolio is 0.53. The only exception, where classification loses to regression, is the equal-weighted Sharpe ratio of the RF long-short portfolio. In that case, the regression model reaches a Sharpe ratio of 2.44 and the classification model reaches 2.40, a difference that is not economically meaningful.

\begin{center}
    [Insert Table~\ref{tab:portfolio_perform} Here]
\end{center}

In the equal-weighted long-short portfolio, the classification stacking alpha ranges from 5.3\% to 5.4\% per month across the CAPM, FF3F, FF5F, and FF6F models. The matched regression stacking portfolio still earns a positive equal-weighted alpha, but it is smaller at 4.1\% to 4.3\% per month. In the value-weighted long-short portfolio, classification remains strong under every factor model. The monthly alpha stays between 4.1\% and 4.2\%. The matched regression alpha, by contrast, falls to between 1.8\% and 2.3\% across the four factor models. Moving from equal- to value-weighted portfolios roughly halves the regression alpha while leaving the classification alpha near 4\%, so the classification advantage widens precisely where portfolio capacity is largest. The same pattern holds across the other machine learning models in value-weighted long-short portfolios. The monthly FF6F alphas for classification are 3.4\% in GBT, 3.0\% in RF, and 3.5\% in NN, while the matched regression counterparts deliver 0.8\%, 0.9\%, and 1.6\%, respectively.

Overall, across both weighting schemes and the full set of factor models, classification alphas are larger than regression alphas in equal-weighted portfolios and far more resilient once portfolios are value-weighted. The matched regression still extracts some information in equal-weighted long-short portfolios, but much of that performance weakens once the portfolios are value-weighted and the factor model becomes more demanding.

Notably, the simple logistic regression, our classification benchmark, outperforms every regression model in value-weighted portfolios and matches or outperforms it in equal-weighted portfolios. This highlights the efficiency of classification in extracting return information. In particular, the logistic regression reaches an equal-weighted Sharpe ratio of 3.16 and a value-weighted Sharpe ratio of 1.57, whereas OLS reaches 2.57 and 1.03. In the value-weighted FF6F model, the logistic regression earns a 2.6\% alpha per month, while OLS earns a 1.1\% alpha. Because both models share the same linear functional form, with no hidden layers or nonlinear interactions, the performance gap cannot be attributed to model complexity. Under the same information set, the evidence points directly to the difference in loss functions. For comparability across studies, we focus on the stacking ensembles in the subsequent analyses because they summarize the information from the three base models in each case.

\subsection{Augmented Factor-Model Regressions}
\label{subsec:factor_model_tests}
Table~\ref{tab:factor_model_tests} reports augmented factor-model regressions for the classification and regression stacking ensembles. Each time-series regression extends a standard factor model with one augmented portfolio. The first specification, labeled \textit{+Benchmark}, adds the corresponding linear benchmark portfolio, namely logistic regression for the classification panel and OLS for the regression panel. It evaluates how much model complexity contributes to capturing pricing information beyond what the linear benchmark already explains \citep{fama1992cross, fama2015five,carhart1997persistence}. If the alpha is negative or insignificant, it means that added model complexity captures no information beyond the pricing factors and the linear benchmark. Relative to the corresponding alpha from Table~\ref{tab:portfolio_perform}, if the alpha shrinks substantially, it means that the increase in portfolio performance driven by the added model complexity is limited.

\begin{center}
    [Insert Table~\ref{tab:factor_model_tests} Here]
\end{center}

The second specification, labeled \textit{+Counterpart}, adds the matched stacking portfolio. This tests the influence of loss functions on portfolio construction. Specifically, for the classification panel we add the regression stacking portfolio as an additional regressor, and for the regression panel we add the classification stacking portfolio. In this spanning test, a positive and significant alpha for one portfolio indicates that its associated loss function extracts additional return-related information left uncaptured by the counterpart portfolio's loss function. Across the comparisons, if controlling for the counterpart portfolio shrinks the focal portfolio's alpha more than the reverse control shrinks the counterpart's alpha, it suggests that the counterpart portfolio provides more overall return-related information.

The classification alpha is large and stable under both augmentation specifications. The pattern is uniform across weighting schemes and factor models. Under the +Benchmark specification, value-weighted classification long-short alphas range from 2.2\% to 2.3\% per month across the CAPM, FF3F, FF5F, and FF6F models. This alpha measures the incremental contribution of the machine learning ensemble relative to the logistic regression, suggesting that the complex classification models extract additional pricing information from firm characteristics that the linear model leaves behind. The +Counterpart specification replaces the logistic control with the regression stacking ensemble, isolating what is distinctive about the training objective itself. Classification stacking alphas rise to 3.4\% to 3.5\% per month once we control for the regression stacking portfolio, consistent with the finding in Table~\ref{tab:portfolio_perform} that the complex regression models underperform the logistic regression. Relative to the standard Fama-French factor models, adding the regression stacking portfolio reduces classification stacking alphas in value-weighted portfolios by only 0.6\% to 0.8\%, or 15\% to 19\% in relative terms.

When the regression stacking portfolio is the dependent variable and the classification stacking portfolio is added as a control, value-weighted regression alphas remain positive at 1.0\% to 1.4\% per month across the four factor models, but they are much smaller than the corresponding classification alphas and shrink materially relative to the uncontrolled alphas ranging from 1.8\% to 2.3\% per month in Table~\ref{tab:portfolio_perform}, and the reduction level is equivalent to 39\% to 44\% in relative terms. In equal-weighted portfolios, the reduction is larger still, so the asymmetry is not confined to one weighting scheme. Read together, the two panels describe an informational relationship rather than two separate horse races. Controlling for the linear benchmark lowers the classification alpha more than controlling for the regression ensemble does, which means the matched regression portfolio captures less of what classification earns than a simple logistic model already captures. The reverse does not hold. Once classification is controlled for, the regression alpha falls sharply from its Table~\ref{tab:portfolio_perform} level. Taken together, we conclude that classification preserves more return-relevant information and is a better practical choice for portfolio construction.

\section{Robustness Tests}
\label{sec:robustness}
We conduct robustness tests to assess whether the performance advantage of classification models over regression models is stable across target granularity, stock subsamples, time periods, and transaction-cost assumptions. To keep the main text focused, we present the multiclass sensitivity analysis and the large-cap subsample and summarize the subperiod and transaction-cost results, with full details in the OSM.

Our baseline uses a 10-class specification, in which stocks are placed into return deciles each month. Given the strong performance of 10-class classification, a natural question is whether the advantage is specific to that class definition. To examine this possibility, we rerun the models over different levels of granularity. Table~\ref{tab:multiclass} reports results for 2-, 3-, and 5-class alternatives. Because the main results use the soft construction, we report both constructions in this section to assess the robustness of the soft construction's advantage.
\begin{center}
    [Insert Table~\ref{tab:multiclass} Here]
\end{center}

Under the 2-class specification, stocks are divided into above-median and below-median return groups. This is the coarsest classification design. The 3-class specification uses terciles, and the 5-class specification uses quintiles. We also include 10-class results for the convenience of comparison. All 16 classification long-short portfolios earn significant alphas that the common factor models cannot explain. 

Classification performance improves sharply as the number of classes rises above 2 and remains strong at finer levels of granularity. For the value-weighted long-short portfolio from the soft construction, the annualized Sharpe ratio rises from 1.33 under 2 classes to 1.76 under 3 classes and 2.33 under 5 classes, then remains high at 2.08 under 10 classes. The corresponding factor-model alphas are also economically large throughout. Relative to the matched regression ensemble, the 2-class soft portfolio is already comparable in value-weighted terms, and  the  advantage is clear from 3 classes onward. In equal-weighted portfolios, the two are comparable at 3 classes, and classification leads from 5 classes onward.

Panels C and D of Table~\ref{tab:multiclass} demonstrate the performance of argmax-based hard portfolio construction. The hard construction remains strong in its own right. All but its coarsest specifications deliver economically large Sharpe ratios, and it matches the regression ensemble from 5 classes onward in value-weighted portfolios and at 10 classes in equal-weighted portfolios. However, consistent with the literature \citep{gneiting2007strictly, hinton2015distilling}, the soft construction outperforms the hard construction at every class count and in both weighting schemes, and the gap is wider in the short legs. Because the two constructions share an identical model and differ only in how the predicted probabilities are used, the gap between them measures what the portfolio construction contributes once predictability is held fixed. The multiclass evidence therefore shows that the classification advantage strengthens with class granularity, and that it is realized in full only when the estimated probabilities are used directly.

Next, we examine model performance within the group of large-cap stocks. Large-cap stocks represent the most competitive and heavily arbitraged segment of the equity market. Any predictability that survives in this universe is less likely to reflect microstructural frictions or short-side impediments concentrated among small and illiquid stocks \citep{stambaugh2015arbitrage}. Restricting the investable universe to the top 50\% of stocks by market capitalization therefore serves as a robustness test and bears directly on whether the strategy can be implemented at scale. For large-cap stocks, we construct the portfolios with the same procedure as in the full sample, and the classification advantage survives the size restriction. We report both the classification results and the regression results in Table~\ref{tab:top_me_perform}.

\begin{center}
    [Insert Table~\ref{tab:top_me_perform} Here]
\end{center}

The classification stacking ensemble achieves an equal-weighted Sharpe ratio of 2.15 and a value-weighted Sharpe ratio of 1.30. The matched regression stacking ensemble achieves equal- and value-weighted Sharpe ratios of 1.55 and 1.25. The value-weighted FF6F alpha remains economically meaningful for classification at 2.0\% per month and is still larger than the matched regression alpha of 1.1\%.

Classification remains ahead of regression for the base models, except for the value-weighted RF pair. The logistic regression benchmark also slips slightly below the matched OLS benchmark in both weighting schemes, at large-cap Sharpe ratios of 1.57 versus 1.63 in equal-weighted portfolios and 1.00 versus 1.07 in value-weighted portfolios.\footnote{This arises because the logistic regression's performance is split evenly between the long and short legs, whereas the OLS performance is concentrated in the long leg. Restricting the universe to large-cap stocks shrinks the short-leg premium, so it removes more of the logistic advantage than of the OLS advantage.} The broader classification advantage nonetheless remains intact for the leading models.

\begin{figure}[H]
    \centering
    \pdfdest name {figs-tabs-fig1} fith
    \includegraphics[width=0.9\textwidth]{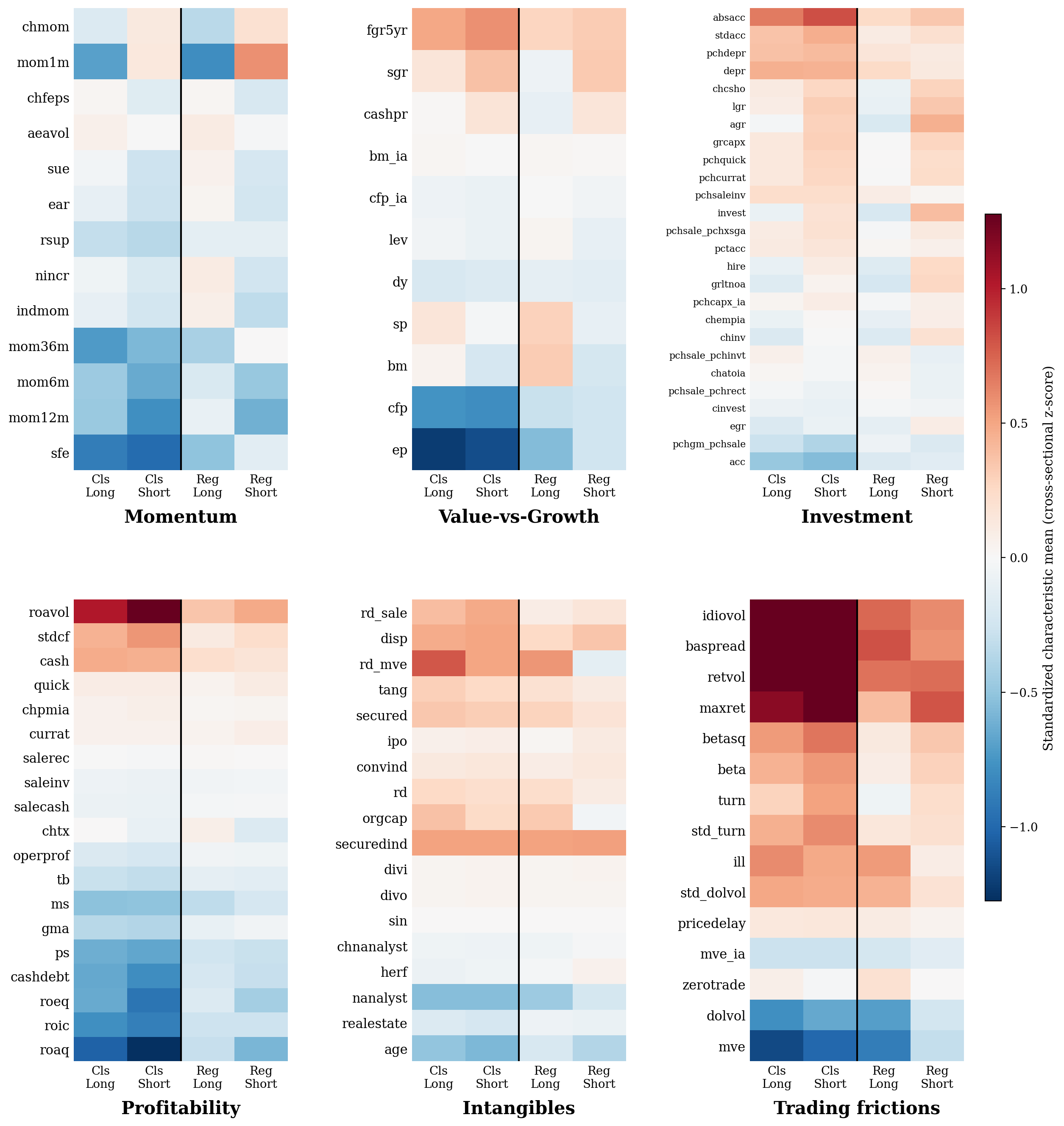}
    \bookmark[level=2,dest=figs-tabs-fig1]{Figure 1: Portfolio Holdings by Anomaly Category}
    \caption{Portfolio Holdings by Anomaly Category}
    \justifying{\footnotesize\noindent This figure reports the average standardized value of each firm characteristic in the classification and regression stacking long and short portfolios. Before portfolio construction, every characteristic is standardized cross-sectionally by month across all stocks to zero mean and unit standard deviation. The standardized values are then averaged within each portfolio over the out-of-sample period from July 1983 to December 2024. The four columns are the classification long and short portfolios and the regression long and short portfolios, separated by the vertical line. Red denotes a positive loading and blue a negative loading. The 102 characteristics are grouped following the spirit of the six anomaly categories of \citet{hou2018replicating}. Classification loads more heavily than regression on the trading-frictions characteristics, such as idiosyncratic volatility, bid-ask spread, return volatility, and maximum daily return.}
    \label{fig:holdings}
\end{figure}

The large-cap results also confirm the wide volatility gap between the two loss functions. The classification value-weighted long-short portfolios carry annualized volatilities ranging from 0.19 to 0.29, which is roughly one and a half times the 0.13 to 0.18 range for their matched regression counterparts. In the full sample, the comparison is narrower, but the qualitative observation is the same, with classification volatilities of 0.21 to 0.34 against regression volatilities of 0.16 to 0.21. The observed volatility gap is consistent with the mechanism we document in the diagnostics reported in the OSM. Classification tilts toward stocks with high idiosyncratic volatility and other return-uncertainty characteristics, while regression instead leans on firm size and liquidity. The higher classification volatility therefore reflects a deliberate tilt toward return-uncertainty stocks. This tilt is rewarded with higher average returns. The classification Sharpe ratio and six-factor alpha also stay above those of the matched regression, even as the advantage narrows in the large-cap subsample, indicating a better risk-return trade-off. We summarize the characteristic holdings of the ensemble portfolios in Figure~\ref{fig:holdings}.

\begin{center}
    [Insert Figure~\ref{fig:holdings} Here]
\end{center}

High portfolio turnover and trading costs can erode strategy returns in practice \citep{gu2020empirical}. Table~\ref{tab:costs} reports maximum drawdown and average monthly turnover for all models. Turnover sums purchases and sales, so replacing a leg in full registers as 200\% rather than 100\%, and the classification stacking ensemble's 144\% corresponds to replacing roughly three-quarters of the book each month.\footnote{The OSM provides the calculation details of maximum drawdown, turnover, and cumulative log returns after transaction costs and short-selling costs.} Trading intensity is comparable across the two loss functions. For the stacking ensemble, the maximum drawdown is smaller under classification in value-weighted portfolios. In equal-weighted portfolios, both loss functions keep the maximum drawdown under 14\% across the four decades, a small fraction of the market's 59\%. Turnover is slightly lower under the classification stacking ensemble than under the regression counterpart in value-weighted portfolios, though it is higher for classification in equal-weighted portfolios. The maximum drawdowns and turnovers indicate that the superior performance of classification does not arise from greater downside risk or from differences in trading costs. Both legs contribute to the spread, with their relative importance varying over time, as the subperiod results in the OSM show.

\begin{center}
    [Insert Table~\ref{tab:costs} Here]
\end{center}

\begin{landscape}
\begin{figure}[H]
    \centering
    \pdfdest name {figs-tabs-fig2} fith
    \makebox[\textwidth][c]{%
    \begin{minipage}{\dimexpr\paperheight-1.6in\relax}
    \centering
    \includegraphics[width=\linewidth]{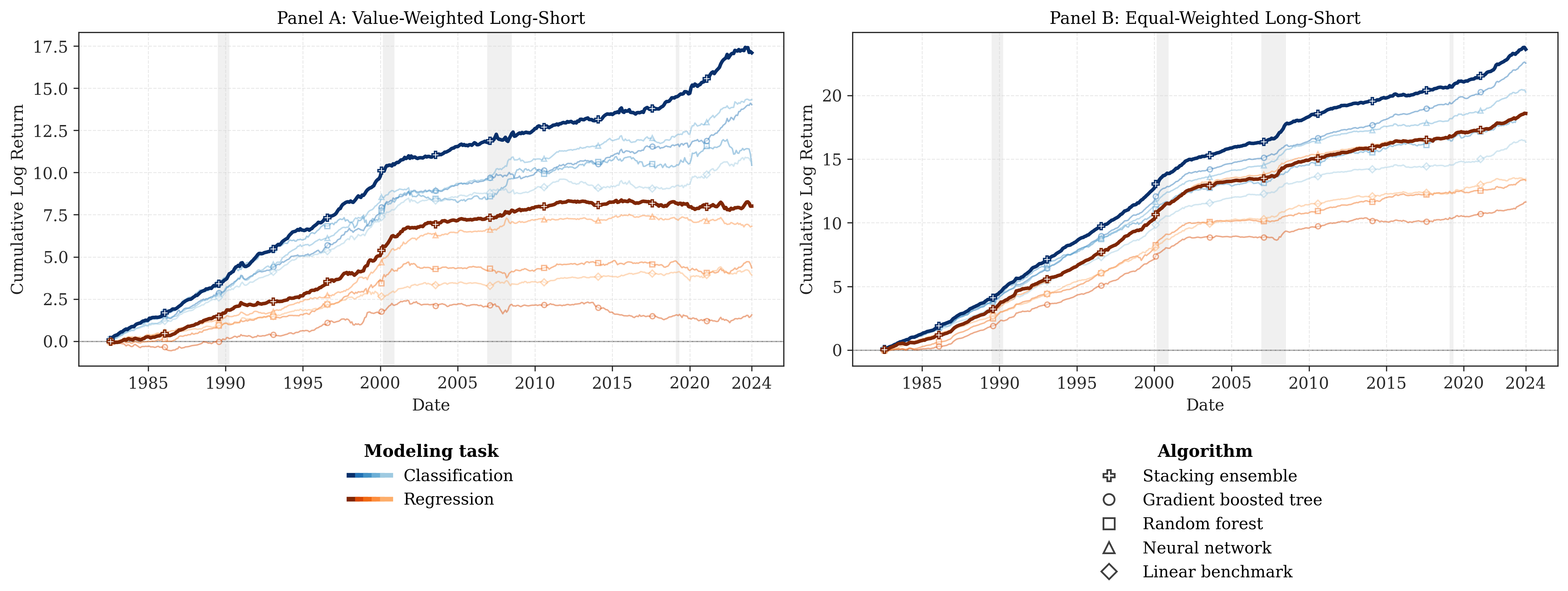}
    \bookmark[level=2,dest=figs-tabs-fig2]{Figure 2: Cumulative Log Returns}
    \caption{Cumulative Log Returns}
    \justifying{\footnotesize\noindent This figure plots the out-of-sample cumulative log returns of the classification and regression stacking long-short portfolios from July 1983 to December 2024, net of a trading cost of 30 basis points per one-way trade applied to the portfolio's average monthly two-way turnover and a 50-basis-point annual borrowing cost on short positions. Panel A reports the performance of value-weighted portfolios. Panel B reports the performance of equal-weighted portfolios. Gray shaded areas indicate National Bureau of Economic Research recession months.}
    \label{fig:timeseries_costs}
    \end{minipage}}
\end{figure}
\end{landscape}

\begin{center}
    [Insert Figure~\ref{fig:timeseries_costs} Here]
\end{center}

Figure~\ref{fig:timeseries_costs} plots the cumulative log returns of the long-short classification and regression stacking portfolios over the full OOS period of 498 months after deducting a trading cost of 30 basis points per one-way trade applied to the portfolio's average monthly two-way turnover, which is a full round-trip charge, and an additional 50-basis-point annual borrowing cost on short positions \citep{davolio2002market,novymarx2016taxonomy}. After the trading costs, the value-weighted Sharpe ratio of the classification stacking long-short portfolio remains 1.85, whereas the matched regression portfolio's Sharpe ratio falls to 1.09. 

We also confirm that the classification advantage is not concentrated in a single period. The stacking long-short portfolio under classification reaches value-weighted Sharpe ratios of 3.73, 2.63, 1.32, and 1.58 across the subperiods of 198307:199311, 199312:200404, 200405:201408, and 201409:202412, compared with 2.07, 2.19, 1.34, and 0.29 for the matched regression counterpart. Classification leads in three of the four windows and is essentially tied in the third. Across National Bureau of Economic Research (NBER) regimes, classification leads clearly in the 458 expansion months, with a value-weighted Sharpe ratio of 2.12 against 1.32, while in the 40 recession months the two methods are close, at 1.87 against 2.15. We report the full subperiod and NBER regime results in the OSM.

On balance, the robustness tests confirm that the classification advantage is not an artifact of a single design choice. The advantage holds from 3 classes upward in value-weighted portfolios and from 5 classes upward in equal-weighted portfolios, among large-cap stocks, and after realistic transaction and short-side borrowing costs. Across the four decade-long subperiods, it holds in three windows with the two methods essentially tied in the third, and across NBER regimes it is carried by the expansion months that make up over 90\% of the sample. Taken together, these tests indicate that the loss-function advantage we document is a stable feature of portfolio construction rather than a property of the baseline specification.

\section{Conclusion}
\label{sec:conclusion}

In this study, we examine the role of common loss functions in portfolio construction. We construct matched pairs of classification and regression models across four machine learning models, comprising GBT, RF, NN, and a stacking ensemble of the three base models. The models are trained on identical firm characteristics and evaluated over the OOS period from 1983 to 2024. The only difference within each matched pair is the loss function. Through our controlled experiment, we show that classification models trained using cross-entropy loss outperform regression models.

The long-short portfolio of the classification stacking ensemble earns an annualized value-weighted Sharpe ratio of 2.08, compared with 1.39 for the matched regression ensemble. Spanning regressions reinforce this finding. The classification portfolio continues to earn a much larger alpha after controlling for the matched regression portfolio, while the regression portfolio's alpha shrinks materially once classification is added to the specification. We supplement the machine learning models with a set of linear benchmarks. The logistic regression trained on decile assignments achieves a value-weighted Sharpe ratio of 1.57, outperforming the complex machine learning regression models.

The advantage of classification persists among large-cap stocks, in NBER expansion months, and after realistic transaction costs. Across the four subperiods, classification leads in three. Under the transaction-cost specification, the classification value-weighted Sharpe ratio remains 1.85, while the Sharpe ratio of the matched regression portfolio falls to 1.09. The result also survives alternative class designs, from 3 classes upward in value-weighted portfolios and from 5 classes upward in equal-weighted portfolios, and it strengthens with class granularity. Consistent with the machine learning literature, the advantage is realized fully when the portfolio construction is directly based on the predicted probability vector instead of the argmax-based label prediction.

Our diagnostics are consistent with the view that the economic performance reflects the alignment between the loss function and the goal of portfolio formation. Cross-entropy trains the model to identify the correct return decile, enabling more precise decile assignments. In contrast, mean squared error underperforms in this setting because its limited capacity is spent fitting return magnitudes instead of distinguishing between return deciles. Our study suggests that choosing the loss function merits attention in portfolio construction.

\clearpage

\phantomsection
\pdfbookmark[1]{References}{references-main}

\newcommand\invisiblesection[1]{%
  \refstepcounter{section}%
  \phantomsection%
  \addcontentsline{toc}{section}{\protect\numberline{\thesection}#1}
  \sectionmark{#1}}


\pdfbookmark[1]{Figures and Tables}{figs-tabs}
\section*{Figures and Tables}

\clearpage

\clearpage


\begin{landscape}
\begin{table}[!htbp]
\scriptsize
\setlength{\tabcolsep}{3pt}

\centering
\begin{minipage}{1.3\textwidth}
\justifying{\noindent This table reports out-of-sample long-short portfolio performance for classification and regression models over July 1983--December 2024 (498 months). For each model, stocks are assigned monthly by the predicted signal into 10 deciles. The long (short) portfolio holds the top (bottom) decile. Panels A and B report, respectively, the value-weighted and equal-weighted long-short portfolio performance. Classification forms each class's portfolio independently: stocks are ranked by their predicted probability of belonging to that class, and the top 10\% of stocks enter that class's portfolio. The market portfolio is a buy-and-hold portfolio over the entire market. The columns report the performance of the market portfolio and the long-short portfolios based on the stacking ensemble and its gradient boosted trees (GBT), random forest (RF), and neural network (NN) base models, alongside the logistic regression and ordinary least squares (OLS) linear benchmarks. Average Return is the annualized average raw return in decimal. Standard Deviation is the annualized standard deviation in decimal. Sharpe Ratio is the annualized Sharpe ratio. $\alpha$ is the monthly abnormal return in decimal from the corresponding factor model. The factor models are the Capital Asset Pricing Model (CAPM) and the Fama-French three-, five-, and six-factor (FF3F, FF5F, and FF6F) models. Statistical significance is based on Newey-West $t$-statistics adjusted with 6 lags: $^{**}p{<}0.01$, $^{*}p{<}0.05$.}
\end{minipage}

\pdfdest name {figs-tabs-tab1} fit
\bookmark[level=2,dest=figs-tabs-tab1]{Table 1: Portfolio Performance}
\caption{Portfolio Performance}
\label{tab:portfolio_perform}
\begin{tabularx}{1.3\textwidth}{l Y Y Y Y Y Y Y Y Y Y Y}
\toprule
 & & \multicolumn{5}{c}{\textbf{Classification}} & \multicolumn{5}{c}{\textbf{Regression}} \\
\cmidrule(lr){3-7}\cmidrule(lr){8-12}
Metric & Market & Stack & GBT & RF & NN & Logistic & Stack & GBT & RF & NN & OLS \\
\midrule
\multicolumn{12}{c}{\textbf{Panel A}: Value-Weighted Long-Short} \\
\midrule
Average Return   & 0.117 & 0.504 & 0.426 & 0.377 & 0.428 & 0.335 & 0.271 & 0.109 & 0.182 & 0.237 & 0.164 \\
Standard Deviation       & 0.157 & 0.243 & 0.234 & 0.336 & 0.228 & 0.214 & 0.195 & 0.182 & 0.209 & 0.182 & 0.159 \\
Sharpe Ratio            & 0.533 & 2.080 & 1.823 & 1.123 & 1.881 & 1.569 & 1.392 & 0.602 & 0.871 & 1.306 & 1.033 \\
CAPM $\alpha$        &             & 0.041** & 0.035** & 0.032** & 0.036** & 0.028** & 0.023** & 0.008** & 0.016** & 0.020** & 0.014** \\
FF3F $\alpha$        &             & 0.042** & 0.035** & 0.031** & 0.036** & 0.028** & 0.023** & 0.008** & 0.014** & 0.020** & 0.014** \\
FF5F $\alpha$        &             & 0.041** & 0.035** & 0.029** & 0.036** & 0.027** & 0.019** & 0.008* & 0.011** & 0.016** & 0.013** \\
FF6F $\alpha$        &             & 0.041** & 0.034** & 0.030** & 0.035** & 0.026** & 0.018** & 0.008* & 0.009** & 0.016** & 0.011** \\
\midrule
\multicolumn{12}{c}{\textbf{Panel B}: Equal-Weighted Long-Short} \\
\midrule
Average Return   & 0.112 & 0.651 & 0.621 & 0.545 & 0.560 & 0.460 & 0.519 & 0.344 & 0.386 & 0.517 & 0.390 \\
Standard Deviation       & 0.204 & 0.173 & 0.161 & 0.227 & 0.155 & 0.146 & 0.166 & 0.157 & 0.158 & 0.164 & 0.152 \\
Sharpe Ratio            & 0.384 & 3.767 & 3.857 & 2.398 & 3.615 & 3.158 & 3.117 & 2.198 & 2.441 & 3.158 & 2.570 \\
CAPM $\alpha$        &             & 0.054** & 0.051** & 0.046** & 0.046** & 0.038** & 0.043** & 0.028** & 0.033** & 0.043** & 0.032** \\
FF3F $\alpha$        &             & 0.054** & 0.051** & 0.045** & 0.046** & 0.038** & 0.043** & 0.028** & 0.031** & 0.043** & 0.032** \\
FF5F $\alpha$        &             & 0.053** & 0.050** & 0.043** & 0.046** & 0.037** & 0.041** & 0.027** & 0.029** & 0.041** & 0.031** \\
FF6F $\alpha$        &             & 0.054** & 0.051** & 0.044** & 0.046** & 0.037** & 0.042** & 0.027** & 0.030** & 0.042** & 0.031** \\
\bottomrule
\end{tabularx}
\end{table}
\end{landscape}
\clearpage


\begin{table}[!htbp]
\footnotesize

\justifying{\noindent This table reports alphas from factor models augmented with regressor portfolios. The dependent variables are the monthly zero-investment portfolio returns of the focal models' long-short portfolios and the monthly excess returns of the long and short portfolios. The regressors include standard factors, augmented with either benchmark models' portfolios or portfolios from the counterpart model using a different loss function. Panel A reports results for the portfolio returns based on the classification stacking ensemble as the dependent variables. Panel B reports results for the portfolio returns based on the regression stacking ensemble as the dependent variables. For the augmented regressors, +Benchmark adds the zero-investment portfolio return (or excess return) of the linear benchmark portfolio as an additional control (the logistic regression for the classification panel and ordinary least squares for the regression panel). +Counterpart adds the opposing model type (regression stacking for the classification panel and classification stacking for the regression panel) as an additional control. $\alpha$ is the monthly abnormal return in decimal from the corresponding factor model. The factor models are the Capital Asset Pricing Model (CAPM) and the Fama-French three-, five-, and six-factor (FF3F, FF5F, and FF6F) models. Statistical significance is based on Newey-West $t$-statistics adjusted with 6 lags: $^{**}p{<}0.01$, $^{*}p{<}0.05$.}

\centering
\pdfdest name {figs-tabs-tab2} fit
\bookmark[level=2,dest=figs-tabs-tab2]{Table 2: Augmented Factor-Model Regressions}
\caption{Augmented Factor-Model Regressions}
\begin{tabularx}{\textwidth}{l Y Y Y Y Y Y}
\toprule
\multicolumn{7}{c}{\textbf{Panel A}: Classification Stacking as Dependent Variable (Benchmark: Logistic, Counterpart: Regression Stacking)} \\
\midrule
 & \multicolumn{2}{c}{\textbf{Long-Short}} & \multicolumn{2}{c}{\textbf{Long}}
 & \multicolumn{2}{c}{\textbf{Short}} \\
\cmidrule(lr){2-3}\cmidrule(lr){4-5}\cmidrule(lr){6-7}
Factor Model & {+Benchmark} & {+Counterpart} & {+Benchmark} & {+Counterpart}
             & {+Benchmark} & {+Counterpart} \\
\midrule
\multicolumn{7}{c}{\textbf{Value-Weighted Portfolios}} \\
\midrule
CAPM $\alpha$   & 0.022** & 0.034** & 0.006** & -0.002 & 0.008** & 0.019** \\
\quad($t$) & (7.98) & (8.61) & (3.13) & (-0.61) & (4.84) & (4.60) \\
FF3F $\alpha$   & 0.022** & 0.034** & 0.007** & 0.002 & 0.009** & 0.020** \\
\quad($t$) & (8.09) & (8.64) & (3.28) & (0.74) & (5.11) & (5.14) \\
FF5F $\alpha$   & 0.022** & 0.035** & 0.008** & 0.009** & 0.009** & 0.019** \\
\quad($t$) & (7.65) & (8.41) & (3.72) & (2.76) & (4.86) & (4.93) \\
FF6F $\alpha$   & 0.023** & 0.035** & 0.009** & 0.011** & 0.009** & 0.018** \\
\quad($t$) & (7.54) & (8.29) & (4.13) & (3.66) & (4.73) & (4.90) \\
\midrule
\multicolumn{7}{c}{\textbf{Equal-Weighted Portfolios}} \\
\midrule
CAPM $\alpha$   & 0.019** & 0.025** & 0.006** & -0.004** & 0.011** & 0.008** \\
\quad($t$) & (7.63) & (9.15) & (7.80) & (-3.02) & (9.45) & (4.90) \\
FF3F $\alpha$   & 0.020** & 0.025** & 0.006** & -0.002 & 0.011** & 0.007** \\
\quad($t$) & (7.77) & (9.38) & (7.29) & (-1.20) & (9.80) & (4.06) \\
FF5F $\alpha$   & 0.020** & 0.026** & 0.006** & 0.002 & 0.011** & 0.007** \\
\quad($t$) & (8.00) & (9.30) & (6.55) & (1.37) & (9.41) & (4.28) \\
FF6F $\alpha$   & 0.020** & 0.026** & 0.006** & 0.003 & 0.011** & 0.007** \\
\quad($t$) & (8.67) & (9.08) & (6.81) & (1.85) & (9.44) & (4.22) \\
\bottomrule
\end{tabularx}
\label{tab:factor_model_tests}
\end{table}

\begin{table}
\footnotesize
\centering
\caption*{\textbf{Table~\ref{tab:factor_model_tests}}. Augmented Factor-Model Regressions (Continued)}
\begin{tabularx}{\textwidth}{l Y Y Y Y Y Y}
\toprule
\multicolumn{7}{c}{\textbf{Panel B}: Regression Stacking as Dependent Variable (Benchmark: OLS, Counterpart: Classification Stacking)} \\
\midrule
 & \multicolumn{2}{c}{\textbf{Long-Short}} & \multicolumn{2}{c}{\textbf{Long}}
 & \multicolumn{2}{c}{\textbf{Short}} \\
\cmidrule(lr){2-3}\cmidrule(lr){4-5}\cmidrule(lr){6-7}
Factor Model & {+Benchmark} & {+Counterpart} & {+Benchmark} & {+Counterpart}
             & {+Benchmark} & {+Counterpart} \\
\midrule
\multicolumn{7}{c}{\textbf{Value-Weighted Portfolios}} \\
\midrule
CAPM $\alpha$   & 0.016** & 0.014** & 0.003 & 0.005** & 0.010** & 0.002 \\
\quad($t$) & (5.05) & (4.80) & (1.74) & (2.66) & (4.28) & (1.22) \\
FF3F $\alpha$   & 0.016** & 0.013** & 0.004* & 0.005** & 0.009** & 0.003 \\
\quad($t$) & (5.21) & (4.71) & (2.41) & (2.83) & (4.37) & (1.59) \\
FF5F $\alpha$   & 0.014** & 0.011** & 0.006** & 0.005** & 0.005** & 0.002 \\
\quad($t$) & (4.67) & (4.06) & (2.84) & (2.72) & (3.32) & (1.10) \\
FF6F $\alpha$   & 0.014** & 0.010** & 0.006** & 0.005** & 0.004** & 0.002 \\
\quad($t$) & (4.56) & (3.83) & (2.77) & (2.61) & (2.75) & (0.94) \\
\midrule
\multicolumn{7}{c}{\textbf{Equal-Weighted Portfolios}} \\
\midrule
CAPM $\alpha$   & 0.019** & 0.010** & 0.002 & 0.006** & 0.003 & -0.003* \\
\quad($t$) & (6.38) & (3.16) & (1.76) & (5.30) & (1.76) & (-2.12) \\
FF3F $\alpha$   & 0.018** & 0.010** & 0.000 & 0.006** & 0.005** & -0.001 \\
\quad($t$) & (6.76) & (2.99) & (0.35) & (4.98) & (2.66) & (-0.67) \\
FF5F $\alpha$   & 0.017** & 0.009** & 0.002 & 0.004** & 0.004** & -0.001 \\
\quad($t$) & (6.61) & (2.67) & (1.10) & (3.04) & (2.68) & (-0.85) \\
FF6F $\alpha$   & 0.018** & 0.009** & 0.004** & 0.005** & 0.004** & -0.001 \\
\quad($t$) & (6.91) & (3.07) & (3.01) & (3.86) & (3.07) & (-0.77) \\
\bottomrule
\end{tabularx}
\end{table}
\clearpage


\begin{landscape}
\begin{table}[!htbp]
\scriptsize
\setlength{\tabcolsep}{3pt}

\centering
\begin{minipage}{1.3\textwidth}
\justifying{\noindent This table reports out-of-sample long-short portfolio performance across the 2-, 3-, 5-, and 10-class specifications over July 1983--December 2024 (498 months). For each model, stocks are assigned monthly by the predicted signal into the 2, 3, 5, or 10 classes indicated in the column headings. The long (short) portfolio holds the top (bottom) class. The soft construction forms each class's portfolio independently: stocks are ranked by their predicted probability of belonging to that class, and the top 10\% of stocks enter that class's portfolio. The hard construction assigns each stock to its predicted class, the class with the highest predicted probability. Panels A and B report the value-weighted and equal-weighted long-short portfolio performance for the soft construction, while Panels C and D report the value-weighted and equal-weighted long-short portfolio performance for the hard construction. L-S indicates long-short portfolios.
Average Return is the annualized average raw return in decimal.
Standard Deviation is the annualized standard deviation in decimal.
Sharpe Ratio is the annualized Sharpe ratio.
$\alpha$ is the monthly abnormal return in decimal from the corresponding factor model. The factor models are the Capital Asset Pricing Model (CAPM) and the Fama-French three-, five-, and six-factor (FF3F, FF5F, and FF6F) models.
Statistical significance is based on Newey-West $t$-statistics adjusted with 6 lags: $^{**}p{<}0.01$, $^{*}p{<}0.05$.}
\end{minipage}

\pdfdest name {figs-tabs-tab3} fit
\bookmark[level=2,dest=figs-tabs-tab3]{Table 3: Multiclass Sensitivity}
\caption{Multiclass Sensitivity}
\label{tab:multiclass}
\begin{tabularx}{1.3\textwidth}{l Y Y Y Y Y Y Y Y Y Y Y Y}
\toprule
 & \multicolumn{3}{c}{\textbf{2-Class}} & \multicolumn{3}{c}{\textbf{3-Class}}
 & \multicolumn{3}{c}{\textbf{5-Class}} & \multicolumn{3}{c}{\textbf{10-Class}} \\
\cmidrule(lr){2-4}\cmidrule(lr){5-7}\cmidrule(lr){8-10}\cmidrule(lr){11-13}
Metric & L-S & Long & Short & L-S & Long & Short & L-S & Long & Short & L-S & Long & Short \\
\midrule
\multicolumn{13}{c}{\textbf{Panel A}: Value-Weighted Soft Construction} \\
\midrule
Average Return   & 0.378 & 0.168 & 0.211 & 0.426 & 0.228 & 0.199 & 0.500 & 0.252 & 0.245 & 0.504 & 0.240 & 0.264 \\
Standard Deviation       & 0.285 & 0.145 & 0.350 & 0.242 & 0.270 & 0.382 & 0.215 & 0.357 & 0.386 & 0.243 & 0.426 & 0.404 \\
Sharpe Ratio            & 1.327 & 0.911 & 0.508 & 1.763 & 0.714 & 0.435 & 2.329 & 0.622 & 0.547 & 2.080 & 0.485 & 0.571 \\
CAPM $\alpha$        & 0.037** & 0.005** & 0.027** & 0.038** & 0.005* & 0.027** & 0.042** & 0.006 & 0.030** & 0.041** & 0.003 & 0.033** \\
FF3F $\alpha$        & 0.034** & 0.004** & 0.024** & 0.037** & 0.008** & 0.024** & 0.042** & 0.009* & 0.027** & 0.042** & 0.007 & 0.030** \\
FF5F $\alpha$        & 0.027** & 0.004** & 0.018** & 0.032** & 0.011** & 0.016** & 0.039** & 0.013** & 0.020** & 0.041** & 0.013** & 0.022** \\
FF6F $\alpha$        & 0.025** & 0.003** & 0.016** & 0.029** & 0.010** & 0.014** & 0.037** & 0.014** & 0.017** & 0.041** & 0.015** & 0.020** \\
\midrule
\multicolumn{13}{c}{\textbf{Panel B}: Equal-Weighted Soft Construction} \\
\midrule
Average Return   & 0.367 & 0.192 & 0.171 & 0.522 & 0.360 & 0.164 & 0.585 & 0.372 & 0.209 & 0.651 & 0.369 & 0.283 \\
Standard Deviation       & 0.267 & 0.149 & 0.318 & 0.174 & 0.298 & 0.357 & 0.163 & 0.357 & 0.351 & 0.173 & 0.374 & 0.337 \\
Sharpe Ratio            & 1.373 & 1.097 & 0.434 & 2.993 & 1.085 & 0.366 & 3.586 & 0.959 & 0.500 & 3.767 & 0.896 & 0.738 \\
CAPM $\alpha$        & 0.034** & 0.007** & 0.021** & 0.043** & 0.016** & 0.022** & 0.048** & 0.017** & 0.025** & 0.054** & 0.017** & 0.031** \\
FF3F $\alpha$        & 0.031** & 0.007** & 0.019** & 0.042** & 0.018** & 0.019** & 0.048** & 0.020** & 0.022** & 0.054** & 0.019** & 0.029** \\
FF5F $\alpha$        & 0.025** & 0.005** & 0.014** & 0.039** & 0.021** & 0.013** & 0.046** & 0.024** & 0.016** & 0.053** & 0.025** & 0.023** \\
FF6F $\alpha$        & 0.022** & 0.005** & 0.011** & 0.038** & 0.023** & 0.010** & 0.046** & 0.027** & 0.014** & 0.054** & 0.028** & 0.021** \\
\bottomrule
\end{tabularx}
\end{table}
\end{landscape}
\clearpage

\begin{landscape}
\begin{table}
\scriptsize
\setlength{\tabcolsep}{3pt}
\centering
\caption*{\textbf{Table~\ref{tab:multiclass}}. Multiclass Sensitivity (Continued)}
\begin{tabularx}{1.3\textwidth}{l Y Y Y Y Y Y Y Y Y Y Y Y}
\toprule
 & \multicolumn{3}{c}{\textbf{2-Class}} & \multicolumn{3}{c}{\textbf{3-Class}}
 & \multicolumn{3}{c}{\textbf{5-Class}} & \multicolumn{3}{c}{\textbf{10-Class}} \\
\cmidrule(lr){2-4}\cmidrule(lr){5-7}\cmidrule(lr){8-10}\cmidrule(lr){11-13}
Metric & L-S & Long & Short & L-S & Long & Short & L-S & Long & Short & L-S & Long & Short \\

\midrule
\multicolumn{13}{c}{\textbf{Panel C}: Value-Weighted Hard Construction} \\
\midrule
Average Return   & 0.104 & 0.132 & -0.024 & 0.157 & 0.168 & -0.015 & 0.193 & 0.192 & -0.000 & 0.230 & 0.195 & 0.035 \\
Standard Deviation       & 0.163 & 0.149 & 0.261 & 0.133 & 0.246 & 0.298 & 0.134 & 0.274 & 0.317 & 0.150 & 0.309 & 0.349 \\
Sharpe Ratio            & 0.639 & 0.627 & -0.218 & 1.180 & 0.562 & -0.163 & 1.439 & 0.585 & -0.106 & 1.535 & 0.522 & 0.005 \\
CAPM $\alpha$        & 0.012** & 0.001** & 0.006** & 0.014** & 0.001 & 0.008** & 0.017** & 0.002 & 0.009** & 0.021** & 0.002 & 0.013** \\
FF3F $\alpha$        & 0.010** & 0.001** & 0.004** & 0.014** & 0.003** & 0.005** & 0.017** & 0.004** & 0.007** & 0.020** & 0.004* & 0.010** \\
FF5F $\alpha$        & 0.006** & 0.000 & 0.000 & 0.011** & 0.005** & 0.000 & 0.013** & 0.006** & 0.001 & 0.017** & 0.008** & 0.004 \\
FF6F $\alpha$        & 0.005** & 0.000 & -0.001 & 0.009** & 0.005** & -0.001 & 0.011** & 0.006** & -0.001 & 0.016** & 0.008** & 0.002 \\
\midrule
\multicolumn{13}{c}{\textbf{Panel D}: Equal-Weighted Hard Construction} \\
\midrule
Average Return   & 0.130 & 0.168 & -0.043 & 0.217 & 0.252 & -0.030 & 0.258 & 0.276 & -0.013 & 0.330 & 0.304 & 0.025 \\
Standard Deviation       & 0.154 & 0.166 & 0.264 & 0.117 & 0.236 & 0.286 & 0.111 & 0.253 & 0.300 & 0.104 & 0.278 & 0.323 \\
Sharpe Ratio            & 0.843 & 0.837 & -0.289 & 1.858 & 0.910 & -0.223 & 2.313 & 0.944 & -0.154 & 3.163 & 0.973 & -0.026 \\
CAPM $\alpha$        & 0.013** & 0.005** & 0.003 & 0.019** & 0.008** & 0.005 & 0.022** & 0.010** & 0.006* & 0.028** & 0.013** & 0.010** \\
FF3F $\alpha$        & 0.011** & 0.004** & 0.001 & 0.018** & 0.009** & 0.003 & 0.021** & 0.011** & 0.004 & 0.028** & 0.015** & 0.007** \\
FF5F $\alpha$        & 0.007** & 0.004** & -0.002 & 0.014** & 0.010** & -0.001 & 0.018** & 0.013** & 0.000 & 0.025** & 0.017** & 0.002 \\
FF6F $\alpha$        & 0.006** & 0.004** & -0.004** & 0.013** & 0.011** & -0.003* & 0.017** & 0.014** & -0.002 & 0.024** & 0.019** & 0.000 \\
\bottomrule
\end{tabularx}
\end{table}
\end{landscape}
\clearpage


\begin{landscape}
\begin{table}[!htbp]
\scriptsize
\setlength{\tabcolsep}{3pt}

\centering
\begin{minipage}{1.3\textwidth}
\justifying{\noindent This table replicates Table~\ref{tab:portfolio_perform}, restricting the investable universe to stocks with above-median market capitalization. The table reports out-of-sample long-short portfolio performance for classification and regression models over July 1983--December 2024 (498 months). For each model, stocks with above-median market capitalization are assigned monthly by the predicted signal into 10 deciles. The long (short) portfolio holds the top (bottom) decile. The columns report the performance of the long-short portfolios based on the stacking ensemble and its gradient boosted trees (GBT), random forest (RF), and neural network (NN) base models, alongside the logistic regression and ordinary least squares (OLS) linear benchmarks. Panels A and B report, respectively, the value-weighted and equal-weighted long-short portfolio performance. Classification forms each class's portfolio independently: stocks are ranked by their predicted probability of belonging to that class, and the top 10\% of stocks enter that class's portfolio. Average Return is the annualized average raw return in decimal. Standard Deviation is the annualized standard deviation in decimal. Sharpe Ratio is the annualized Sharpe ratio. $\alpha$ is the monthly abnormal return in decimal from the corresponding factor model. The factor models are the Capital Asset Pricing Model (CAPM) and the Fama-French three-, five-, and six-factor (FF3F, FF5F, and FF6F) models. Statistical significance is based on Newey-West $t$-statistics adjusted with 6 lags: $^{**}p{<}0.01$, $^{*}p{<}0.05$.}
\end{minipage}

\pdfdest name {figs-tabs-tab4} fit
\bookmark[level=2,dest=figs-tabs-tab4]{Table 4: Large-Cap Performance: Top 50\% by Market Capitalization}
\caption{Large-Cap Performance: Top 50\% by Market Capitalization}
\label{tab:top_me_perform}
\begin{tabularx}{1.3\textwidth}{l Y Y Y Y Y Y Y Y Y Y}
\toprule
 & \multicolumn{5}{c}{\textbf{Classification}} & \multicolumn{5}{c}{\textbf{Regression}} \\
\cmidrule(lr){2-6}\cmidrule(lr){7-11}
Metric & Stack & GBT & RF & NN & Logistic & Stack & GBT & RF & NN & OLS \\
\midrule
\multicolumn{11}{c}{\textbf{Panel A}: Value-Weighted Long-Short} \\
\midrule
Average Return   & 0.282 & 0.241 & 0.146 & 0.275 & 0.194 & 0.179 & 0.084 & 0.103 & 0.155 & 0.152 \\
Standard Deviation       & 0.218 & 0.211 & 0.291 & 0.205 & 0.194 & 0.143 & 0.130 & 0.175 & 0.144 & 0.141 \\
Sharpe Ratio            & 1.296 & 1.140 & 0.502 & 1.340 & 1.002 & 1.252 & 0.651 & 0.588 & 1.075 & 1.074 \\
CAPM $\alpha$        & 0.024** & 0.022** & 0.012** & 0.023** & 0.017** & 0.015** & 0.007** & 0.011** & 0.013** & 0.012** \\
FF3F $\alpha$        & 0.024** & 0.021** & 0.013** & 0.023** & 0.016** & 0.015** & 0.007** & 0.008** & 0.013** & 0.012** \\
FF5F $\alpha$        & 0.023** & 0.019** & 0.011** & 0.020** & 0.014** & 0.013** & 0.006** & 0.005* & 0.010** & 0.011** \\
FF6F $\alpha$        & 0.020** & 0.017** & 0.009* & 0.019** & 0.013** & 0.011** & 0.005* & 0.003 & 0.009** & 0.009** \\
\midrule
\multicolumn{11}{c}{\textbf{Panel B}: Equal-Weighted Long-Short} \\
\midrule
Average Return   & 0.333 & 0.293 & 0.207 & 0.298 & 0.239 & 0.217 & 0.115 & 0.145 & 0.230 & 0.224 \\
Standard Deviation       & 0.155 & 0.145 & 0.209 & 0.140 & 0.152 & 0.140 & 0.121 & 0.170 & 0.132 & 0.138 \\
Sharpe Ratio            & 2.146 & 2.019 & 0.991 & 2.126 & 1.567 & 1.547 & 0.955 & 0.849 & 1.735 & 1.631 \\
CAPM $\alpha$        & 0.028** & 0.025** & 0.016** & 0.025** & 0.020** & 0.020** & 0.010** & 0.014** & 0.020** & 0.018** \\
FF3F $\alpha$        & 0.028** & 0.025** & 0.017** & 0.025** & 0.020** & 0.019** & 0.009** & 0.012** & 0.019** & 0.018** \\
FF5F $\alpha$        & 0.026** & 0.023** & 0.016** & 0.023** & 0.017** & 0.015** & 0.007** & 0.007** & 0.016** & 0.016** \\
FF6F $\alpha$        & 0.024** & 0.021** & 0.015** & 0.021** & 0.016** & 0.013** & 0.006** & 0.005** & 0.014** & 0.014** \\
\bottomrule
\end{tabularx}
\end{table}
\end{landscape}
\clearpage


\begin{landscape}
\begin{table}[!htbp]
\scriptsize
\setlength{\tabcolsep}{3pt}

\centering
\begin{minipage}{1.3\textwidth}
\justifying{\noindent This table reports maximum drawdown (MaxDD) and average monthly turnover, both in decimal, of classification and regression models' long-short portfolios and the market portfolio over July 1983--December 2024 (498 months). Turnover sums purchases and sales, so replacing a leg in full registers as 200\%, and the long-short turnover is the average of the two legs. For each model, stocks are assigned monthly by the predicted signal into 10 deciles. The long (short) portfolio holds the top (bottom) decile. Classification forms each class's portfolio independently: stocks are ranked by their predicted probability of belonging to that class, and the top 10\% of stocks enter that class's portfolio. The market portfolio is a buy-and-hold portfolio over the entire market. The columns report the results of the market portfolio and the portfolios based on the stacking ensemble and its gradient boosted trees (GBT), random forest (RF), and neural network (NN) base models, alongside the logistic regression and ordinary least squares (OLS) linear benchmarks.}
\end{minipage}

\pdfdest name {figs-tabs-tab5} fit
\bookmark[level=2,dest=figs-tabs-tab5]{Table 5: Maximum Drawdown and Turnover}
\caption{Maximum Drawdown and Turnover}
\begin{tabularx}{1.3\textwidth}{l Y Y Y Y Y Y Y Y Y Y Y}
\toprule
 & & \multicolumn{5}{c}{\textbf{Classification}} & \multicolumn{5}{c}{\textbf{Regression}} \\
\cmidrule(lr){3-7}\cmidrule(lr){8-12}
Metric & Market & Stack & GBT & RF & NN & Logistic & Stack & GBT & RF & NN & OLS \\
\midrule
Value-weighted MaxDD         & -0.516 & -0.314 & -0.283 & -0.758 & -0.301 & -0.382 & -0.354 & -0.536 & -0.482 & -0.373 & -0.387 \\
Equal-weighted MaxDD         & -0.591 & -0.136 & -0.112 & -0.287 & -0.175 & -0.164 & -0.085 & -0.234 & -0.273 & -0.099 & -0.155 \\
Value-weighted Turnover      & 0.088 & 1.440 & 1.428 & 1.725 & 1.344 & 1.362 & 1.483 & 1.403 & 1.399 & 1.414 & 1.446 \\
Equal-weighted Turnover      & 0.122 & 1.381 & 1.358 & 1.683 & 1.275 & 1.253 & 1.242 & 1.224 & 1.209 & 1.208 & 1.301 \\
\bottomrule
\end{tabularx}
\label{tab:costs}
\end{table}
\end{landscape}

\begin{appendices}
\renewcommand{\appendixtocname}{Appendix} 
\phantomsection
\pdfbookmark[1]{Appendix}{appendix}
\centering\section*{\appendixtocname}
\addtocontents{toc}{\protect\contentsline{section}{Appendix}{\thepage}{appendix}\protected@file@percent}
\makeatletter
\addtocontents{toc}{\let\protect\l@*section}
\makeatother

\setcounter{table}{0}
\renewcommand{\thetable}{A\arabic{table}}
\renewcommand*{\theHtable}{\thetable}


\begin{table}[!htbp]
\small
\justifying{\noindent This table details the hyperparameter search grids for all models. The hyperparameters are determined through tuning on the validation datasets. GBT and RF are gradient boosted trees and random forest models, respectively. NN is a fully connected feedforward network. For the tree models (RF and GBT), each number in the search grid represents a candidate maximum tree depth. For the NN model, the primary architectural choices concern the number of hidden layers and the number of neurons per layer. Each pair of parentheses encloses a single candidate architecture. Each number within it represents the neuron count in one hidden layer. We keep each model's usual search space while isolating its iterative capacity, namely the number of trees and the number of epochs.}

\centering
\pdfdest name {figs-tabs-tabA1} fit
\bookmark[level=2,dest=figs-tabs-tabA1]{Table A1: Hyperparameter Search Grids}
\caption{Hyperparameter Search Grids}
\begin{tabularx}{\textwidth}{lXX}
\toprule
Model & Hyperparameter & Choice \\
\midrule
RF and GBT & Depth & 1, 2, 4, 8 \\
\multirow{6}{*}{NN} & 1 Layer & (8), (16), (32), (64), (128) \\
& 2 Layers & (128, 64), (64, 32), (32, 16), (16, 8) \\
& 3 Layers & (128, 64, 32), (64, 32, 16), (32, 16, 8) \\
& 4 Layers & (128, 64, 32, 16), (64, 32, 16, 8) \\
& 5 Layers & (128, 64, 32, 16, 8) \\
\bottomrule
\end{tabularx}
\label{tab:arch_search}
\end{table}
\clearpage

\begin{table}[!htbp]
\small

\justifying{\noindent This table summarizes the selected hyperparameters for each model type across all 42 training windows. For each model and each of its hyperparameters, we report the most frequently selected value and its count across the 42 windows. GBT and RF are gradient boosted trees and random forest models, respectively. NN is a fully connected feedforward network. The classification models use 10 decile classes for the target variable. The number of trees and the number of epochs are fixed caps of iterative training.}

\centering
\pdfdest name {figs-tabs-tabA2} fit
\bookmark[level=2,dest=figs-tabs-tabA2]{Table A2: Selected Hyperparameters}
\caption{Selected Hyperparameters}
\begin{tabularx}{\textwidth}{l l l Y c}
\toprule
Task & Model & Parameter & Most Common (Count) & \# Windows \\
\midrule
\multirow{7}{*}{Classification} & \multirow{2}{*}{GBT} & Max Depth & 4 (30), 2 (9), 1 (2), 8 (1) & 42 \\
 &  & Trees & 1000 (42) & 42 \\
 & \multirow{2}{*}{RF} & Max Depth & 8 (42) & 42 \\
 &  & Trees & 1000 (42) & 42 \\
& \multirow{5}{*}{NN} & Hidden Layers & [16, 8] (16), [16] (9), [32] (6), [32, 16] (4), [32, 16, 8] (4), [8] (3) & 42 \\
 &  & Max Epochs & 100 (42) & 42 \\
 &  & Activation & Tanh (42) & 42 \\
\midrule
\multirow{7}{*}{Regression} & \multirow{2}{*}{GBT} & Max Depth & 1 (15), 2 (13), 4 (13), 8 (1) & 42 \\
 &  & Trees & 1000 (42) & 42 \\
 & \multirow{2}{*}{RF} & Max Depth & 8 (20), 1 (12), 4 (8), 2 (2) & 42 \\
 &  & Trees & 1000 (42) & 42 \\
& \multirow{5}{*}{NN} & Hidden Layers & [8] (14), [32] (8), [16] (8), [128] (5), [16, 8] (5), [128, 64, 32, 16] (1), [64, 32, 16, 8] (1) & 42 \\
 &  & Max Epochs & 100 (42) & 42 \\
 &  & Activation & Tanh (42) & 42 \\
\bottomrule
\end{tabularx}
\label{tab:selected_models}
\end{table}

\end{appendices}
\clearpage
\clearpage
\setcounter{section}{0}
\setcounter{figure}{0}
\setcounter{table}{0}
\renewcommand{\thesection}{OSM.\arabic{section}}
\renewcommand{\thesubsection}{OSM.\arabic{section}.\arabic{subsection}}
\renewcommand{\theHsection}{OSM.\arabic{section}}
\renewcommand{\theHsubsection}{OSM.\arabic{section}.\arabic{subsection}}
\renewcommand{\thefigure}{OSM\arabic{figure}}
\renewcommand{\thetable}{OSM\arabic{table}}

\setcounter{figure}{0}
\renewcommand{\thefigure}{OSM\arabic{figure}}
\renewcommand*{\theHfigure}{\thefigure}

\renewcommand{\thetable}{OSM\arabic{table}}
\renewcommand*{\theHtable}{\thetable}
\renewcommand{\thesection}{OSM.\arabic{section}}
\renewcommand{\thesubsection}{OSM.\arabic{section}.\arabic{subsection}}

\phantomsection
\pdfbookmark[0]{Online Supplemental Material}{osm-title}

\pagenumbering{arabic}
\vspace*{0.9in}
\newcounter{osmfnsave}
\setcounter{osmfnsave}{\value{footnote}}
\renewcommand{\thefootnote}{\fnsymbol{footnote}}
\setcounter{footnote}{0}
\begin{center}
{\Large\bfseries Online Supplemental Material}\\[10pt]
{\large\itshape Machine Learning Classification and Portfolio Construction: Does the Loss Function Matter?}\\[16pt]
{\large Yang Bai}\footnote{Assistant Professor of Finance, College of Business and Economics, California State University, Fullerton, California, \href{mailto:yabai@fullerton.edu}{yabai@fullerton.edu}.}\\[6pt]
{\large Kuntara Pukthuanthong}\footnote{Professor of Finance and Robert J. Trulaske, Jr. Professor of Finance, Robert J. Trulaske, Sr. College of Business, University of Missouri, Columbia, Missouri, \href{mailto:pukthuanthongk@missouri.edu}{pukthuanthongk@missouri.edu}.}
\\[14pt]
{\normalsize August 21, 2026}\\[2pt]
{\normalsize First Draft: May 31, 2020}
\end{center}
\renewcommand{\thefootnote}{\arabic{footnote}}
\setcounter{footnote}{\value{osmfnsave}}
\onehalfspacing
\justifying

This document is the Online Supplemental Material for the study. Section~\ref{sec:osm_data} describes the sample construction and reports summary statistics for the firm characteristics. Section~\ref{sec:osm_models} provides formal specifications of the models and the expanding-window training protocol described in the main paper. Section~\ref{sec:osm_optimizer} presents a robustness analysis comparing neural network optimizers. Section~\ref{sec:osm_subperiods} reports out-of-sample (OOS) performance over four approximately equal-length windows and by National Bureau of Economic Research (NBER) business cycle regime. Section~\ref{sec:osm_costs} describes the calculation of the maximum drawdown, the turnover formula, the trading-cost haircut, and the short-side borrowing cost used throughout the paper. Sections~\ref{sec:osm8_diagnostics} and \ref{sec:osm9_chars} present supplemental empirical results on model diagnostics and the portfolio characteristic profiles, and Section~\ref{sec:osm_relative} examines alternative regression target definitions.

\section{Data}
\label{sec:osm_data}

In this section, we report the summary statistics of our data. We obtain the stock returns from the Center for Research in Security Prices and financial data from Compustat. Our sample covers common stocks listed on the New York Stock Exchange, American Stock Exchange, and National Association of Securities Dealers Automated Quotations from July 1962 to December 2024. We exclude observations with missing current returns and restrict the sample to common shares with share codes 10, 11, or 12. For factor-model tests and the risk-free rate, we use data from Kenneth French's data library \citep{osm:fama1992cross,osm:fama2015five}. The final sample contains 3,361,673 firm-month observations.

The predictor set consists of 102 firm characteristics reconstructed following \citet{osm:green2017characteristics}, plus the two-digit Standard Industrial Classification industry indicators. For model estimation, we transform each numeric characteristic each month into a cross-sectional normalized rank in $[-1, 1]$, consistent with prior literature \citep{osm:freyberger2020dissecting,osm:gu2020empirical,osm:howard2024choices}.\footnote{The choices of statistical transformation of the characteristics do not affect the main conclusions. The z-score normalization penalizes the regression models and enlarges the performance gap.} Table~\ref{tab:char_stats} reports the summary statistics of the 102 firm characteristics used in the prediction models.

\begin{center}
  [Insert Table~\ref{tab:char_stats} Here]
\end{center}

\section{Model Specifications}
\label{sec:osm_models}

\subsection{Prediction Process}

Firm characteristics and industry indicators serve as inputs to each machine learning model. During training, the optimization process calibrates model parameters on in-sample data. For classification, it aligns the predicted decile probabilities with the realized cross-sectional return distribution. For regression, it aligns the predicted return with the realized return. The OOS classification output is a 10-dimensional probability vector for each stock in our main analysis. The predicted probability of top-decile membership forms the long-leg signal, and the predicted probability of bottom-decile membership forms the short-leg signal, as described in Section~2.6 of the main paper. For regression, the output is a scalar predicted return, and stocks are ranked accordingly.

\subsection{Gradient Boosted Trees}

Gradient boosted trees (GBT) construct an additive ensemble by sequentially fitting trees to the negative gradient of the loss function \citep{osm:friedman2001greedy}. GBT takes the firm characteristics and industry indicators as its input features.
The ensemble sums $B$ trees,
\begin{align}
\mathbf{z} = \sum_{b=1}^{B} \boldsymbol{\phi}_b(\mathbf{x}),
\end{align}
\noindent where $\boldsymbol{\phi}_b$ is the round-$b$ tree, and each round fits $\boldsymbol{\phi}_b$ to the pseudo-residual, namely the negative gradient of the loss evaluated at the ensemble $\mathbf{z}^{(b-1)}$ after $b-1$ rounds. For classification, each tree outputs a vector over the $|D|=10$ deciles, $\boldsymbol{\phi}_b:\mathbb{R}^{K}\to\mathbb{R}^{|D|}$, where $K$ is the number of characteristics, so $\mathbf{z}$ is a vector of class scores that the softmax maps to probabilities $\mathbf{q}=\operatorname{softmax}(\mathbf{z})$, the loss is the cross-entropy, and the round-$b$ pseudo-residual for stock $i$ in month $t$ is $\mathbf{g}_{i,t,b}=\mathbf{y}_{i,t}-\mathbf{q}^{(b-1)}_{i,t}$. For regression, each tree instead outputs a scalar, $\boldsymbol{\phi}_b:\mathbb{R}^{K}\to\mathbb{R}$, so the ensemble output is the return forecast $\widehat r_{i,t}$, the loss is the mean squared error, and the round-$b$ pseudo-residual is $r_{i,t}-\widehat r^{(b-1)}_{i,t}$. We search over maximum tree depths of 1, 2, 4, and 8, and we grow 1000 trees in each candidate with early stopping after 5 rounds.

\subsection{Random Forest}

A random forest (RF) builds a large ensemble of decision trees using bootstrap aggregating \citep{osm:breiman2001random}. It uses the same firm characteristics and industry indicators as inputs.
Each tree $\boldsymbol{\phi}(\mathbf{x})$ recursively partitions the feature space into $J$ nonoverlapping leaves $R_m$ and assigns each leaf a response $\boldsymbol{\gamma}_m$:
\begin{align}
\boldsymbol{\phi}(\mathbf{x}) = \sum_{m=1}^{J} \boldsymbol{\gamma}_m \cdot \mathbf{I}(\mathbf{x} \in R_m),
\end{align}
\noindent where $\mathbf{I}(\cdot)$ is the indicator function, equal to 1 when $\mathbf{x}$ lies in leaf $R_m$ and 0 otherwise. At each internal node, a random subset of features is considered before the best split is selected. This decorrelates the individual trees and reduces ensemble variance.
For classification, each leaf response $\boldsymbol{\gamma}_m$ is a class-probability vector, and the forest produces the probability vector directly as the average across its trees, $\mathbf{q}=\frac{1}{B}\sum_{b=1}^{B}\boldsymbol{\phi}_b(\mathbf{x})$, without a softmax.
For regression, the leaf response is a scalar and the forest averages the predicted returns.
We search over maximum tree depths of 1, 2, 4, and 8, following Appendix Table~A1 of the main paper. We grow 1000 trees in each candidate forest. The feature sampling is set to the square root of the feature count for classification and to one-third for regression, while the row sampling is set to 0.632 \citep{osm:breiman2001random}.

\subsection{Neural Network}

Our neural network (NN) is a fully connected feedforward architecture following \citet{osm:gu2020empirical}. The input layer contains the firm characteristics and industry indicators. Hidden layers apply, elementwise, the $\tanh$ activation function
\begin{align*}
\tanh(\xi) = \frac{\exp(\xi)-\exp(-\xi)}{\exp(\xi)+\exp(-\xi)}.
\end{align*}
A grid search over the candidate architectures listed in Appendix Table~A1 of the main paper selects the best layer-neuron specification on the validation dataset.

Formally, an NN with $L$ hidden layers maps the characteristic vector $\mathbf{x}\in\mathbb{R}^{K}$ through the input and the hidden layers
\begin{align}
\mathbf{h}^{0} = \mathbf{x}, \qquad \mathbf{h}^{l} = \tanh\!\left(\mathbf{U}_{l}\mathbf{h}^{l-1} + \mathbf{a}_{l}\right)\ \ (l=1,\dots,L),
\end{align}
\noindent where $\mathbf{x}$ is the input vector, $\mathbf{U}_{l}\in\mathbb{R}^{N_l\times N_{l-1}}$ is the layer-$l$ weight matrix, $\mathbf{a}_{l}\in\mathbb{R}^{N_l}$ is the bias vector, and $N_l$ is the number of neurons in layer $l$ with $N_0=K$. A linear output layer then maps the last hidden layer $\mathbf{h}^{L}$ to the prediction.

For classification in decile-portfolio construction, the output layer has $|D|=10$ units producing the score vector $\mathbf{z}=\mathbf{U}_{L+1}\mathbf{h}^{L}+\mathbf{a}_{L+1}$, which the softmax maps to probabilities,
\begin{align}
q(d) = \frac{\exp(z_d)}{\sum_{d' \in D} \exp(z_{d'})},
\end{align}
\noindent where $z_d$ is the $d$-th entry of $\mathbf{z}$. For regression, the output layer is instead a single linear unit producing a scalar return forecast $\widehat r_{i,t}=\mathbf{U}_{L+1}\mathbf{h}^{L}+a_{L+1}$. We train the NN for up to 100 epochs and select among the candidate architectures on the validation set according to their loss, and the selected model generates the OOS predictions. The optimizer is Adam with a learning rate of 0.01. We also stop training early after 5 epochs without improvement in the loss. Adam materially strengthens the NN under both loss functions, and it strengthens the regression side more, raising the regression NN's value-weighted Sharpe ratio by more than half. Our comparison therefore evaluates classification against regression at the strongest configuration we found for the regression side. As a consistency check, we also estimate every specification under Adadelta, the optimizer used in the replication literature. The classification-regression gap is wider under Adadelta, at 1.83 versus 1.11 in value-weighted Sharpe ratios against 2.08 versus 1.39 under Adam, so the results reported in the main paper present the more conservative version of the comparison. The conclusion is not an artifact of the optimizer. Table~\ref{tab:optimizer} reports the full comparison.

\subsection{Stacking Ensemble}

The stacking ensemble combines three base models, namely GBT, RF, and NN, using equal-weight averaging. The classification models average per-class probability predictions and assign stocks to portfolios based on the averaged probabilities across the base models. The regression models average the predicted returns and assign stocks to portfolios based on the averaged return across the base models.

\subsection{Experimental Design and Training Windows}

In our setting, the models are updated each June. For OOS predictions in period $t$, the training set includes all stock-month observations from July 1962 through the end of period $t-2$. The 12 months of period $t-1$ serve as a held-out validation set for selecting hyperparameters. This cycle repeats 42 times, covering the full OOS evaluation period of July 1983 through December 2024.

\section{Optimizer Comparison}
\label{sec:osm_optimizer}

\begin{center}
    [Insert Table~\ref{tab:optimizer} Here]
\end{center}

Table~\ref{tab:optimizer} reports the optimizer consistency check. Under Adadelta, we train every NN specification with a learning rate of 1.0 and the implementation-default decay parameters ($\rho = 0.9$, $\varepsilon = 10^{-6}$), holding every other design choice fixed. Adam with a learning rate of 0.01 strengthens the NNs under both loss functions, and the gains concentrate on the regression side. The regression NN's value-weighted Sharpe ratio rises from 0.85 to 1.31, a gain of more than half, while the classification NN rises from 1.65 to 1.88. As a result, the classification-regression gap is wider under Adadelta than under Adam in every comparison in the table. For the value-weighted stacking ensembles, the Sharpe ratios are 1.83 against 1.11 under Adadelta and 2.08 against 1.39 under Adam. The results reported in the main paper therefore present the conservative version of the comparison, and the classification advantage is not an artifact of the optimizer choice.

\section{Subperiod Analysis}
\label{sec:osm_subperiods}

Table~\ref{tab:subperiod} reports performance across four nonoverlapping subperiods of approximately a decade each that span the full OOS evaluation period from July 1983 to December 2024. The four periods are 198307:199311, 199312:200404, 200405:201408, and 201409:202412.

\begin{center}
    [Insert Table~\ref{tab:subperiod} Here]
\end{center}

We find that the classification long-short portfolio earns positive value-weighted Sharpe ratios across all four subperiods, at 3.73, 2.63, 1.32, and 1.58 in successive decades. The broadly declining pattern over the first three decades is consistent with the literature on attenuation in anomaly profits as markets become more competitive \citep{osm:ChenZimmermann2021, osm:chordia2014have, osm:green2017characteristics, osm:mclean2016does}. The matched regression value-weighted Sharpe ratios are 2.07, 2.19, 1.34, and 0.29. Classification leads in three of the four subperiods and is essentially tied with regression in 200405:201408. Its advantage is widest at the two ends of the sample, and in the most recent decade the regression Sharpe ratio falls to 0.29 while classification maintains 1.58.

Consistent with the diagnostics in Table~\ref{tab:model_diag}, the short leg is the primary performance driver in the first and last subperiods for classification, with Sharpe ratios of 1.32 and 0.71 against 0.49 and 0.38 for the long leg, while the long leg leads in the middle two decades. Neither leg dominates throughout. The prominence of the short leg at the sample ends is consistent with evidence that short-sale impediments allow overpricing to persist \citep{osm:nagel2005short, osm:stambaugh2012short}, although that evidence is established for characteristic- and regression-based sorts rather than for classification portfolios, so we read it as context rather than as a direct implication for our setting.

We also examine NBER expansions and recessions. During the 458 expansion months in our OOS window, the classification portfolio achieves a value-weighted Sharpe ratio of 2.12, whereas the matched regression counterpart reaches 1.32. During the 40 recession months, the two methods perform comparably, at 1.87 and 2.15. With only 40 observations, the recession split is measured with substantial noise. The expansion months, which constitute over 90\% of the sample, carry the classification advantage.

\section{Costs of Implementation Details}
\label{sec:osm_costs}

This section describes the maximum drawdown calculation, the turnover formula, the trading-cost haircut, and the short-side borrowing cost used throughout the paper. Each definition matches the implementation behind the after-cost results and the cumulative log-return figure reported in the main paper and in Table~\ref{tab:aftercost}.

\subsection{Maximum Drawdown}

Maximum drawdown measures the largest peak-to-trough decline in cumulative portfolio value over the OOS window. Formally, it is defined as
\begin{align}
\text{MaxDD} = \min_{t}\frac{V_t - V_t^{\max}}{V_t^{\max}},\qquad V_t^{\max}=\max_{s\le t}V_s,
\end{align}
\noindent where $V_t$ is the cumulative portfolio value in month $t$ and $V_t^{\max}$ is its running peak through month $t$. A value closer to zero indicates smaller peak-to-trough losses and therefore more resilient performance under adverse market conditions.

\subsection{Turnover}

Portfolio turnover is computed following \citet{osm:gu2020empirical} and \citet{osm:neely2014forecasting}. For a portfolio with stock weights $w_{j,t}$, monthly turnover is defined as
\begin{align}
\text{Turnover} = \frac{1}{T}\sum_{t=1}^{T}\left(\sum_{j}\left|w_{j,t+1}-\frac{w_{j,t}(1+r_{j,t+1})}{1+\sum_k w_{k,t}r_{k,t+1}}\right|\right),
\end{align}
\noindent where $w_{j,t}$ is the weight of stock $j$ at the end of month $t$ and $r_{j,t+1}$ is its return in month $t+1$. The denominator adjusts for the drift in portfolio weights due to returns before rebalancing. Equal-weighted and value-weighted turnover are computed separately for each leg. For the long-short portfolio, turnover is the average of the two legs rather than netting weights across the long and short sides. A zero-investment portfolio has no net capital base, so its turnover is expressed per dollar of gross exposure, and averaging the two legs places the long-short portfolio on the same 0\% to 200\% scale as the individual legs while capturing the cost of rebalancing each leg independently.

\subsection{Trading-Cost Haircut}

The after-cost return is computed as a scalar haircut applied to the full-period mean return of each portfolio:
\begin{align}
\bar{r}^{\,\text{net}} = \bar{r} - \overline{\text{Turnover}} \times \frac{c}{10{,}000},
\end{align}
\noindent where $\bar{r}$ is the average monthly excess return, $\overline{\text{Turnover}}$ is the time-series mean two-way turnover, and $c$ is the trading cost per one-way trade in basis points, so that $c$ times two-way turnover charges the full round trip. The baseline specification sets $c = 30$ basis points (TC30), consistent with \citet{osm:novymarx2016taxonomy}. Multiplying average turnover by the per-unit cost and subtracting once from the mean return is equivalent to deducting the same cost each period when turnover is stable.

\subsection{Short-Side Borrowing Cost}

An additional borrowing cost is applied to all strategies that include a short leg. The adjustment is a flat monthly haircut equal to the annualized borrow rate divided by 12:
\begin{align}
\bar{r}^{\,\text{net,short}} = \bar{r}^{\,\text{net}} - \frac{c_b}{10{,}000 \times 12},
\end{align}
\noindent where $c_b$ is the annual borrow cost in basis points. The baseline specification sets $c_b = 50$ basis points annually (BC50), consistent with \citet{osm:davolio2002market}.\footnote{\citet{osm:davolio2002market} shows that it is appropriate to apply a 24-basis-point deduction annually to value-weighted portfolios and a 60-basis-point deduction annually to equal-weighted portfolios. Given the period that \citet{osm:davolio2002market} examines, we err on the conservative side and deduct 50 basis points from both the value-weighted and equal-weighted portfolios.} This rate is applied uniformly across all stocks in the short leg and does not vary with stock-level borrow availability. The borrowing friction is modeled as a carry cost rather than as a hard short-sale quantity constraint. We impose no locate-failure rule, short-sale cap, or no-trade constraint. The combined TC30+BC50 specification therefore deducts from the mean return 30 basis points times average two-way turnover, plus an additional monthly short-side charge of $50/12$ basis points, equal to the 50-basis-point annual borrowing cost spread evenly across the year.

\subsection{After-Cost Portfolio Performance}

Table~\ref{tab:aftercost} reports post-adjustment return, standard deviation, Sharpe ratio, and factor-model alphas for the stacking ensemble portfolios under TC30+BC50, separately for value-weighted and equal-weighted portfolios. After transaction costs, the classification stacking portfolio still earns a value-weighted Sharpe ratio of 1.85, whereas the matched regression portfolio falls to 1.09 \citep{osm:davolio2002market, osm:novymarx2016taxonomy}. The classification strategy therefore remains economically attractive after realistic implementation frictions.

\begin{center}
    [Insert Table~\ref{tab:aftercost} Here]
\end{center}

\subsection{Construction of After-Cost Cumulative Return Figures}

Figure~2 of the main paper plots cumulative log returns of the long-short classification and regression stacking portfolios after TC30+BC50 cost deductions. We construct cost-adjusted cumulative log-return series by deducting from each monthly portfolio return a trading-cost haircut equal to 30 basis points per one-way trade times the portfolio's reported average monthly two-way turnover, a full round-trip charge. For short-exposed strategies, including the standalone short leg and the long-short portfolio, we additionally deduct a 50-basis-point annual borrowing-cost charge allocated monthly \citep{osm:davolio2002market}. For weighting scheme $w \in \{\text{EW}, \text{VW}\}$, where EW and VW denote equal-weighted and value-weighted portfolios, let $\tau_{\mathrm{L},w}$, $\tau_{\mathrm{S},w}$, and $\tau_{\mathrm{L\text{-}S},w}$ denote the average monthly two-way turnover of the long leg, short leg, and long-short portfolio. The net monthly returns for the legs and long-short portfolio are
\begin{align}
r^{\text{net}}_{\mathrm{L},t,w} &= r_{\mathrm{L},t,w} - 0.003\,\tau_{\mathrm{L},w}, \\
r^{\text{net}}_{\mathrm{S},t,w} &= r_{\mathrm{S},t,w} - 0.003\,\tau_{\mathrm{S},w} - \frac{0.005}{12}, \\
r^{\text{net}}_{\mathrm{L\text{-}S},t,w} &= r_{\mathrm{L\text{-}S},t,w} - 0.003\,\tau_{\mathrm{L\text{-}S},w} - \frac{0.005}{12},
\end{align}
\noindent where the short-leg return $r_{\mathrm{S},t,w}$ is the return from short selling the bottom-class stocks before cost deduction, and the long-leg and long-short portfolio returns $r_{\mathrm{L},t,w}$ and $r_{\mathrm{L\text{-}S},t,w}$ are the respective holding returns. The short leg and the long-short portfolio both include the borrowing-cost term because they hold short positions. The plotted objects in Figure~2 of the main paper are cumulative log returns, defined as
\begin{equation}
\text{Cumulative Log Return}_t = \sum_{t' \le t} \ln(1 + r^{\text{net}}_{t'}),
\end{equation}
\noindent where $r^{\text{net}}_{t'}$ is the after-cost return of the corresponding portfolio at time $t'$. The figure therefore shows net implementation performance under a restrictive TC30+BC50 specification. This construction is directly comparable to the TC30+BC50 results in Table~\ref{tab:aftercost}.

\section{Model Diagnostics}
\label{sec:osm8_diagnostics}

Table~\ref{tab:model_diag} reports predictive accuracy, confusion matrices, feature importance, and Fama-MacBeth regressions for the classification and regression stacking ensembles. Because the classification ensemble can assign stocks under either the hard or the soft rule, the diagnostics span three sets of decile assignments, those of the hard construction, the soft construction, and regression. Although our main results are consistent across all five model pairs, we focus on the stacking ensemble for this diagnostic analysis because it summarizes the information from the three base models in each case. We report both construction results to show that the classification advantage is robust and that the soft construction improves on the hard construction.

\begin{center}
    [Insert Table~\ref{tab:model_diag} Here]
\end{center}

\subsection{Confusion Matrices and Portfolio Construction}

To visualize classification patterns, we present row-normalized confusion matrices, where each row sums to 100\% and shows the proportion of actual class $d$ stocks that the model assigns to each predicted class $d'$. The diagonal elements of the confusion matrices correspond to class-specific recall rates, and off-diagonal patterns show the model's error structure, particularly its tendency to confuse neighboring deciles. The confusion matrices thus expose systematic misclassification biases. For instance, misclassified stocks may cluster in adjacent classes, indicating reasonable errors, or scatter randomly, suggesting model failure. 

Panels A through C of Table~\ref{tab:model_diag} report the models' row-normalized confusion matrices. Their diagonal entries directly describe the purity of the long and short legs. The classification ensemble concentrates much of its accuracy in the extreme deciles that matter for portfolio formation. In Panel B, recall is especially high in Class 1, and it is materially higher in Class 10 than in most of the middle deciles. This concentration aligns with the decile-assignment objective of a long-short classification strategy.

Panel C shows the soft construction confusion matrix. For each class $d$, the soft construction determines membership in class $d$ by ranking stocks on $q_{i,t}(d)$ independently. As shown in Panel C, such an independent ranking reduces the overassignment of stocks to some middle classes under the hard construction, thereby improving the balance between true-positive and false-positive assignments. This improvement matters for portfolio construction, because the objective is to include stocks with high return potential in the long portfolio and stocks with low return potential in the short portfolio. The goal is to maximize correct assignments in the extreme portfolios while minimizing the inclusion of stocks that belong to other deciles. The improved class-level trade-off is thus consistent with the stronger economic performance we document in the previous sections.

By contrast, the regression confusion matrix in Panel A is substantially flatter. Diagonal recall reaches only 19.43\% for Class 1 and 16.90\% for Class 10. Middle deciles exhibit near-uniform assignment around 10\%, indicating the model barely distinguishes among them. This diffuse pattern shows that weak decile separation produces noisy extreme portfolios contaminated by stocks drawn from across the return distribution, which is consistent with the inferior performance.

\subsection{Overall Accuracy}

To comprehensively evaluate the statistical performance, we employ multiple metrics that capture different dimensions of predictive accuracy. Balanced accuracy averages each class's sensitivity and specificity in a one-versus-rest sense and then averages these values across classes, which generally adjusts for class imbalance and ensures that the performance on minority classes is not overshadowed by more populous classes, although we do not have class imbalance in our setting. Cohen's kappa measures agreement between predicted and actual classifications while accounting for chance agreement, providing a more robust assessment than raw accuracy alone. Macro precision and macro recall average precision and recall uniformly across all classes. The macro F1-score is the harmonic mean of macro precision and macro recall and offers a single summary statistic that balances both concerns. Together, these metrics provide a holistic view of model performance.

Panels D and E report aggregate and by-class classification statistics. OOS accuracy under the hard construction is 16.57\%, well above the 10\% benchmark from random assignment across 10 classes. Balanced accuracy reaches 53.65\%, which is above the 50\% random benchmark. Cohen's kappa is also positive. The soft construction is slightly stronger on each of these metrics. However, the matched regression ensemble performs much worse on the same decile-classification exercise. Its OOS accuracy is 12.45\%, near the 10\% random baseline. Its balanced accuracy of 51.36\%, the macro average of per-class sensitivity and specificity, exceeds 50\% only modestly, confirming that the regression loss does not induce sharp separation across return deciles. This weak separation is consistent with its inferior portfolio performance documented earlier. Notably, precision increases markedly from regression to the soft construction, both overall and by class. Precision for the top and bottom portfolios, the most economically important deciles, jumps from 16.90\% and 19.40\% in regression to 20.98\% and 29.68\% in the soft construction. This precision gain means that the soft construction includes fewer false positives in its long-short portfolios, translating into the superior Sharpe ratios and alphas we observe. Across all metrics, the statistical performance aligns with the corresponding economic performance.

\subsection{Feature Importance}

We report average feature importance at the level of stacking ensembles across the windows and the submodels.\footnote{The feature importance is calculated and normalized to sum to 1 per training window. Then, we average the decimal feature importance across the 42 windows. A categorical variable can have multiple feature importance measures through its subcategories. We collect them together via summation for easier interpretation.} Panel F shows that the two stacking ensembles place weight on materially different signals. The classification model assigns its highest importance to the industry indicator sich2 and to return-uncertainty variables such as idiovol, retvol, baspread, and maxret. The regression ensemble also ranks sich2 first, but its next most important signal is firm size (mve), followed by short-term reversal (mom1m), the number of consecutive earnings increases (nincr), and signals tied to dividend yield (dy) and analyst coverage (nanalyst). Idiosyncratic volatility is the classification ensemble's second most important feature yet does not enter the regression ensemble's top 10, whereas firm size is the regression ensemble's second most important feature yet ranks only sixth for classification.

\subsection{Fama-MacBeth Regression of Model Correctness}

Over the OOS window, we estimate 60-month rolling Fama-MacBeth regressions to analyze the cross-sectional relationship between decile prediction outcomes and firm characteristics. This approach examines whether a particular modeling structure captures a stable relationship between the prediction target and the predictors. Specifically, if a model yields a high adjusted $R^2$ when regressing the stock-level prediction outcome on firm characteristics, it suggests a tighter statistical mapping from those characteristics to the prediction task. 

We define the dependent variable as a binary indicator of whether the model correctly classifies a stock into its actual return decile. The independent variables are the same 102 firm characteristics used in model training, along with industry indicators. The average window adjusted $R^2$ across the OOS period serves as a summary measure of how well the model's correctness is explained by observable characteristics.\footnote{In the soft construction under classification, for simplicity, when the model assigns a stock to multiple different classes, we regard the assignment as a correct assignment if the realized class is among those assigned classes.} For easier interpretation and to stay consistent with the prediction modeling setup, we standardize the characteristics each date before running the regressions.

Panel G shows that classification output is more systematically connected to observable firm characteristics. The average window adjusted $R^2$ in the Fama-MacBeth regression is 0.02 for the hard construction, 0.11 for the soft construction, and 0.01 for regression. The higher OOS Fama-MacBeth $R^2$ under the soft construction is consistent with a tighter statistical relation between firm characteristics and correct decile assignment than under regression or the hard construction. 

Panel G also identifies which characteristics are related to a correct decile assignment. Idiosyncratic volatility (idiovol) is a clear case. Under the soft construction, a stock with higher idiosyncratic volatility bears a strong and significant relation to correct placement, whereas the same characteristic bears no significant relation under regression. Maximum daily return (maxret) shows the same pattern, significant under classification and insignificant under regression. These are among the characteristics with the highest feature importance in the classification model, and its long leg overweights them, as the characteristic profile in the next section confirms. Consistent with the feature importance results, classification reads the signals that mark a stock's return uncertainty and uses them to separate the extreme deciles.

\section{Portfolio Characteristic Profile}
\label{sec:osm9_chars}

Table~\ref{tab:char_profile} reports portfolio profiles of firm characteristics for the long and short legs of the two main stacking portfolios. All characteristics are cross-sectionally z-score normalized each month. Statistical tests are Newey-West $t$-tests adjusted with 6 lags of whether the monthly mean of the classification-minus-regression difference equals zero, computed separately for the long-leg and short-leg differences (Diff. Long and Diff. Short).
\begin{center}
    [Insert Table~\ref{tab:char_profile} Here]
\end{center}

The two approaches select stocks with systematically different characteristic profiles. We interpret these differences in six categories, namely momentum, value versus growth, investment, profitability, intangibles, and trading frictions \citep{osm:hou2018replicating}.

The largest long-leg differences all come from the trading-frictions group. Relative to the regression long leg, the classification long leg loads 0.80 standard deviations more on idiosyncratic volatility (idiovol), 0.80 more on bid-ask spread (baspread), 0.83 more on return volatility (retvol), and 0.77 more on maximum daily return (maxret). These large differences indicate that classification tilts more strongly toward stocks with greater return uncertainty. These four rank second through fifth in the classification model's feature importance, and each is significantly related to correct decile placement under the soft construction, as the Fama-MacBeth regressions in Table~\ref{tab:model_diag} show. What classification holds, what it weights, and which stocks it classifies correctly therefore trace to one set of return-uncertainty signals. The classification long leg also tilts toward low-profitability stocks. The largest differences relative to regression appear in return on assets (roaq), return on invested capital (roic), return on equity (roeq), and gross profitability (gma). The regression long leg has substantially higher profitability on all four measures.

Classification assigns less weight to recent winners. The 6-month, 12-month, and 36-month momentum measures are all lower in the classification long leg than in the regression long leg. The contrast is economically meaningful, as classification tilts toward recent underperformers whereas regression tilts toward recent winners. The characteristics that improve classification accuracy are tied more closely to tail return states and reversal-type behavior than to the signals emphasized by regression.

The remaining categories reinforce the same tilt toward hard-to-value firms. On the value-versus-growth dimension, the classification long leg loads lower than the regression long leg on every valuation ratio, by 0.65 standard deviations on earnings-to-price (ep), 0.49 on cash-flow-to-price (cfp), 0.28 on book-to-market (bm), and 0.14 on sales-to-price (sp). At the same time, it loads higher on forward growth, by 0.22 on the long-term earnings growth forecast (fgr5yr) and 0.21 on sales growth (sgr). Classification therefore favors growth firms with rich valuations and high expected expansion. The investment category tells the same story through the asset side of the balance sheet. The classification long leg loads higher on asset growth (agr) by 0.17, on the change in shares outstanding (chcsho) by 0.20, on long-term debt growth (lgr) by 0.19, and on industry-adjusted capital expenditure growth (grcapx) by 0.14, and it carries larger absolute accruals (absacc) by 0.42, a marker of lower earnings quality. These tilts identify aggressively expanding firms whose reported earnings rest more heavily on accruals than on realized cash flow.

The intangibles category sharpens the same profile. The classification long leg loads higher on research intensity, by 0.30 standard deviations on research and development to sales (rd\_sale) and 0.23 on research and development to market equity (rd\_mve), and it tilts toward younger firms, with firm age (age) lower by 0.30. The differences in organizational capital and several other intangibles proxies are economically small or statistically indistinguishable from zero, so the intangibles tilt operates mainly through research spending and firm youth. Taken together, the value, investment, and intangibles differences describe the same firm that the trading-frictions and profitability results point to under the classification approach, namely young, rapidly growing, research-intensive companies with rich valuations, weak current profitability, and highly uncertain returns. The matched regression leans on older, more profitable, and more established firms.

This portfolio composition also helps explain the large-cap results in Section 4 of the main paper. Firm size is the regression ensemble's second most important signal, carrying more than four times the weight it has in the classification ensemble. The return relevance of size and liquidity is concentrated among small, hard-to-arbitrage firms \citep{osm:stambaugh2015arbitrage}, so once the investable universe is restricted to large-cap stocks, the regression ensemble loses much of the information it relies on. Classification instead leans on idiosyncratic volatility and related return-uncertainty signals, which separate extreme-decile stocks across the entire size distribution and therefore remain informative among large caps. The characteristic profiles thus trace the value-weighted and large-cap performance gap back to the specific information each objective extracts.

\section{Alternative Regression Target Definitions}
\label{sec:osm_relative}
A concern about the performance comparison is whether the regression underperformance reflects the choice of prediction target rather than the loss function. We therefore reestimate every regression model with two relative targets, namely the stock return in excess of the market and the stock return in excess of its two-digit Standard Industrial Classification industry peers. Table~\ref{tab:relative_target} reports the resulting long-short portfolio performance, with the market-relative target in Panel A and the industry-relative target in Panel B. The relative targets do not close the gap that Table 1 of the main paper documents. For the value-weighted stacking ensemble, the market-relative target delivers an annualized Sharpe ratio of 1.51 and a statistically significant monthly six-factor alpha of 2.0\%, and the industry-relative target delivers a Sharpe ratio of 1.62 and an alpha of 2.2\%. Both remain below the matched classification ensemble's Sharpe ratio of 2.08 and alpha of 4.1\%.

\begin{center}
    [Insert Table~\ref{tab:relative_target} Here]
\end{center}

The pattern is consistent across weighting schemes and models. The relative targets raise the equal-weighted Sharpe ratio only modestly, to 3.23 under the market target and 3.32 under the industry target, just above the raw-target regression at 3.12. Both still stay well below the classification ensemble at 3.77. Across all five model pairs and under both relative targets, every relative-target regression earns a smaller six-factor alpha than its classification counterpart. The stronger relative-target stacking ensemble reaches a value-weighted Sharpe ratio of 1.62 against 2.08 for classification, and its six-factor alpha of 2.2\% per month remains below the classification ensemble's 4.1\%. Two Sharpe ratios run the other way, the value-weighted NN under the industry-relative target at 1.91 against 1.88 and the equal-weighted RF under the market-relative target at 2.44 against 2.40, the latter being the same pair identified in Table 1 of the main paper. The relative targets therefore narrow but do not close the gap, confirming that the advantage stems from the loss function rather than the prediction target.

\newpage
\phantomsection
\pdfbookmark[1]{References}{references-osm}

\clearpage
\phantomsection
\pdfbookmark[1]{Tables}{osm-figs-tabs}
\section*{Tables}


\begin{table}[!htbp]
\small
\justifying{\noindent This table reports summary statistics for the 102 firm characteristics used in the prediction models, computed over the full sample of 3,361,673 firm-month observations from July 1962 to December 2024. Mean, Std, P25, Median, and P75 are the cross-sectional mean, standard deviation, 25th percentile, median, and 75th percentile. Characteristics follow \citet{green2017characteristics}. Variables are listed alphabetically. This table continues on the following two pages.}

\centering
\pdfdest name {osm-figs-tabs-tab1} fit
\bookmark[level=2,dest=osm-figs-tabs-tab1]{Table OSM1: Firm Characteristic Summary Statistics}
\caption{Firm Characteristic Summary Statistics}
\begin{tabularx}{\textwidth}{l Y Y Y Y Y}
\toprule
Variable & Mean & Std & P25 & Median & P75 \\
\midrule
absacc & 0.099 & 0.119 & 0.030 & 0.065 & 0.122 \\
acc & -0.026 & 0.146 & -0.079 & -0.021 & 0.046 \\
aeavol & 0.859 & 2.064 & -0.241 & 0.302 & 1.162 \\
age & 15.291 & 13.275 & 5.000 & 11.000 & 21.000 \\
agr & 0.291 & 1.168 & -0.011 & 0.082 & 0.233 \\
baspread & 0.055 & 0.068 & 0.022 & 0.036 & 0.062 \\
beta & 1.093 & 0.652 & 0.634 & 1.022 & 1.469 \\
betasq & 1.624 & 1.830 & 0.408 & 1.049 & 2.162 \\
bm & 0.747 & 0.721 & 0.305 & 0.578 & 0.976 \\
bm\_ia & 20.553 & 681.321 & -0.383 & 0.026 & 0.605 \\
cash & 0.176 & 0.223 & 0.025 & 0.079 & 0.233 \\
cashdebt & -0.077 & 1.712 & 0.010 & 0.122 & 0.274 \\
cashpr & -0.407 & 52.557 & -7.991 & -0.403 & 5.161 \\
cfp & 0.015 & 0.320 & -0.048 & 0.041 & 0.115 \\
cfp\_ia & 12.958 & 300.418 & -0.161 & 0.019 & 0.149 \\
chatoia & -0.005 & 0.240 & -0.076 & 0.003 & 0.083 \\
chcsho & 0.237 & 1.241 & 0.000 & 0.008 & 0.091 \\
chempia & -0.133 & 0.952 & -0.184 & -0.063 & 0.024 \\
chfeps & 0.039 & 2.566 & -0.010 & 0.000 & 0.010 \\
chinv & 0.015 & 0.059 & -0.001 & 0.000 & 0.024 \\
chmom & 0.015 & 0.588 & -0.257 & 0.000 & 0.269 \\
chnanalyst & 0.025 & 1.569 & 0.000 & 0.000 & 0.000 \\
chpmia & 0.471 & 8.857 & -0.234 & -0.004 & 0.127 \\
chtx & 0.001 & 0.013 & -0.001 & 0.000 & 0.003 \\
cinvest & -0.192 & 13.387 & -0.033 & -0.002 & 0.024 \\
convind & 0.128 & 0.334 & 0.000 & 0.000 & 0.000 \\
currat & 3.378 & 5.490 & 1.215 & 1.970 & 3.233 \\
depr & 0.275 & 0.476 & 0.096 & 0.152 & 0.275 \\
disp & 0.172 & 0.459 & 0.018 & 0.045 & 0.125 \\
divi & 0.039 & 0.193 & 0.000 & 0.000 & 0.000 \\
\bottomrule
\end{tabularx}
\label{tab:char_stats}
\end{table}

\begin{table}
\small

\centering
\caption*{\textbf{Table~\ref{tab:char_stats}}. Firm Characteristic Summary Statistics (Continued)}
\begin{tabularx}{\textwidth}{l Y Y Y Y Y}
\toprule
Variable & Mean & Std & P25 & Median & P75 \\
\midrule
divo & 0.030 & 0.170 & 0.000 & 0.000 & 0.000 \\
dolvol & 11.257 & 3.087 & 9.034 & 11.107 & 13.498 \\
dy & 0.017 & 0.033 & 0.000 & 0.001 & 0.026 \\
ear & 0.003 & 0.085 & -0.035 & 0.001 & 0.040 \\
egr & 0.206 & 1.987 & -0.036 & 0.080 & 0.216 \\
ep & -0.034 & 0.383 & -0.018 & 0.047 & 0.086 \\
fgr5yr & 16.726 & 11.958 & 10.000 & 14.670 & 20.000 \\
gma & 0.367 & 0.388 & 0.114 & 0.305 & 0.538 \\
grcapx & 1.320 & 5.223 & -0.345 & 0.175 & 1.056 \\
grltnoa & 0.094 & 0.173 & 0.017 & 0.058 & 0.130 \\
herf & 0.067 & 0.082 & 0.025 & 0.042 & 0.075 \\
hire & 0.093 & 0.346 & -0.021 & 0.009 & 0.125 \\
idiovol & 0.066 & 0.038 & 0.038 & 0.056 & 0.083 \\
ill & 0.000 & 0.000 & 0.000 & 0.000 & 0.000 \\
indmom & 0.089 & 0.292 & -0.089 & 0.068 & 0.221 \\
invest & 0.097 & 0.232 & 0.001 & 0.044 & 0.127 \\
ipo & 0.057 & 0.232 & 0.000 & 0.000 & 0.000 \\
lev & 2.161 & 4.607 & 0.226 & 0.658 & 1.855 \\
lgr & 0.308 & 1.047 & -0.046 & 0.078 & 0.291 \\
maxret & 0.076 & 0.074 & 0.032 & 0.054 & 0.092 \\
mom12m & 0.091 & 0.620 & -0.246 & 0.032 & 0.315 \\
mom1m & 0.001 & 0.178 & -0.071 & 0.000 & 0.072 \\
mom36m & 0.259 & 0.972 & -0.306 & 0.105 & 0.568 \\
mom6m & 0.029 & 0.397 & -0.167 & 0.011 & 0.189 \\
ms & 3.625 & 1.683 & 2.000 & 4.000 & 5.000 \\
mve & 11.812 & 2.295 & 10.127 & 11.652 & 13.372 \\
mve\_ia & -284.146 & 8938.081 & -1931.951 & -403.813 & -59.135 \\
nanalyst & 4.955 & 6.668 & 0.000 & 2.000 & 7.000 \\
nincr & 0.938 & 1.290 & 0.000 & 1.000 & 1.000 \\
operprof & 0.812 & 1.615 & 0.281 & 0.602 & 1.081 \\
orgcap & 0.137 & 0.473 & 0.006 & 0.015 & 0.038 \\
pchcapx\_ia & 3.251 & 51.788 & -1.394 & -0.574 & 0.200 \\
pchcurrat & 0.202 & 1.303 & -0.156 & -0.004 & 0.170 \\
pchdepr & 0.110 & 0.585 & -0.100 & 0.022 & 0.166 \\
pchgm\_pchsale & -0.104 & 1.226 & -0.081 & -0.002 & 0.069 \\
pchquick & 0.251 & 1.552 & -0.182 & -0.003 & 0.207 \\
pchsale\_pchinvt & -0.066 & 0.882 & -0.153 & 0.012 & 0.174 \\
pchsale\_pchrect & -0.061 & 0.799 & -0.146 & -0.000 & 0.140 \\
\bottomrule
\end{tabularx}
\end{table}

\begin{table}
\small

\centering
\caption*{\textbf{Table~\ref{tab:char_stats}}. Firm Characteristic Summary Statistics (Continued)}
\begin{tabularx}{\textwidth}{l Y Y Y Y Y}
\toprule
Variable & Mean & Std & P25 & Median & P75 \\
\midrule
pchsale\_pchxsga & 0.032 & 0.443 & -0.085 & -0.000 & 0.086 \\
pchsaleinv & 0.162 & 1.086 & -0.128 & 0.009 & 0.176 \\
pctacc & -0.677 & 5.842 & -1.182 & -0.272 & 0.655 \\
pricedelay & 0.138 & 0.977 & -0.054 & 0.058 & 0.280 \\
ps & 4.104 & 1.766 & 3.000 & 4.000 & 5.000 \\
quick & 2.672 & 4.944 & 0.877 & 1.302 & 2.311 \\
rd & 0.115 & 0.319 & 0.000 & 0.000 & 0.000 \\
rd\_mve & 0.069 & 0.135 & 0.006 & 0.028 & 0.076 \\
rd\_sale & 1.073 & 9.604 & 0.005 & 0.032 & 0.128 \\
realestate & 0.281 & 0.210 & 0.115 & 0.243 & 0.409 \\
retvol & 0.033 & 0.026 & 0.016 & 0.026 & 0.041 \\
roaq & -0.011 & 0.073 & -0.007 & 0.005 & 0.018 \\
roavol & 0.034 & 0.075 & 0.006 & 0.013 & 0.032 \\
roeq & -0.010 & 0.208 & -0.014 & 0.021 & 0.043 \\
roic & -0.171 & 1.352 & -0.008 & 0.063 & 0.134 \\
rsup & -0.044 & 3.711 & -0.006 & 0.012 & 0.051 \\
salecash & 48.637 & 159.410 & 2.306 & 9.251 & 33.742 \\
saleinv & 26.564 & 70.873 & 4.555 & 7.561 & 17.002 \\
salerec & 11.806 & 49.673 & 3.705 & 5.903 & 8.979 \\
secured & 0.551 & 0.517 & 0.005 & 0.519 & 1.000 \\
securedind & 0.390 & 0.488 & 0.000 & 0.000 & 1.000 \\
sfe & -10.527 & 184.405 & 0.009 & 0.042 & 0.073 \\
sgr & 0.244 & 0.825 & -0.007 & 0.100 & 0.254 \\
sin & 0.007 & 0.086 & 0.000 & 0.000 & 0.000 \\
sp & 2.160 & 3.588 & 0.400 & 0.984 & 2.357 \\
std\_dolvol & 0.848 & 0.414 & 0.519 & 0.776 & 1.105 \\
std\_turn & 5.495 & 35.151 & 0.872 & 1.968 & 4.373 \\
stdacc & 11.967 & 74.438 & 0.083 & 0.143 & 0.305 \\
stdcf & 24.554 & 164.033 & 0.090 & 0.160 & 0.375 \\
sue & -0.009 & 0.492 & -0.005 & 0.000 & 0.006 \\
tang & 0.541 & 0.161 & 0.465 & 0.549 & 0.617 \\
tb & -0.130 & 1.522 & -0.699 & -0.085 & 0.432 \\
turn & 1.228 & 3.623 & 0.220 & 0.563 & 1.344 \\
zerotrade & 1.297 & 3.284 & 0.000 & 0.000 & 0.000 \\
\bottomrule
\end{tabularx}
\end{table}
\clearpage

\begin{table}[!htbp]
\small
\justifying{\noindent This table replicates the out-of-sample long-short portfolio performance of Table 1 of the main paper for classification and regression models under the optimizer Adadelta for the stacking ensembles and the neural networks over July 1983--December 2024 (498 months).
For ease of comparison, the corresponding main results from Table 1 of the main paper under Adam with a learning rate of 0.01 are also included.
For each model, stocks are assigned monthly by the predicted signal into 10 deciles. The long (short) portfolio holds the top (bottom) decile.
Panels A and B report, respectively, the value-weighted and equal-weighted long-short portfolio performance for classification and regression models.
Classification forms each class's portfolio independently: stocks are ranked by their predicted probability of belonging to that class, and the top 10\% of stocks enter that class's portfolio.
Average Return is the annualized average raw return in decimal.
Standard Deviation is the annualized standard deviation in decimal.
Sharpe Ratio is the annualized Sharpe ratio.
$\alpha$ is the monthly abnormal return in decimal from the corresponding factor model. The factor models are the Capital Asset Pricing Model (CAPM) and the Fama-French three-, five-, and six-factor (FF3F, FF5F, and FF6F) models.
Statistical significance is based on Newey-West $t$-statistics adjusted with 6 lags: $^{**}p{<}0.01$, $^{*}p{<}0.05$.}
\pdfdest name {osm-figs-tabs-optimizer} fit
\bookmark[level=2,dest=osm-figs-tabs-optimizer]{Table OSM2: Optimizer Comparison for Neural-Network-Dependent Models}
\caption{Optimizer Comparison for Neural-Network-Dependent Models}
\centering
\label{tab:optimizer}
\begin{tabularx}{\textwidth}{l Y Y Y Y Y Y Y Y}
\toprule
 & \multicolumn{4}{c}{\textbf{Adadelta}} & \multicolumn{4}{c}{\textbf{Adam}} \\
\cmidrule(lr){2-5}\cmidrule(lr){6-9}
 & \multicolumn{2}{c}{Classification} & \multicolumn{2}{c}{Regression} & \multicolumn{2}{c}{Classification} & \multicolumn{2}{c}{Regression} \\
\cmidrule(lr){2-3}\cmidrule(lr){4-5}\cmidrule(lr){6-7}\cmidrule(lr){8-9}
 & Stacking & NN & Stacking & NN & Stacking & NN & Stacking & NN \\
\midrule
\multicolumn{9}{c}{\textbf{Panel A}: Value-Weighted Long-Short} \\
\midrule
Average Return     & 0.468 & 0.384 & 0.204 & 0.132 & 0.504 & 0.428 & 0.271 & 0.237 \\
Standard Deviation & 0.253 & 0.232 & 0.180 & 0.156 & 0.243 & 0.228 & 0.195 & 0.182 \\
Sharpe Ratio       & 1.831 & 1.652 & 1.109 & 0.849 & 2.080 & 1.881 & 1.392 & 1.306 \\
CAPM $\alpha$      & 0.039** & 0.033** & 0.016** & 0.010** & 0.041** & 0.036** & 0.023** & 0.020** \\
FF3F $\alpha$      & 0.040** & 0.034** & 0.016** & 0.011** & 0.042** & 0.036** & 0.023** & 0.020** \\
FF5F $\alpha$      & 0.039** & 0.031** & 0.015** & 0.009** & 0.041** & 0.036** & 0.019** & 0.016** \\
FF6F $\alpha$      & 0.039** & 0.030** & 0.014** & 0.008** & 0.041** & 0.035** & 0.018** & 0.016** \\
\midrule
\multicolumn{9}{c}{\textbf{Panel B}: Equal-Weighted Long-Short} \\
\midrule
Average Return     & 0.612 & 0.516 & 0.456 & 0.372 & 0.651 & 0.560 & 0.519 & 0.517 \\
Standard Deviation & 0.166 & 0.156 & 0.159 & 0.145 & 0.173 & 0.155 & 0.166 & 0.164 \\
Sharpe Ratio       & 3.616 & 3.269 & 2.834 & 2.598 & 3.767 & 3.615 & 3.117 & 3.158 \\
CAPM $\alpha$      & 0.050** & 0.042** & 0.038** & 0.030** & 0.054** & 0.046** & 0.043** & 0.043** \\
FF3F $\alpha$      & 0.050** & 0.042** & 0.037** & 0.030** & 0.054** & 0.046** & 0.043** & 0.043** \\
FF5F $\alpha$      & 0.050** & 0.042** & 0.036** & 0.029** & 0.053** & 0.046** & 0.041** & 0.041** \\
FF6F $\alpha$      & 0.051** & 0.042** & 0.036** & 0.029** & 0.054** & 0.046** & 0.042** & 0.042** \\
\bottomrule
\end{tabularx}
\end{table}
\clearpage

\begin{table}[!htbp]
\small
\justifying{\noindent This table reports out-of-sample performance separately by subperiods and National Bureau of Economic Research (NBER) business cycle regimes for the classification stacking ensemble and the regression stacking ensemble.
Panels A--D split the 198307:202412 sample into four approximately decade-length windows.
Panels E and F separate NBER recession from expansion months.
Within each panel, columns show Long-Short, Long, and Short portfolios for the classification and regression stacking ensembles, separately for value-weighted and equal-weighted schemes. For each model, stocks are assigned monthly by the predicted signal into 10 deciles. The long (short) portfolio holds the top (bottom) decile. Classification forms each class's portfolio independently: stocks are ranked by their predicted probability of belonging to that class, and the top 10\% of stocks enter that class's portfolio.
Average Return is the annualized average raw return in decimal.
Standard Deviation is the annualized standard deviation in decimal.
Sharpe Ratio is the annualized Sharpe ratio.
$\alpha$ is the monthly abnormal return in decimal from the corresponding factor model. The factor models are the Capital Asset Pricing Model (CAPM) and the Fama-French three-, five-, and six-factor (FF3F, FF5F, and FF6F) models.
Statistical significance is based on Newey-West $t$-statistics adjusted with 6 lags: $^{**}p{<}0.01$, $^{*}p{<}0.05$.}
\pdfdest name {osm-figs-tabs-tab4} fit
\bookmark[level=2,dest=osm-figs-tabs-tab4]{Table OSM3: Subperiod and NBER Regime Analysis}
\caption{Subperiod and NBER Regime Analysis}
\centering
\label{tab:subperiod}
\begin{tabularx}{\textwidth}{l Y Y Y Y Y Y}
\toprule
\multicolumn{7}{c}{\textbf{Panel A}: 198307:199311} \\
\midrule
 & \multicolumn{3}{c}{\textbf{Classification}} & \multicolumn{3}{c}{\textbf{Regression}} \\
 \cmidrule(lr){2-4}\cmidrule(lr){5-7}
Metric & Long-Short & Long & Short & Long-Short & Long & Short \\
\midrule
\multicolumn{7}{c}{\textbf{Value-Weighted}} \\
\midrule
Average Return    & 0.596 & 0.204 & 0.392 & 0.295 & 0.200 & 0.095 \\
Standard Deviation        & 0.160 & 0.289 & 0.249 & 0.143 & 0.207 & 0.188 \\
Sharpe Ratio             & 3.733 & 0.485 & 1.321 & 2.069 & 0.659 & 0.169 \\
CAPM $\alpha$         & 0.049** & 0.004 & 0.034** & 0.024** & 0.004 & 0.009** \\
\quad$t$ & (10.12) & (0.60) & (7.75) & (5.62) & (1.65) & (2.68) \\
FF3F $\alpha$         & 0.049** & 0.008 & 0.031** & 0.026** & 0.006** & 0.009** \\
\quad$t$ & (11.18) & (1.94) & (7.56) & (6.23) & (3.04) & (2.94) \\
FF5F $\alpha$         & 0.048** & 0.013** & 0.025** & 0.026** & 0.007** & 0.008** \\
\quad$t$ & (10.08) & (2.85) & (7.42) & (6.13) & (2.64) & (2.75) \\
FF6F $\alpha$         & 0.047** & 0.013** & 0.024** & 0.025** & 0.007* & 0.008* \\
\quad$t$ & (9.17) & (2.83) & (6.59) & (5.55) & (2.46) & (2.55) \\
\midrule
\multicolumn{7}{c}{\textbf{Equal-Weighted}} \\
\midrule
Average Return    & 0.748 & 0.377 & 0.371 & 0.599 & 0.397 & 0.202 \\
Standard Deviation        & 0.117 & 0.267 & 0.232 & 0.140 & 0.253 & 0.199 \\
Sharpe Ratio             & 6.396 & 1.175 & 1.326 & 4.273 & 1.320 & 0.697 \\
CAPM $\alpha$         & 0.062** & 0.020** & 0.032** & 0.049** & 0.021** & 0.017** \\
\quad$t$ & (16.53) & (2.84) & (6.81) & (10.07) & (3.44) & (4.46) \\
FF3F $\alpha$         & 0.064** & 0.022** & 0.031** & 0.050** & 0.023** & 0.016** \\
\quad$t$ & (17.68) & (4.00) & (6.87) & (9.89) & (4.59) & (4.75) \\
FF5F $\alpha$         & 0.062** & 0.026** & 0.026** & 0.050** & 0.025** & 0.014** \\
\quad$t$ & (17.74) & (5.42) & (7.41) & (10.95) & (6.68) & (4.58) \\
FF6F $\alpha$         & 0.062** & 0.026** & 0.025** & 0.050** & 0.026** & 0.013** \\
\quad$t$ & (17.44) & (5.35) & (6.79) & (10.80) & (6.64) & (4.22) \\
\bottomrule
\end{tabularx}
\end{table}
\clearpage

\begin{table}[!htbp]
\small
\caption*{\textbf{Table~\ref{tab:subperiod}}. Subperiod and NBER Regime Analysis (Continued)}
\begin{tabularx}{\textwidth}{l Y Y Y Y Y Y}
\toprule
\multicolumn{7}{c}{\textbf{Panel B}: 199312:200404} \\
\midrule
 & \multicolumn{3}{c}{\textbf{Classification}} & \multicolumn{3}{c}{\textbf{Regression}} \\
 \cmidrule(lr){2-4}\cmidrule(lr){5-7}
Metric & Long-Short & Long & Short & Long-Short & Long & Short \\
\midrule
\multicolumn{7}{c}{\textbf{Value-Weighted}} \\
\midrule
Average Return    & 0.628 & 0.371 & 0.256 & 0.539 & 0.346 & 0.193 \\
Standard Deviation        & 0.239 & 0.553 & 0.527 & 0.246 & 0.300 & 0.323 \\
Sharpe Ratio             & 2.628 & 0.599 & 0.411 & 2.193 & 1.019 & 0.475 \\
CAPM $\alpha$         & 0.051** & 0.012 & 0.033** & 0.047** & 0.017** & 0.023** \\
\quad$t$ & (6.96) & (1.43) & (3.20) & (6.66) & (3.45) & (5.16) \\
FF3F $\alpha$         & 0.053** & 0.021** & 0.025** & 0.045** & 0.019** & 0.020** \\
\quad$t$ & (6.94) & (2.69) & (3.17) & (6.42) & (3.66) & (5.09) \\
FF5F $\alpha$         & 0.056** & 0.033** & 0.016 & 0.043** & 0.024** & 0.013** \\
\quad$t$ & (6.45) & (3.93) & (1.64) & (5.01) & (3.77) & (3.43) \\
FF6F $\alpha$         & 0.057** & 0.038** & 0.013 & 0.043** & 0.025** & 0.011** \\
\quad$t$ & (5.67) & (4.81) & (1.35) & (4.25) & (3.44) & (2.91) \\
\midrule
\multicolumn{7}{c}{\textbf{Equal-Weighted}} \\
\midrule
Average Return    & 0.891 & 0.604 & 0.287 & 0.802 & 0.613 & 0.188 \\
Standard Deviation        & 0.178 & 0.467 & 0.430 & 0.191 & 0.371 & 0.350 \\
Sharpe Ratio             & 5.000 & 1.209 & 0.575 & 4.204 & 1.546 & 0.425 \\
CAPM $\alpha$         & 0.074** & 0.036** & 0.032** & 0.068** & 0.040** & 0.022** \\
\quad$t$ & (12.56) & (4.45) & (3.66) & (11.57) & (5.73) & (3.36) \\
FF3F $\alpha$         & 0.075** & 0.040** & 0.028** & 0.068** & 0.041** & 0.020** \\
\quad$t$ & (12.64) & (6.12) & (5.62) & (11.26) & (6.33) & (5.38) \\
FF5F $\alpha$         & 0.076** & 0.048** & 0.020** & 0.067** & 0.046** & 0.014** \\
\quad$t$ & (10.81) & (5.50) & (3.79) & (8.80) & (4.90) & (3.37) \\
FF6F $\alpha$         & 0.078** & 0.056** & 0.016** & 0.070** & 0.054** & 0.010** \\
\quad$t$ & (9.39) & (6.21) & (3.16) & (8.01) & (5.66) & (2.89) \\
\bottomrule
\end{tabularx}
\end{table}
\clearpage

\begin{table}[!htbp]
\small
\caption*{\textbf{Table~\ref{tab:subperiod}}. Subperiod and NBER Regime Analysis (Continued)}
\begin{tabularx}{\textwidth}{l Y Y Y Y Y Y}
\toprule
\multicolumn{7}{c}{\textbf{Panel C}: 200405:201408} \\
\midrule
 & \multicolumn{3}{c}{\textbf{Classification}} & \multicolumn{3}{c}{\textbf{Regression}} \\
 \cmidrule(lr){2-4}\cmidrule(lr){5-7}
Metric & Long-Short & Long & Short & Long-Short & Long & Short \\
\midrule
\multicolumn{7}{c}{\textbf{Value-Weighted}} \\
\midrule
Average Return    & 0.299 & 0.211 & 0.088 & 0.188 & 0.214 & -0.026 \\
Standard Deviation        & 0.226 & 0.404 & 0.359 & 0.141 & 0.246 & 0.229 \\
Sharpe Ratio             & 1.323 & 0.488 & 0.205 & 1.336 & 0.812 & -0.177 \\
CAPM $\alpha$         & 0.024** & 0.002 & 0.020** & 0.015** & 0.007** & 0.006 \\
\quad$t$ & (5.91) & (0.35) & (4.83) & (4.47) & (2.82) & (1.71) \\
FF3F $\alpha$         & 0.024** & 0.002 & 0.020** & 0.015** & 0.007** & 0.005 \\
\quad$t$ & (5.79) & (0.39) & (5.22) & (4.85) & (2.92) & (1.65) \\
FF5F $\alpha$         & 0.019** & 0.003 & 0.013** & 0.016** & 0.009** & 0.005 \\
\quad$t$ & (4.00) & (0.74) & (3.17) & (4.25) & (3.70) & (1.35) \\
FF6F $\alpha$         & 0.020** & 0.004 & 0.013** & 0.016** & 0.009** & 0.004 \\
\quad$t$ & (3.88) & (0.85) & (3.52) & (3.97) & (3.73) & (1.10) \\
\midrule
\multicolumn{7}{c}{\textbf{Equal-Weighted}} \\
\midrule
Average Return    & 0.482 & 0.313 & 0.169 & 0.343 & 0.291 & 0.052 \\
Standard Deviation        & 0.156 & 0.332 & 0.297 & 0.145 & 0.286 & 0.239 \\
Sharpe Ratio             & 3.079 & 0.900 & 0.520 & 2.359 & 0.968 & 0.158 \\
CAPM $\alpha$         & 0.040** & 0.013 & 0.024** & 0.028** & 0.013* & 0.012** \\
\quad$t$ & (6.41) & (1.66) & (6.33) & (5.27) & (2.00) & (4.16) \\
FF3F $\alpha$         & 0.039** & 0.013 & 0.024** & 0.028** & 0.013* & 0.012** \\
\quad$t$ & (6.58) & (1.88) & (6.47) & (5.78) & (2.29) & (4.01) \\
FF5F $\alpha$         & 0.040** & 0.018* & 0.019** & 0.031** & 0.019** & 0.010** \\
\quad$t$ & (5.80) & (2.42) & (4.82) & (5.06) & (2.78) & (3.62) \\
FF6F $\alpha$         & 0.040** & 0.019** & 0.018** & 0.032** & 0.020** & 0.009** \\
\quad$t$ & (6.65) & (3.17) & (4.82) & (5.35) & (3.59) & (3.42) \\
\bottomrule
\end{tabularx}
\end{table}
\clearpage

\begin{table}[!htbp]
\small
\caption*{\textbf{Table~\ref{tab:subperiod}}. Subperiod and NBER Regime Analysis (Continued)}
\begin{tabularx}{\textwidth}{l Y Y Y Y Y Y}
\toprule
\multicolumn{7}{c}{\textbf{Panel D}: 201409:202412} \\
\midrule
 & \multicolumn{3}{c}{\textbf{Classification}} & \multicolumn{3}{c}{\textbf{Regression}} \\
 \cmidrule(lr){2-4}\cmidrule(lr){5-7}
Metric & Long-Short & Long & Short & Long-Short & Long & Short \\
\midrule
\multicolumn{7}{c}{\textbf{Value-Weighted}} \\
\midrule
Average Return    & 0.493 & 0.173 & 0.320 & 0.060 & 0.083 & -0.023 \\
Standard Deviation        & 0.313 & 0.418 & 0.429 & 0.204 & 0.232 & 0.232 \\
Sharpe Ratio             & 1.578 & 0.376 & 0.709 & 0.295 & 0.290 & -0.170 \\
CAPM $\alpha$         & 0.042** & -0.003 & 0.043** & 0.005 & -0.006* & 0.008* \\
\quad$t$ & (4.90) & (-0.44) & (5.25) & (0.96) & (-2.04) & (2.32) \\
FF3F $\alpha$         & 0.043** & 0.003 & 0.038** & 0.005 & -0.005 & 0.007* \\
\quad$t$ & (4.99) & (0.49) & (5.96) & (0.94) & (-1.68) & (2.13) \\
FF5F $\alpha$         & 0.043** & 0.005 & 0.035** & 0.002 & -0.005 & 0.004 \\
\quad$t$ & (4.66) & (0.96) & (5.81) & (0.44) & (-1.69) & (1.30) \\
FF6F $\alpha$         & 0.041** & 0.006 & 0.033** & -0.001 & -0.006* & 0.003 \\
\quad$t$ & (4.20) & (1.01) & (5.18) & (-0.14) & (-2.06) & (0.87) \\
\midrule
\multicolumn{7}{c}{\textbf{Equal-Weighted}} \\
\midrule
Average Return    & 0.482 & 0.180 & 0.302 & 0.328 & 0.185 & 0.143 \\
Standard Deviation        & 0.200 & 0.396 & 0.360 & 0.146 & 0.312 & 0.277 \\
Sharpe Ratio             & 2.408 & 0.413 & 0.794 & 2.250 & 0.540 & 0.459 \\
CAPM $\alpha$         & 0.037** & -0.003 & 0.038** & 0.026** & 0.000 & 0.023** \\
\quad$t$ & (6.07) & (-0.42) & (4.24) & (6.21) & (-0.02) & (4.38) \\
FF3F $\alpha$         & 0.038** & 0.002 & 0.033** & 0.026** & 0.004 & 0.019** \\
\quad$t$ & (5.71) & (0.41) & (4.49) & (5.98) & (1.10) & (4.36) \\
FF5F $\alpha$         & 0.037** & 0.004 & 0.030** & 0.025** & 0.006 & 0.017** \\
\quad$t$ & (5.48) & (0.87) & (4.46) & (6.07) & (1.60) & (4.34) \\
FF6F $\alpha$         & 0.036** & 0.006 & 0.027** & 0.025** & 0.008* & 0.015** \\
\quad$t$ & (4.95) & (1.27) & (4.26) & (5.92) & (2.15) & (4.42) \\
\bottomrule
\end{tabularx}
\end{table}
\clearpage

\begin{table}[!htbp]
\small
\caption*{\textbf{Table~\ref{tab:subperiod}}. Subperiod and NBER Regime Analysis (Continued)}
\begin{tabularx}{\textwidth}{l Y Y Y Y Y Y}
\toprule
\multicolumn{7}{c}{\textbf{Panel E}: NBER Recession (40~months, 198307:202412)} \\
\midrule
 & \multicolumn{3}{c}{\textbf{Classification}} & \multicolumn{3}{c}{\textbf{Regression}} \\
 \cmidrule(lr){2-4}\cmidrule(lr){5-7}
Metric & Long-Short & Long & Short & Long-Short & Long & Short \\
\midrule
\multicolumn{7}{c}{\textbf{Value-Weighted}} \\
\midrule
Average Return    & 0.615 & 0.062 & 0.553 & 0.508 & 0.038 & 0.470 \\
Standard Deviation        & 0.329 & 0.768 & 0.606 & 0.236 & 0.406 & 0.414 \\
Sharpe Ratio             & 1.868 & 0.041 & 0.864 & 2.148 & 0.019 & 1.063 \\
CAPM $\alpha$         & 0.057** & 0.037* & 0.015 & 0.042** & 0.020** & 0.017* \\
\quad$t$ & (4.59) & (2.04) & (0.87) & (4.64) & (2.80) & (2.08) \\
FF3F $\alpha$         & 0.052** & 0.026 & 0.020 & 0.043** & 0.020* & 0.018* \\
\quad$t$ & (4.06) & (1.50) & (1.15) & (4.50) & (2.40) & (2.02) \\
FF5F $\alpha$         & 0.049** & 0.037* & 0.007 & 0.038** & 0.021* & 0.012 \\
\quad$t$ & (2.73) & (2.01) & (0.48) & (4.67) & (2.33) & (1.40) \\
FF6F $\alpha$         & 0.047* & 0.025 & 0.017 & 0.041** & 0.018* & 0.018* \\
\quad$t$ & (2.51) & (1.75) & (1.55) & (4.06) & (2.50) & (2.39) \\
\midrule
\multicolumn{7}{c}{\textbf{Equal-Weighted}} \\
\midrule
Average Return    & 1.011 & 0.489 & 0.522 & 0.791 & 0.419 & 0.372 \\
Standard Deviation        & 0.259 & 0.592 & 0.475 & 0.228 & 0.462 & 0.435 \\
Sharpe Ratio             & 3.904 & 0.774 & 1.034 & 3.470 & 0.841 & 0.785 \\
CAPM $\alpha$         & 0.090** & 0.065** & 0.019 & 0.067** & 0.054** & 0.008 \\
\quad$t$ & (6.44) & (3.34) & (1.66) & (5.85) & (3.62) & (0.76) \\
FF3F $\alpha$         & 0.091** & 0.059** & 0.027* & 0.068** & 0.049** & 0.014 \\
\quad$t$ & (6.27) & (2.98) & (2.17) & (6.16) & (3.35) & (1.28) \\
FF5F $\alpha$         & 0.092** & 0.072** & 0.015 & 0.064** & 0.055** & 0.004 \\
\quad$t$ & (5.59) & (4.10) & (1.27) & (5.19) & (3.63) & (0.47) \\
FF6F $\alpha$         & 0.087** & 0.059** & 0.021* & 0.064** & 0.046** & 0.012 \\
\quad$t$ & (5.11) & (5.03) & (2.03) & (4.78) & (4.01) & (1.81) \\
\bottomrule
\end{tabularx}
\end{table}
\clearpage

\begin{table}[!htbp]
\small
\caption*{\textbf{Table~\ref{tab:subperiod}}. Subperiod and NBER Regime Analysis (Continued)}
\begin{tabularx}{\textwidth}{l Y Y Y Y Y Y}
\toprule
\multicolumn{7}{c}{\textbf{Panel F}: NBER Expansion (458~months, 198307:202412)} \\
\midrule
 & \multicolumn{3}{c}{\textbf{Classification}} & \multicolumn{3}{c}{\textbf{Regression}} \\
 \cmidrule(lr){2-4}\cmidrule(lr){5-7}
Metric & Long-Short & Long & Short & Long-Short & Long & Short \\
\midrule
\multicolumn{7}{c}{\textbf{Value-Weighted}} \\
\midrule
Average Return    & 0.495 & 0.256 & 0.239 & 0.250 & 0.226 & 0.024 \\
Standard Deviation        & 0.234 & 0.383 & 0.382 & 0.190 & 0.231 & 0.227 \\
Sharpe Ratio             & 2.117 & 0.580 & 0.538 & 1.319 & 0.833 & -0.042 \\
CAPM $\alpha$         & 0.041** & 0.002 & 0.033** & 0.021** & 0.005* & 0.010** \\
\quad$t$ & (11.25) & (0.69) & (8.00) & (6.16) & (2.45) & (4.49) \\
FF3F $\alpha$         & 0.041** & 0.007* & 0.028** & 0.021** & 0.006** & 0.008** \\
\quad$t$ & (11.16) & (2.36) & (7.97) & (6.10) & (3.08) & (3.90) \\
FF5F $\alpha$         & 0.041** & 0.013** & 0.022** & 0.018** & 0.008** & 0.005* \\
\quad$t$ & (9.98) & (3.66) & (5.83) & (5.22) & (3.21) & (2.45) \\
FF6F $\alpha$         & 0.040** & 0.015** & 0.020** & 0.016** & 0.008** & 0.003 \\
\quad$t$ & (8.27) & (3.90) & (5.45) & (4.04) & (2.59) & (1.57) \\
\midrule
\multicolumn{7}{c}{\textbf{Equal-Weighted}} \\
\midrule
Average Return    & 0.620 & 0.358 & 0.262 & 0.495 & 0.368 & 0.127 \\
Standard Deviation        & 0.161 & 0.350 & 0.323 & 0.158 & 0.295 & 0.252 \\
Sharpe Ratio             & 3.861 & 0.928 & 0.706 & 3.125 & 1.133 & 0.369 \\
CAPM $\alpha$         & 0.051** & 0.014** & 0.032** & 0.041** & 0.017** & 0.019** \\
\quad$t$ & (16.01) & (3.52) & (8.30) & (12.00) & (4.66) & (6.83) \\
FF3F $\alpha$         & 0.051** & 0.018** & 0.028** & 0.040** & 0.019** & 0.016** \\
\quad$t$ & (15.97) & (5.46) & (9.55) & (12.19) & (6.14) & (7.96) \\
FF5F $\alpha$         & 0.051** & 0.022** & 0.023** & 0.039** & 0.021** & 0.012** \\
\quad$t$ & (14.96) & (6.02) & (8.08) & (11.94) & (5.98) & (6.54) \\
FF6F $\alpha$         & 0.051** & 0.026** & 0.020** & 0.040** & 0.025** & 0.010** \\
\quad$t$ & (12.97) & (5.98) & (7.34) & (10.76) & (6.06) & (5.40) \\
\bottomrule
\end{tabularx}
\end{table}
\clearpage


\begin{landscape}
\begin{table}[!htbp]
\scriptsize
\setlength{\tabcolsep}{3pt}
\centering
\begin{minipage}{1.3\textwidth}
\justifying{\noindent This table reports after-cost portfolio performance for the stacking ensemble portfolios under TC30+BC50. The adjustment deducts 30 basis points per one-way trade multiplied by average monthly two-way turnover and additionally imposes a 50-basis-point annual borrowing cost for short positions. For each model, stocks are assigned monthly by the predicted signal into 10 deciles. The long (short) portfolio holds the top (bottom) decile. Classification forms each class's portfolio independently: stocks are ranked by their predicted probability of belonging to that class, and the top 10\% of stocks enter that class's portfolio. L-S indicates long-short portfolios.
Average Return is the annualized average raw return in decimal.
Standard Deviation is the annualized standard deviation in decimal.
Sharpe Ratio is the annualized Sharpe ratio.
$\alpha$ is the monthly abnormal return in decimal from the corresponding factor model. The factor models are the Capital Asset Pricing Model (CAPM) and the Fama-French three-, five-, and six-factor (FF3F, FF5F, and FF6F) models.
Statistical significance is based on Newey-West $t$-statistics adjusted with 6 lags: $^{**}p{<}0.01$, $^{*}p{<}0.05$.}
\end{minipage}
\pdfdest name {osm-figs-tabs-tab3} fit
\bookmark[level=2,dest=osm-figs-tabs-tab3]{Table OSM4: After-Cost Portfolio Performance}
\caption{After-Cost Portfolio Performance}
\begin{tabularx}{1.3\textwidth}{l Y Y Y Y Y Y Y Y Y Y Y Y}
\toprule
 & \multicolumn{6}{c}{\textbf{Value-Weighted}}
 & \multicolumn{6}{c}{\textbf{Equal-Weighted}} \\
\cmidrule(lr){2-7}\cmidrule(lr){8-13}
 & \multicolumn{3}{c}{\textbf{Classification}} & \multicolumn{3}{c}{\textbf{Regression}}
 & \multicolumn{3}{c}{\textbf{Classification}} & \multicolumn{3}{c}{\textbf{Regression}} \\
\cmidrule(lr){2-4}\cmidrule(lr){5-7}\cmidrule(lr){8-10}\cmidrule(lr){11-13}
Metric & L-S & Long & Short & L-S & Long & Short & L-S & Long & Short & L-S & Long & Short \\
\midrule
Average Return   & 0.448 & 0.185 & 0.215 & 0.213 & 0.155 & 0.009 & 0.597 & 0.318 & 0.234 & 0.469 & 0.327 & 0.102 \\
Standard Deviation       & 0.243 & 0.426 & 0.404 & 0.195 & 0.249 & 0.249 & 0.173 & 0.374 & 0.337 & 0.166 & 0.311 & 0.272 \\
Sharpe Ratio            & 1.845 & 0.435 & 0.533 & 1.092 & 0.622 & 0.036 & 3.450 & 0.850 & 0.692 & 2.818 & 1.053 & 0.374 \\
CAPM $\alpha$        & 0.037** & -0.001 & 0.028** & 0.018** & 0.001 & 0.007 & 0.049** & 0.013** & 0.027** & 0.039** & 0.015** & 0.014** \\
FF3F $\alpha$        & 0.037** & 0.002 & 0.025** & 0.018** & 0.002 & 0.006 & 0.049** & 0.015** & 0.024** & 0.039** & 0.017** & 0.013** \\
FF5F $\alpha$        & 0.036** & 0.009* & 0.018** & 0.015** & 0.003 & 0.001* & 0.049** & 0.020** & 0.018** & 0.037** & 0.020** & 0.008** \\
FF6F $\alpha$        & 0.036** & 0.011** & 0.015** & 0.013** & 0.004 & 0.000** & 0.049** & 0.024** & 0.016** & 0.038** & 0.022** & 0.006 \\
\bottomrule
\end{tabularx}
\label{tab:aftercost}
\end{table}
\end{landscape}
\clearpage


\begin{landscape}
\begin{table}[!htbp]
\scriptsize
\setlength{\tabcolsep}{3pt}
\centering
\makebox[\textwidth][c]{%
\begin{minipage}{1.3\textwidth}
\small
\justifying{\noindent This table reports model diagnostics for the 10-class stacking ensembles. Panel A is the row-normalized confusion matrix in percent for the regression stacking ensemble. Predicted class is the within-date decile rank of the predicted return. Panel B is the row-normalized confusion matrix in percent for the hard construction, where each stock is assigned to the class with the highest predicted probability. Panel C is the soft construction confusion matrix in percent, where each class's portfolio is formed by ranking all stocks on that class's predicted probability and selecting the top-decile entries independently. Under the soft construction, a stock can appear in multiple class portfolios. Panel D reports overall modeling performance statistics. Panel E reports per-class statistics for all three methods. Panel F reports top-10 feature importance. Panel G reports 60-month rolling-window Fama-MacBeth regressions of a binary correctness indicator on all 102 standardized firm characteristics with industry indicators. $t$-statistics in Panel G are Newey-West $t$-statistics adjusted with 6 lags. Statistical significance: $^{**}p{<}0.01$, $^{*}p{<}0.05$.}
\pdfdest name {osm-figs-tabs-tab5} fit
\bookmark[level=2,dest=osm-figs-tabs-tab5]{Table OSM5: Model Diagnostics: Classification Performance and Feature Importance}
\caption{Model Diagnostics: Classification Performance and Feature Importance}
\label{tab:model_diag}
\end{minipage}}
\begin{tabularx}{1.3\textwidth}{l Y Y Y Y Y Y Y Y Y Y}
\toprule
\multicolumn{11}{c}{\textbf{Panel A}: Regression Confusion Matrix (\%)} \\
\midrule
\multirow{2}{*}{Actual $\downarrow$ / Predicted $\rightarrow$}
  & \multicolumn{10}{c}{Predicted Class} \\
\cmidrule(lr){2-11}
 & 1 & 2 & 3 & 4 & 5 & 6 & 7 & 8 & 9 & 10 \\
\midrule
Class  1 & 19.430 & 11.668 & 9.037 & 7.818 & 7.375 & 6.933 & 6.951 & 7.119 & 8.352 & 15.317 \\
Class  2 & 13.266 & 11.314 & 10.266 & 9.609 & 9.174 & 9.080 & 8.826 & 8.854 & 9.333 & 10.279 \\
Class  3 & 9.859 & 10.448 & 10.577 & 10.406 & 10.373 & 10.241 & 9.974 & 9.797 & 9.770 & 8.554 \\
Class  4 & 8.108 & 9.782 & 10.460 & 10.849 & 10.843 & 10.796 & 10.697 & 10.501 & 10.036 & 7.928 \\
Class  5 & 7.434 & 9.586 & 10.395 & 10.818 & 11.081 & 11.043 & 11.015 & 10.883 & 10.043 & 7.701 \\
Class  6 & 6.812 & 9.267 & 10.369 & 10.956 & 11.309 & 11.365 & 11.274 & 11.032 & 10.185 & 7.432 \\
Class  7 & 6.888 & 9.203 & 10.319 & 10.877 & 11.120 & 11.258 & 11.191 & 11.139 & 10.325 & 7.680 \\
Class  8 & 7.292 & 9.251 & 10.109 & 10.572 & 10.762 & 10.977 & 11.110 & 11.022 & 10.574 & 8.330 \\
Class  9 & 8.631 & 9.562 & 9.795 & 9.925 & 9.943 & 10.204 & 10.492 & 10.680 & 10.822 & 9.947 \\
Class 10 & 12.367 & 9.915 & 8.644 & 8.158 & 8.032 & 8.057 & 8.435 & 8.964 & 10.533 & 16.896 \\
\midrule
\multicolumn{11}{c}{\textbf{Panel B}: Hard Construction Confusion Matrix (\%)} \\
\midrule
\multirow{2}{*}{Actual $\downarrow$ / Predicted $\rightarrow$}
  & \multicolumn{10}{c}{Predicted Class} \\
\cmidrule(lr){2-11}
 & 1 & 2 & 3 & 4 & 5 & 6 & 7 & 8 & 9 & 10 \\
\midrule
Class  1 & 54.523 & 9.199 & 1.861 & 1.185 & 1.432 & 2.689 & 1.335 & 2.459 & 6.543 & 18.776 \\
Class  2 & 32.316 & 12.241 & 3.298 & 2.577 & 3.069 & 9.862 & 4.135 & 5.692 & 10.160 & 16.650 \\
Class  3 & 21.550 & 10.804 & 3.716 & 3.288 & 4.654 & 18.349 & 6.733 & 7.194 & 10.220 & 13.492 \\
Class  4 & 16.288 & 9.135 & 3.544 & 3.372 & 6.048 & 24.685 & 8.375 & 7.642 & 9.648 & 11.262 \\
Class  5 & 14.340 & 8.097 & 3.254 & 3.348 & 6.869 & 28.435 & 9.008 & 7.566 & 8.760 & 10.323 \\
Class  6 & 12.846 & 7.600 & 3.118 & 3.307 & 6.506 & 30.145 & 9.764 & 8.006 & 9.110 & 9.599 \\
Class  7 & 13.232 & 7.891 & 3.116 & 3.130 & 5.920 & 28.632 & 9.937 & 8.413 & 9.700 & 10.029 \\
Class  8 & 14.980 & 8.645 & 3.111 & 2.909 & 5.120 & 24.385 & 9.278 & 8.785 & 11.218 & 11.569 \\
Class  9 & 20.248 & 10.262 & 3.086 & 2.674 & 4.010 & 16.623 & 7.064 & 8.210 & 12.878 & 14.945 \\
Class 10 & 38.512 & 9.575 & 2.327 & 1.846 & 2.301 & 5.832 & 2.693 & 3.952 & 9.679 & 23.281 \\
\bottomrule
\end{tabularx}
\end{table}
\end{landscape}

\clearpage

\begin{landscape}
\begin{table}[!htbp]
\scriptsize
\setlength{\tabcolsep}{3pt}
\centering
\caption*{\textbf{Table~\ref{tab:model_diag}}. Model Diagnostics: Classification Performance and Feature Importance (Continued)}
\begin{tabularx}{1.3\textwidth}{l Y Y Y Y Y Y Y Y Y Y}
\toprule
\multicolumn{11}{c}{\textbf{Panel C}: Soft Construction Confusion Matrix (\%)} \\
\midrule
\multirow{2}{*}{Actual $\downarrow$ / Predicted $\rightarrow$}
  & \multicolumn{10}{c}{Predicted Class} \\
\cmidrule(lr){2-11}
 & 1 & 2 & 3 & 4 & 5 & 6 & 7 & 8 & 9 & 10 \\
\midrule
Class  1 & 32.361 & 19.663 & 6.968 & 1.315 & 1.087 & 0.777 & 0.850 & 1.627 & 7.605 & 27.747 \\
Class  2 & 15.519 & 18.089 & 13.415 & 5.885 & 4.821 & 4.608 & 4.751 & 6.327 & 12.260 & 14.325 \\
Class  3 & 8.166 & 11.435 & 13.162 & 10.379 & 9.445 & 9.338 & 9.293 & 9.864 & 10.896 & 8.022 \\
Class  4 & 5.158 & 7.449 & 10.993 & 12.827 & 12.735 & 12.611 & 12.314 & 11.529 & 9.011 & 5.371 \\
Class  5 & 4.318 & 5.676 & 9.469 & 13.853 & 14.425 & 14.055 & 13.647 & 12.133 & 7.759 & 4.664 \\
Class  6 & 3.551 & 5.030 & 8.898 & 13.991 & 14.610 & 14.797 & 14.538 & 12.963 & 7.742 & 3.880 \\
Class  7 & 3.804 & 5.360 & 8.908 & 13.229 & 13.832 & 14.205 & 14.375 & 13.431 & 8.672 & 4.183 \\
Class  8 & 4.814 & 6.664 & 9.576 & 11.830 & 12.069 & 12.597 & 12.938 & 13.207 & 10.904 & 5.400 \\
Class  9 & 7.996 & 10.304 & 10.933 & 8.893 & 8.709 & 8.876 & 9.298 & 11.158 & 14.802 & 9.031 \\
Class 10 & 22.445 & 16.334 & 8.499 & 3.044 & 2.906 & 2.607 & 2.613 & 3.829 & 12.443 & 25.280 \\
\bottomrule
\end{tabularx}
\end{table}
\end{landscape}

\clearpage

\begin{table}[!htbp]
\footnotesize
\centering
\caption*{\textbf{Table~\ref{tab:model_diag}}. Model Diagnostics: Classification Performance and Feature Importance (Continued)}
\begin{tabularx}{\textwidth}{l Y Y Y}
\toprule
\multicolumn{4}{c}{\textbf{Panel D}: Overall Statistics} \\
\midrule
Metric & \textbf{Hard Construction} & \textbf{Soft Construction} & \textbf{Regression} \\
\midrule
Stock-Month Obs. & 2,643,604 & 2,643,604 & 2,643,604 \\
Accuracy & 16.571\% & 16.931\% & 12.454\% \\
Balanced Accuracy & 53.652\% & 54.050\% & 51.364\% \\
Cohen's Kappa & 7.303\% & 7.701\% & 2.727\% \\
Macro Precision & 14.846\% & 16.931\% & 12.453\% \\
Macro Recall & 16.575\% & 17.333\% & 12.455\% \\
Macro F1 & 14.073\% & 17.057\% & 12.454\% \\
\bottomrule
\end{tabularx}
\end{table}

\clearpage

\begin{landscape}
\begin{table}[!htbp]
\scriptsize
\setlength{\tabcolsep}{3pt}
\centering
\caption*{\textbf{Table~\ref{tab:model_diag}}. Model Diagnostics: Classification Performance and Feature Importance (Continued)}
\begin{tabularx}{1.3\textwidth}{c Y Y Y Y Y Y Y Y Y Y Y Y}
\toprule
\multicolumn{13}{c}{\textbf{Panel E}: Per-Class Statistics (\%)} \\
\midrule
 & \multicolumn{4}{c}{\textbf{Hard Construction}} & \multicolumn{4}{c}{\textbf{Soft Construction}} & \multicolumn{4}{c}{\textbf{Regression}} \\
\cmidrule(lr){2-5}\cmidrule(lr){6-9}\cmidrule(lr){10-13}
Class & Recall & Prec. & F1 & Bal. Acc. & Recall & Prec. & F1 & Bal. Acc. & Recall & Prec. & F1 & Bal. Acc. \\
\midrule
1 & 54.523 & 22.810 & 32.164 & 67.021 & 32.361 & 29.678 & 30.961 & 62.309 & 19.430 & 19.398 & 19.414 & 55.234 \\
2 & 12.241 & 13.101 & 12.656 & 51.609 & 18.089 & 14.972 & 16.383 & 54.409 & 11.314 & 11.316 & 11.315 & 50.730 \\
3 & 3.716 & 12.210 & 5.698 & 50.374 & 13.162 & 12.192 & 12.659 & 51.743 & 10.577 & 10.579 & 10.578 & 50.322 \\
4 & 3.372 & 12.202 & 5.284 & 50.338 & 12.827 & 13.566 & 13.186 & 51.581 & 10.849 & 10.851 & 10.850 & 50.472 \\
5 & 6.869 & 14.960 & 9.415 & 51.265 & 14.425 & 16.529 & 15.406 & 52.499 & 11.081 & 11.083 & 11.082 & 50.600 \\
6 & 30.145 & 15.892 & 20.812 & 56.211 & 14.797 & 17.501 & 16.036 & 52.720 & 11.365 & 11.367 & 11.366 & 50.761 \\
7 & 9.937 & 14.543 & 11.807 & 51.725 & 14.375 & 16.532 & 15.379 & 52.472 & 11.191 & 11.194 & 11.193 & 50.664 \\
8 & 8.785 & 12.935 & 10.464 & 51.107 & 13.207 & 14.052 & 13.616 & 51.794 & 11.022 & 11.023 & 11.022 & 50.568 \\
9 & 12.878 & 13.151 & 13.013 & 51.715 & 14.802 & 13.303 & 14.013 & 52.638 & 10.822 & 10.824 & 10.823 & 50.458 \\
10 & 23.281 & 16.651 & 19.416 & 55.160 & 25.280 & 20.980 & 22.930 & 58.331 & 16.896 & 16.899 & 16.897 & 53.828 \\
\bottomrule
\end{tabularx}
\end{table}
\end{landscape}

\clearpage

\begin{table}[!htbp]
\footnotesize
\centering
\caption*{\textbf{Table~\ref{tab:model_diag}}. Model Diagnostics: Classification Performance and Feature Importance (Continued)}
\begin{tabularx}{\textwidth}{c l Y c l Y}
\toprule
\multicolumn{6}{c}{\textbf{Panel F}: Top-10 Average Normalized Feature Importance} \\
\midrule
\multicolumn{3}{c}{\textbf{Classification}}& \multicolumn{3}{c}{\textbf{Regression}} \\
\cmidrule(lr){1-3}\cmidrule(lr){4-6}
Rank & Feature & Avg. Importance & Rank & Feature & Avg. Importance \\
\midrule
1 & sich2 & 0.2568 & 1 & sich2 & 0.2104 \\
2 & idiovol & 0.1298 & 2 & mve & 0.0952 \\
3 & retvol & 0.0606 & 3 & mom1m & 0.0828 \\
4 & baspread & 0.0581 & 4 & nincr & 0.0474 \\
5 & maxret & 0.0299 & 5 & dy & 0.0427 \\
6 & mve & 0.0229 & 6 & nanalyst & 0.0386 \\
7 & dy & 0.0225 & 7 & mom12m & 0.0326 \\
8 & mom12m & 0.0189 & 8 & chnanalyst & 0.0194 \\
9 & ep & 0.0181 & 9 & maxret & 0.0157 \\
10 & roaq & 0.0143 & 10 & agr & 0.0155 \\
\bottomrule
\end{tabularx}
\end{table}

\clearpage

\begin{table}[!htbp]
\small
\caption*{\textbf{Table~\ref{tab:model_diag}}. Model Diagnostics: Classification Performance and Feature Importance (Continued)}
\centering
\begin{tabularx}{\textwidth}{l Y Y Y Y Y Y}
\toprule
\multicolumn{7}{c}{\textbf{Panel G}: Fama-MacBeth Regression of Correctness Indicator on All 102 Characteristics} \\
\midrule
 & \multicolumn{2}{c}{\textbf{Hard Construction}} & \multicolumn{2}{c}{\textbf{Soft Construction}} & \multicolumn{2}{c}{\textbf{Regression}} \\
\cmidrule(lr){2-3}\cmidrule(lr){4-5}\cmidrule(lr){6-7}
Characteristic & Coef & $t$-stat & Coef & $t$-stat & Coef & $t$-stat \\
\midrule
absacc & 0.0021** & 2.752 & 0.0051** & 5.473 & 0.0015* & 2.159 \\
acc & -0.0100** & -7.468 & -0.0273** & -9.377 & -0.0051** & -4.839 \\
aeavol & -0.0016** & -2.879 & -0.0063** & -9.775 & -0.0001 & -0.121 \\
age & -0.0030** & -3.592 & 0.0149** & 7.389 & 0.0005 & 0.859 \\
agr & 0.0023* & 2.166 & 0.0041** & 3.013 & 0.0034** & 3.663 \\
baspread & 0.0228** & 10.760 & 0.0112** & 3.471 & 0.0090** & 7.164 \\
beta & -0.0442 & -1.534 & -0.0550 & -1.898 & -0.0069 & -0.514 \\
betasq & 0.0433 & 1.508 & 0.0576* & 2.014 & 0.0063 & 0.472 \\
bm & -0.0090** & -8.201 & -0.0255** & -14.374 & -0.0053** & -5.036 \\
bm\_ia & -0.0000 & -0.012 & -0.0005 & -0.262 & -0.0003 & -0.301 \\
cash & -0.0064** & -6.491 & -0.0086** & -6.969 & -0.0035** & -4.946 \\
cashdebt & -0.0052** & -4.047 & -0.0269** & -12.288 & -0.0056** & -6.009 \\
cashpr & -0.0005 & -0.555 & -0.0123** & -8.327 & -0.0002 & -0.259 \\
cfp & 0.0002 & 0.111 & -0.0042 & -1.820 & -0.0023* & -2.091 \\
cfp\_ia & -0.0064** & -4.997 & -0.0203** & -11.137 & -0.0021* & -2.281 \\
chatoia & 0.0006 & 0.828 & 0.0025** & 3.374 & 0.0001 & 0.249 \\
chcsho & -0.0016** & -2.908 & -0.0043** & -4.681 & -0.0007 & -1.382 \\
chempia & 0.0018 & 1.464 & 0.0084** & 5.708 & 0.0011 & 1.071 \\
chfeps & 0.0019** & 3.222 & 0.0062** & 8.740 & 0.0006 & 1.324 \\
chinv & 0.0013 & 1.592 & 0.0070** & 6.591 & 0.0002 & 0.202 \\
chmom & 0.0057** & 5.590 & 0.0076** & 5.507 & -0.0010 & -1.116 \\
chnanalyst & -0.0001 & -0.194 & -0.0016* & -2.173 & -0.0005 & -1.169 \\
chpmia & 0.0003 & 0.353 & -0.0005 & -0.380 & 0.0004 & 0.486 \\
chtx & 0.0031** & 6.478 & 0.0061** & 9.202 & 0.0018** & 3.779 \\
cinvest & -0.0015** & -3.422 & -0.0024** & -4.384 & -0.0007 & -1.609 \\
convind\_\_1.0 & 0.0020* & 2.352 & -0.0011 & -0.912 & -0.0007 & -0.842 \\
currat & -0.0124** & -6.559 & -0.0127** & -6.108 & -0.0045** & -3.409 \\
depr & -0.0008 & -1.141 & -0.0001 & -0.083 & -0.0014** & -2.815 \\
disp & -0.0139** & -12.728 & -0.0498** & -15.328 & -0.0074** & -9.092 \\
divi\_\_1.0 & -0.0072** & -4.697 & -0.0316** & -16.146 & -0.0035** & -2.612 \\
\bottomrule
\end{tabularx}
\end{table}

\clearpage

\begin{table}[!htbp]
\small
\caption*{\textbf{Table~\ref{tab:model_diag}}. Model Diagnostics: Classification Performance and Feature Importance (Continued)}
\centering
\begin{tabularx}{\textwidth}{l Y Y Y Y Y Y}
\toprule
 & \multicolumn{2}{c}{\textbf{Hard Construction}} & \multicolumn{2}{c}{\textbf{Soft Construction}} & \multicolumn{2}{c}{\textbf{Regression}} \\
\cmidrule(lr){2-3}\cmidrule(lr){4-5}\cmidrule(lr){6-7}
Characteristic & Coef & $t$-stat & Coef & $t$-stat & Coef & $t$-stat \\
\midrule
divo\_\_1.0 & 0.0007 & 0.477 & 0.0046** & 3.014 & -0.0012 & -0.967 \\
dolvol & 0.0247** & 6.582 & 0.0597** & 7.884 & 0.0035 & 0.943 \\
dy & 0.0112** & 12.175 & 0.0436** & 25.753 & 0.0047** & 6.854 \\
ear & 0.0010* & 1.960 & 0.0020** & 3.021 & -0.0004 & -0.911 \\
egr & -0.0012 & -1.900 & -0.0053** & -5.621 & -0.0018** & -2.854 \\
ep & -0.0036** & -2.729 & 0.0038 & 1.357 & -0.0004 & -0.416 \\
fgr5yr & -0.0058** & -6.320 & -0.0335** & -11.905 & -0.0040** & -4.172 \\
gma & -0.0026* & -2.274 & -0.0096** & -5.453 & -0.0012 & -1.141 \\
grcapx & -0.0014* & -2.359 & -0.0057** & -7.641 & -0.0014* & -2.421 \\
grltnoa & 0.0005 & 0.714 & 0.0039** & 4.023 & -0.0007 & -1.159 \\
herf & -0.0226 & -1.496 & -0.0111 & -0.397 & -0.0261* & -2.227 \\
hire & -0.0036** & -3.253 & -0.0104** & -8.985 & -0.0016 & -1.715 \\
idiovol & 0.0013 & 0.797 & -0.0612** & -18.845 & -0.0024 & -1.940 \\
ill & -0.0050 & -1.229 & -0.0644** & -9.869 & -0.0000 & -0.017 \\
indmom & -0.0005 & -0.269 & 0.0165** & 4.121 & -0.0017 & -1.028 \\
invest & 0.0009 & 1.081 & 0.0016 & 1.358 & 0.0024** & 3.361 \\
ipo\_\_1.0 & -0.0031 & -1.623 & -0.0764** & -33.242 & -0.0052** & -3.597 \\
lev & 0.0041* & 2.092 & 0.0105** & 3.554 & -0.0038** & -2.958 \\
lgr & -0.0013 & -1.622 & -0.0017 & -1.615 & -0.0010 & -1.562 \\
maxret & 0.0001 & 0.094 & -0.0128** & -6.012 & 0.0000 & 0.022 \\
mom12m & -0.0016 & -1.416 & -0.0038* & -2.023 & -0.0051** & -4.974 \\
mom1m & 0.0027** & 3.271 & 0.0180** & 13.548 & 0.0043** & 2.810 \\
mom36m & -0.0043** & -5.174 & -0.0080** & -6.579 & -0.0009 & -1.609 \\
mom6m & -0.0126** & -11.473 & -0.0276** & -18.785 & -0.0069** & -7.730 \\
ms & 0.0012 & 1.705 & 0.0110** & 8.768 & 0.0014* & 2.491 \\
mve & -0.0462** & -11.728 & -0.1708** & -18.353 & -0.0142** & -5.174 \\
mve\_ia & 0.0164** & 11.538 & 0.0802** & 23.374 & 0.0072** & 6.690 \\
nanalyst & -0.0119** & -8.635 & -0.0043 & -1.642 & -0.0032** & -3.091 \\
nincr & 0.0029** & 5.495 & 0.0078** & 11.788 & 0.0003 & 0.619 \\
operprof & -0.0029** & -3.512 & -0.0069** & -5.422 & 0.0000 & 0.033 \\
\bottomrule
\end{tabularx}
\end{table}

\clearpage

\begin{table}[!htbp]
\small
\caption*{\textbf{Table~\ref{tab:model_diag}}. Model Diagnostics: Classification Performance and Feature Importance (Continued)}
\centering
\begin{tabularx}{\textwidth}{l Y Y Y Y Y Y}
\toprule
 & \multicolumn{2}{c}{\textbf{Hard Construction}} & \multicolumn{2}{c}{\textbf{Soft Construction}} & \multicolumn{2}{c}{\textbf{Regression}} \\
\cmidrule(lr){2-3}\cmidrule(lr){4-5}\cmidrule(lr){6-7}
Characteristic & Coef & $t$-stat & Coef & $t$-stat & Coef & $t$-stat \\
\midrule
orgcap & 0.0018 & 1.871 & 0.0041* & 2.451 & 0.0010 & 1.494 \\
pchcapx\_ia & 0.0026** & 3.236 & 0.0018 & 1.871 & 0.0022** & 3.653 \\
pchcurrat & -0.0012 & -0.831 & -0.0039* & -2.018 & -0.0013 & -0.893 \\
pchdepr & 0.0006 & 1.096 & 0.0014 & 1.851 & 0.0011* & 2.277 \\
pchgm\_pchsale & -0.0016** & -3.109 & -0.0035** & -5.716 & -0.0017** & -4.177 \\
pchquick & 0.0012 & 0.796 & 0.0060** & 3.088 & 0.0014 & 0.966 \\
pchsale\_pchinvt & 0.0017 & 0.699 & 0.0150** & 5.326 & -0.0010 & -0.458 \\
pchsale\_pchrect & 0.0002 & 0.357 & 0.0016* & 2.227 & 0.0005 & 1.154 \\
pchsale\_pchxsga & -0.0001 & -0.267 & 0.0001 & 0.168 & 0.0002 & 0.383 \\
pchsaleinv & -0.0009 & -0.358 & -0.0077* & -2.478 & 0.0008 & 0.368 \\
pctacc & 0.0117** & 6.565 & 0.0357** & 7.426 & 0.0056** & 5.420 \\
pricedelay & 0.0016** & 3.206 & 0.0039** & 5.214 & 0.0003 & 0.827 \\
ps & 0.0036** & 4.608 & 0.0147** & 12.930 & 0.0039** & 5.779 \\
quick & 0.0007 & 0.370 & -0.0169** & -6.483 & -0.0008 & -0.650 \\
rd\_\_1.0 & -0.0042** & -5.046 & -0.0063** & -5.739 & -0.0034** & -4.806 \\
rd\_mve & -0.0037** & -2.694 & -0.0034 & -1.863 & -0.0028* & -2.525 \\
rd\_sale & 0.0014 & 1.040 & 0.0002 & 0.081 & -0.0021 & -1.721 \\
realestate & -0.0025** & -2.985 & -0.0060** & -5.809 & -0.0016* & -2.511 \\
retvol & 0.0087** & 4.113 & 0.0245** & 6.263 & 0.0072** & 3.923 \\
roaq & -0.0104** & -9.790 & -0.0349** & -25.392 & -0.0053** & -4.615 \\
roavol & -0.0042** & -4.516 & -0.0232** & -13.252 & -0.0036** & -4.624 \\
roeq & 0.0019 & 1.944 & 0.0088** & 8.455 & -0.0006 & -0.823 \\
roic & -0.0018* & -2.070 & -0.0155** & -12.692 & -0.0014 & -1.889 \\
rsup & -0.0027** & -4.896 & -0.0087** & -13.988 & -0.0019** & -3.522 \\
salecash & -0.0050** & -4.039 & -0.0065** & -3.326 & -0.0040** & -3.987 \\
saleinv & -0.0016* & -2.377 & 0.0120** & 11.405 & -0.0017** & -2.763 \\
salerec & -0.0021** & -2.895 & -0.0055** & -5.605 & -0.0015** & -2.633 \\
secured & -0.0032** & -3.952 & -0.0041** & -2.950 & -0.0027** & -4.354 \\
securedind\_\_1.0 & -0.0021** & -2.641 & -0.0057** & -5.068 & 0.0011 & 1.625 \\
sfe & -0.0060** & -7.124 & -0.0192** & -12.696 & -0.0006 & -0.830 \\
\bottomrule
\end{tabularx}
\end{table}

\clearpage

\begin{table}[!htbp]
\small
\caption*{\textbf{Table~\ref{tab:model_diag}}. Model Diagnostics: Classification Performance and Feature Importance (Continued)}
\centering
\begin{tabularx}{\textwidth}{l Y Y Y Y Y Y}
\toprule
 & \multicolumn{2}{c}{\textbf{Hard Construction}} & \multicolumn{2}{c}{\textbf{Soft Construction}} & \multicolumn{2}{c}{\textbf{Regression}} \\
\cmidrule(lr){2-3}\cmidrule(lr){4-5}\cmidrule(lr){6-7}
Characteristic & Coef & $t$-stat & Coef & $t$-stat & Coef & $t$-stat \\
\midrule
sgr & -0.0013 & -1.470 & -0.0091** & -7.746 & -0.0024** & -3.324 \\
sin\_\_1.0 & -0.0012 & -0.313 & 0.0165** & 4.261 & 0.0035 & 1.075 \\
sp & -0.0085** & -6.446 & -0.0512** & -22.869 & -0.0021 & -1.816 \\
std\_dolvol & 0.0041** & 3.365 & 0.0102** & 5.046 & 0.0038** & 3.208 \\
std\_turn & -0.0113** & -7.374 & -0.0316** & -12.481 & -0.0069** & -3.992 \\
stdacc & 0.0016 & 0.796 & 0.0028 & 1.214 & 0.0000 & 0.016 \\
stdcf & -0.0041 & -1.789 & -0.0087** & -2.713 & -0.0004 & -0.290 \\
sue & -0.0030** & -5.397 & -0.0076** & -11.072 & -0.0018** & -3.493 \\
tang & -0.0023** & -3.655 & -0.0057** & -5.039 & -0.0028** & -3.634 \\
tb & -0.0016** & -3.057 & -0.0029** & -3.498 & -0.0008 & -1.523 \\
turn & 0.0044* & 2.068 & -0.0207** & -5.249 & 0.0018 & 1.125 \\
zerotrade & -0.0139** & -6.204 & -0.0327** & -9.629 & -0.0065** & -3.304 \\
\midrule
Avg.\ Adj.\ $R^2$ & 0.0213 &  & 0.1096 &  & 0.0081 &  \\
Avg.\ \# Stocks & 5308 &                                & 5308 &                                & 5308 &  \\
\# Windows & 498 &                                & 498 &                                & 498 &  \\
\bottomrule
\end{tabularx}
\end{table}

\clearpage


\newpage
\begin{table}[!htbp]
\footnotesize
\setlength{\tabcolsep}{3pt}
\justifying{\noindent This table reports the average standardized value of each firm characteristic in the classification and regression stacking long and short portfolios and how it differs between the two approaches. Before portfolio construction, every characteristic is standardized cross-sectionally by month across all stocks to zero mean and unit standard deviation.
Diff. Long is the classification-minus-regression difference in the long portfolio. Diff. Short
is the classification-minus-regression difference in the short portfolio. Newey-West $t$-statistics adjusted with 6 lags test whether the mean monthly difference is zero. Statistical significance: $^{**}p{<}0.01$, $^{*}p{<}0.05$.}

\pdfdest name {osm-figs-tabs-tab6} fit
\bookmark[level=2,dest=osm-figs-tabs-tab6]{Table OSM6: Portfolio Characteristic Profile: Classification versus Regression}
\caption{Portfolio Characteristic Profile: Classification versus Regression}
\centering
\label{tab:char_profile}
\begin{tabularx}{\textwidth}{l Y Y Y Y Y Y Y Y}
\toprule
 & \multicolumn{2}{c}{\textbf{Classification}} & \multicolumn{2}{c}{\textbf{Regression}} & \multicolumn{4}{c}{\textbf{Difference}} \\
\cmidrule(lr){2-3}\cmidrule(lr){4-5}\cmidrule(lr){6-9}
Char. & Long & Short & Long & Short & Diff. Long & $t$-stat & Diff. Short & $t$-stat \\
\midrule
{absacc} & 0.664 & 0.827 & 0.248 & 0.346 & 0.415** & (23.57) & 0.481** & (16.55) \\
{acc} & -0.483 & -0.549 & -0.186 & -0.140 & -0.298** & (-14.99) & -0.409** & (-15.84) \\
{aeavol} & 0.060 & -0.002 & 0.109 & -0.012 & -0.049** & (-5.03) & 0.010 & (0.99) \\
{age} & -0.505 & -0.574 & -0.205 & -0.371 & -0.301** & (-16.10) & -0.203** & (-12.14) \\
{agr} & -0.025 & 0.292 & -0.192 & 0.450 & 0.166** & (8.52) & -0.158** & (-4.60) \\
{baspread} & 1.605 & 1.689 & 0.810 & 0.573 & 0.795** & (24.94) & 1.117** & (21.76) \\
{beta} & 0.448 & 0.554 & 0.092 & 0.293 & 0.356** & (12.06) & 0.261** & (8.75) \\
{betasq} & 0.544 & 0.683 & 0.125 & 0.340 & 0.419** & (15.39) & 0.343** & (11.93) \\
{bm} & 0.048 & -0.216 & 0.325 & -0.220 & -0.277** & (-13.89) & 0.005 & (0.17) \\
{bm\_ia} & 0.022 & -0.006 & 0.023 & 0.018 & -0.002 & (-0.13) & -0.024* & (-2.56) \\
{cash} & 0.478 & 0.452 & 0.210 & 0.167 & 0.268** & (9.71) & 0.285** & (8.16) \\
{cashdebt} & -0.652 & -0.792 & -0.213 & -0.292 & -0.439** & (-26.00) & -0.500** & (-17.66) \\
{cashpr} & 0.010 & 0.166 & -0.101 & 0.159 & 0.111** & (9.64) & 0.008 & (0.39) \\
{cfp} & -0.768 & -0.797 & -0.281 & -0.254 & -0.486** & (-16.93) & -0.543** & (-14.65) \\
{cfp\_ia} & -0.067 & -0.082 & -0.005 & -0.047 & -0.062** & (-3.73) & -0.035 & (-1.62) \\
{chatoia} & 0.029 & -0.026 & 0.049 & -0.083 & -0.019* & (-2.16) & 0.058** & (4.52) \\
{chcsho} & 0.114 & 0.267 & -0.081 & 0.280 & 0.195** & (13.25) & -0.013 & (-1.09) \\
{chempia} & -0.087 & 0.012 & -0.108 & 0.081 & 0.021 & (1.94) & -0.069** & (-4.17) \\
{chfeps} & 0.023 & -0.150 & 0.029 & -0.202 & -0.007 & (-0.24) & 0.051 & (1.55) \\
{chinv} & -0.181 & -0.006 & -0.178 & 0.196 & -0.002 & (-0.24) & -0.201** & (-14.33) \\
{chmom} & -0.171 & 0.123 & -0.343 & 0.199 & 0.172** & (9.14) & -0.077** & (-3.35) \\
{chnanalyst} & -0.059 & -0.068 & -0.052 & -0.016 & -0.008 & (-0.66) & -0.051** & (-4.99) \\
{chpmia} & 0.054 & 0.071 & 0.023 & 0.038 & 0.030* & (2.18) & 0.034 & (1.65) \\
{chtx} & 0.002 & -0.095 & 0.073 & -0.175 & -0.071** & (-8.17) & 0.080** & (6.48) \\
{cinvest} & -0.070 & -0.090 & -0.024 & -0.042 & -0.046** & (-4.75) & -0.048** & (-3.98) \\
{convind} & 0.127 & 0.145 & 0.099 & 0.138 & 0.029** & (12.23) & 0.007** & (3.01) \\
{currat} & 0.055 & 0.053 & 0.043 & 0.085 & 0.011 & (0.99) & -0.031* & (-2.54) \\
{depr} & 0.455 & 0.447 & 0.242 & 0.124 & 0.213** & (13.57) & 0.323** & (16.92) \\
{disp} & 0.476 & 0.507 & 0.251 & 0.352 & 0.221** & (4.22) & 0.155** & (3.57) \\
{divi} & 0.036 & 0.045 & 0.036 & 0.048 & 0.000 & (0.11) & -0.002 & (-1.92) \\
{divo} & 0.039 & 0.040 & 0.037 & 0.039 & 0.002 & (1.43) & 0.000 & (0.29) \\
{dolvol} & -0.784 & -0.653 & -0.699 & -0.239 & -0.086* & (-2.06) & -0.415** & (-11.07) \\
{dy} & -0.206 & -0.174 & -0.112 & -0.133 & -0.094** & (-5.16) & -0.041* & (-2.06) \\
{ear} & -0.109 & -0.274 & 0.031 & -0.234 & -0.140** & (-12.34) & -0.041** & (-3.37) \\
\bottomrule
\end{tabularx}
\end{table}
\clearpage

\begin{table}[!htbp]
\footnotesize
\caption*{\textbf{Table~\ref{tab:char_profile}}. Portfolio Characteristic Profile: Classification versus Regression (Continued)}
\setlength{\tabcolsep}{3pt}
\begin{tabularx}{\textwidth}{l Y Y Y Y Y Y Y Y}
\toprule
 & \multicolumn{2}{c}{\textbf{Classification}} & \multicolumn{2}{c}{\textbf{Regression}} & \multicolumn{4}{c}{\textbf{Difference}} \\
\cmidrule(lr){2-3}\cmidrule(lr){4-5}\cmidrule(lr){6-9}
Char. & Long & Short & Long & Short & Diff. Long & $t$-stat & Diff. Short & $t$-stat \\
\midrule
{egr} & -0.181 & -0.086 & -0.116 & 0.095 & -0.065** & (-7.48) & -0.182** & (-9.44) \\
{ep} & -1.208 & -1.136 & -0.558 & -0.256 & -0.650** & (-24.28) & -0.881** & (-23.40) \\
{fgr5yr} & 0.496 & 0.586 & 0.271 & 0.328 & 0.221** & (4.02) & 0.256** & (4.49) \\
{gma} & -0.354 & -0.373 & -0.091 & -0.046 & -0.263** & (-13.71) & -0.326** & (-15.60) \\
{grcapx} & 0.135 & 0.303 & -0.008 & 0.271 & 0.143** & (9.69) & 0.031 & (1.89) \\
{grltnoa} & -0.160 & 0.042 & -0.213 & 0.267 & 0.053** & (3.60) & -0.225** & (-8.58) \\
{herf} & -0.077 & -0.052 & -0.028 & 0.053 & -0.049** & (-4.17) & -0.105** & (-7.18) \\
{hire} & -0.091 & 0.101 & -0.167 & 0.256 & 0.076** & (5.19) & -0.155** & (-6.76) \\
{idiovol} & 1.530 & 1.674 & 0.734 & 0.601 & 0.795** & (22.53) & 1.073** & (24.74) \\
{ill} & 0.608 & 0.482 & 0.543 & 0.093 & 0.065 & (1.55) & 0.390** & (8.00) \\
{indmom} & -0.104 & -0.236 & 0.070 & -0.324 & -0.173** & (-6.69) & 0.088** & (3.46) \\
{invest} & -0.087 & 0.187 & -0.204 & 0.392 & 0.117** & (7.32) & -0.205** & (-7.45) \\
{ipo} & 0.063 & 0.087 & 0.025 & 0.119 & 0.038** & (10.67) & -0.031** & (-10.39) \\
{lev} & -0.049 & -0.086 & 0.039 & -0.105 & -0.087** & (-4.08) & 0.019 & (0.80) \\
{lgr} & 0.091 & 0.316 & -0.095 & 0.343 & 0.186** & (15.85) & -0.027 & (-1.22) \\
{maxret} & 1.157 & 1.649 & 0.390 & 0.802 & 0.767** & (31.14) & 0.847** & (17.77) \\
{mom12m} & -0.473 & -0.779 & -0.093 & -0.615 & -0.380** & (-9.45) & -0.163** & (-4.62) \\
{mom1m} & -0.698 & 0.135 & -0.798 & 0.582 & 0.100** & (3.82) & -0.447** & (-17.42) \\
{mom36m} & -0.723 & -0.570 & -0.415 & 0.003 & -0.309** & (-11.04) & -0.572** & (-17.20) \\
{mom6m} & -0.461 & -0.640 & -0.197 & -0.486 & -0.264** & (-7.99) & -0.154** & (-4.80) \\
{ms} & -0.522 & -0.512 & -0.324 & -0.218 & -0.198** & (-11.54) & -0.294** & (-13.86) \\
{mve} & -1.149 & -0.998 & -0.879 & -0.309 & -0.270** & (-6.68) & -0.689** & (-19.70) \\
{mve\_ia} & -0.271 & -0.271 & -0.220 & -0.143 & -0.051** & (-4.72) & -0.128** & (-9.09) \\
{nanalyst} & -0.541 & -0.545 & -0.468 & -0.226 & -0.073** & (-2.62) & -0.319** & (-11.43) \\
{nincr} & -0.059 & -0.195 & 0.103 & -0.248 & -0.162** & (-8.43) & 0.053** & (4.21) \\
{operprof} & -0.187 & -0.211 & -0.042 & -0.050 & -0.145** & (-15.55) & -0.161** & (-16.61) \\
{orgcap} & 0.370 & 0.249 & 0.339 & -0.037 & 0.031 & (1.24) & 0.286** & (9.54) \\
{pchcapx\_ia} & 0.038 & 0.094 & -0.015 & 0.072 & 0.053** & (3.91) & 0.021 & (1.30) \\
{pchcurrat} & 0.131 & 0.267 & -0.004 & 0.231 & 0.135** & (8.34) & 0.036 & (1.74) \\
{pchdepr} & 0.374 & 0.402 & 0.151 & 0.111 & 0.223** & (15.69) & 0.290** & (19.61) \\
{pchgm\_pchsale} & -0.274 & -0.385 & -0.060 & -0.181 & -0.213** & (-21.98) & -0.205** & (-14.30) \\
{pchquick} & 0.139 & 0.274 & 0.000 & 0.228 & 0.139** & (8.77) & 0.047* & (2.25) \\
{pchsale\_pchinvt} & 0.066 & -0.023 & 0.065 & -0.106 & 0.002 & (0.19) & 0.083** & (6.26) \\
{pchsale\_pchrect} & -0.020 & -0.071 & 0.011 & -0.088 & -0.031** & (-2.91) & 0.017 & (1.30) \\
\bottomrule
\end{tabularx}
\end{table}
\clearpage

\begin{table}[!htbp]
\footnotesize
\caption*{\textbf{Table~\ref{tab:char_profile}}. Portfolio Characteristic Profile: Classification versus Regression (Continued)}
\setlength{\tabcolsep}{3pt}
\begin{tabularx}{\textwidth}{l Y Y Y Y Y Y Y Y}
\toprule
 & \multicolumn{2}{c}{\textbf{Classification}} & \multicolumn{2}{c}{\textbf{Regression}} & \multicolumn{4}{c}{\textbf{Difference}} \\
\cmidrule(lr){2-3}\cmidrule(lr){4-5}\cmidrule(lr){6-9}
Char. & Long & Short & Long & Short & Diff. Long & $t$-stat & Diff. Short & $t$-stat \\
\midrule
{pchsale\_pchxsga} & 0.105 & 0.193 & -0.011 & 0.126 & 0.117** & (9.56) & 0.067** & (3.61) \\
{pchsaleinv} & 0.224 & 0.221 & 0.090 & 0.021 & 0.135** & (9.43) & 0.200** & (13.89) \\
{pctacc} & 0.115 & 0.152 & 0.028 & 0.068 & 0.087** & (6.38) & 0.084** & (5.85) \\
{pricedelay} & 0.137 & 0.146 & 0.100 & 0.046 & 0.037** & (2.95) & 0.099** & (8.82) \\
{ps} & -0.619 & -0.666 & -0.258 & -0.280 & -0.361** & (-20.60) & -0.385** & (-14.39) \\
{quick} & 0.092 & 0.093 & 0.043 & 0.100 & 0.049** & (3.81) & -0.007 & (-0.49) \\
{rd} & 0.259 & 0.214 & 0.224 & 0.107 & 0.034** & (8.29) & 0.107** & (14.67) \\
{rd\_mve} & 0.789 & 0.504 & 0.562 & -0.127 & 0.227** & (8.51) & 0.631** & (13.87) \\
{rd\_sale} & 0.393 & 0.480 & 0.094 & 0.154 & 0.299** & (17.14) & 0.326** & (12.88) \\
{realestate} & -0.172 & -0.214 & -0.063 & -0.089 & -0.109** & (-7.08) & -0.125** & (-8.96) \\
{retvol} & 1.515 & 1.787 & 0.689 & 0.715 & 0.827** & (33.24) & 1.072** & (22.15) \\
{roaq} & -1.032 & -1.420 & -0.294 & -0.582 & -0.738** & (-27.63) & -0.837** & (-19.84) \\
{roavol} & 1.027 & 1.288 & 0.355 & 0.487 & 0.672** & (17.45) & 0.801** & (16.32) \\
{roeq} & -0.641 & -0.931 & -0.177 & -0.432 & -0.463** & (-23.39) & -0.499** & (-15.40) \\
{roic} & -0.785 & -0.869 & -0.262 & -0.265 & -0.523** & (-23.89) & -0.604** & (-20.44) \\
{rsup} & -0.305 & -0.356 & -0.128 & -0.124 & -0.176** & (-14.26) & -0.232** & (-14.55) \\
{salecash} & -0.077 & -0.087 & -0.023 & -0.012 & -0.054** & (-6.42) & -0.075** & (-9.41) \\
{saleinv} & -0.062 & -0.079 & -0.040 & -0.032 & -0.022** & (-3.22) & -0.047** & (-5.40) \\
{salerec} & -0.001 & -0.010 & 0.010 & 0.009 & -0.011 & (-1.48) & -0.019 & (-1.64) \\
{secured} & 0.348 & 0.317 & 0.283 & 0.170 & 0.064** & (3.43) & 0.148** & (4.17) \\
{securedind} & 0.515 & 0.511 & 0.515 & 0.526 & 0.001 & (0.09) & -0.015 & (-1.19) \\
{sfe} & -0.884 & -0.984 & -0.518 & -0.135 & -0.366** & (-6.30) & -0.850** & (-20.78) \\
{sgr} & 0.154 & 0.378 & -0.060 & 0.330 & 0.214** & (12.27) & 0.048* & (2.16) \\
{sin} & 0.004 & 0.005 & 0.006 & 0.007 & -0.002** & (-4.96) & -0.002** & (-5.82) \\
{sp} & 0.155 & -0.020 & 0.291 & -0.109 & -0.136** & (-6.57) & 0.089** & (4.36) \\
{std\_dolvol} & 0.495 & 0.478 & 0.446 & 0.188 & 0.049 & (1.65) & 0.291** & (13.30) \\
{std\_turn} & 0.450 & 0.600 & 0.143 & 0.208 & 0.307** & (9.39) & 0.392** & (7.68) \\
{stdacc} & 0.368 & 0.466 & 0.102 & 0.201 & 0.266** & (15.91) & 0.265** & (13.43) \\
{stdcf} & 0.444 & 0.559 & 0.114 & 0.229 & 0.330** & (17.64) & 0.330** & (13.14) \\
{sue} & -0.034 & -0.265 & 0.053 & -0.214 & -0.087** & (-6.00) & -0.051* & (-2.40) \\
{tang} & 0.302 & 0.254 & 0.199 & 0.112 & 0.102** & (4.57) & 0.142** & (4.77) \\
{tb} & -0.280 & -0.313 & -0.119 & -0.134 & -0.162** & (-12.88) & -0.179** & (-10.28) \\
{turn} & 0.283 & 0.509 & -0.057 & 0.220 & 0.340** & (8.79) & 0.289** & (5.26) \\
{zerotrade} & 0.073 & -0.015 & 0.194 & 0.004 & -0.120** & (-4.38) & -0.019 & (-0.72) \\
\bottomrule
\end{tabularx}
\end{table}
\clearpage

\newpage
\begin{table}[!htbp]
\footnotesize
\justifying{\noindent This table reports out-of-sample long-short portfolio performance for regression models reestimated with relative prediction targets over July 1983--December 2024 (498 months). The prediction target is the stock return in excess of the market in Panel A and in excess of the two-digit Standard Industrial Classification (SIC) industry in Panel B, rather than the raw excess return used in Table 1 of the main paper. For each model, stocks are assigned monthly by the predicted relative return into 10 deciles. The long (short) portfolio holds the top (bottom) decile. The columns report the performance of the long-short portfolios based on the stacking ensemble and its gradient boosted trees (GBT), random forest (RF), and neural network (NN) base models, alongside ordinary least squares (OLS). Average Return is the annualized average raw return in decimal. Standard Deviation is the annualized standard deviation in decimal. Sharpe Ratio is the annualized Sharpe ratio. $\alpha$ is the monthly abnormal return in decimal from the corresponding factor model. The factor models are the Capital Asset Pricing Model (CAPM) and the Fama-French three-, five-, and six-factor (FF3F, FF5F, and FF6F) models. Statistical significance is based on Newey-West $t$-statistics adjusted with 6 lags: $^{**}p{<}0.01$, $^{*}p{<}0.05$.}

\pdfdest name {osm-figs-tabs-tab7} fit
\bookmark[level=2,dest=osm-figs-tabs-tab7]{Table OSM7: Alternative Return Definitions for Machine Learning Regressions}
\caption{Alternative Return Definitions for Machine Learning Regressions}
\label{tab:relative_target}
\centering
\setlength{\tabcolsep}{3pt}
\begin{tabularx}{\textwidth}{l *{10}{>{\centering\arraybackslash}X}}
\toprule
\multicolumn{11}{c}{\textbf{Panel A}: Stock Return Relative to Market Return} \\
\midrule
 & \multicolumn{5}{c}{\textbf{Value-Weighted}} & \multicolumn{5}{c}{\textbf{Equal-Weighted}} \\
\cmidrule(lr){2-6}\cmidrule(lr){7-11}
Metric & Stack & GBT & RF & NN & OLS & Stack & GBT & RF & NN & OLS \\
\midrule
Average Return     & 0.316 & 0.178 & 0.187 & 0.331 & 0.158 & 0.564 & 0.427 & 0.387 & 0.540 & 0.389 \\
Standard Deviation & 0.209 & 0.198 & 0.207 & 0.206 & 0.159 & 0.174 & 0.158 & 0.159 & 0.165 & 0.152 \\
Sharpe Ratio       & 1.513 & 0.899 & 0.904 & 1.608 & 0.995 & 3.234 & 2.703 & 2.436 & 3.269 & 2.552 \\
CAPM $\alpha$      & 0.027** & 0.015** & 0.016** & 0.028** & 0.013** & 0.047** & 0.035** & 0.033** & 0.045** & 0.032** \\
FF3F $\alpha$      & 0.026** & 0.013** & 0.014** & 0.028** & 0.013** & 0.046** & 0.034** & 0.031** & 0.044** & 0.032** \\
FF5F $\alpha$      & 0.022** & 0.011** & 0.009** & 0.026** & 0.013** & 0.044** & 0.033** & 0.030** & 0.044** & 0.032** \\
FF6F $\alpha$      & 0.020** & 0.008** & 0.008** & 0.026** & 0.010** & 0.045** & 0.033** & 0.030** & 0.044** & 0.031** \\
\midrule
\multicolumn{11}{c}{\textbf{Panel B}: Stock Return Relative to 2-digit SIC Return} \\
\midrule
Average Return     & 0.348 & 0.181 & 0.171 & 0.370 & 0.150 & 0.560 & 0.444 & 0.386 & 0.555 & 0.388 \\
Standard Deviation & 0.215 & 0.199 & 0.207 & 0.193 & 0.154 & 0.169 & 0.161 & 0.166 & 0.163 & 0.150 \\
Sharpe Ratio       & 1.620 & 0.910 & 0.824 & 1.912 & 0.970 & 3.317 & 2.763 & 2.332 & 3.406 & 2.583 \\
CAPM $\alpha$      & 0.030** & 0.016** & 0.014** & 0.031** & 0.012** & 0.047** & 0.037** & 0.032** & 0.046** & 0.032** \\
FF3F $\alpha$      & 0.028** & 0.014** & 0.012** & 0.030** & 0.012** & 0.046** & 0.036** & 0.031** & 0.046** & 0.031** \\
FF5F $\alpha$      & 0.024** & 0.010** & 0.008** & 0.028** & 0.011** & 0.044** & 0.034** & 0.029** & 0.045** & 0.030** \\
FF6F $\alpha$      & 0.022** & 0.009** & 0.007** & 0.027** & 0.010** & 0.045** & 0.035** & 0.030** & 0.045** & 0.031** \\
\bottomrule
\end{tabularx}
\end{table}
\clearpage

\end{document}